\documentclass[aps,
twocolumn,superscriptaddress,floatfix,
]{revtex4-1}
\usepackage{amsmath,mathtools,amsthm,amssymb,pifont}
\usepackage[utf8]{inputenc}
\usepackage[T1]{fontenc}
\usepackage[american]{babel}
\usepackage{graphicx,bbold,titlesec}
\usepackage{braket}
\usepackage{float}
\usepackage{color}
\usepackage{MnSymbol}
\usepackage{hyperref}
\usepackage[dvipsnames]{xcolor}
\definecolor{BlueRome}{HTML}{4287f5}
\definecolor{C1}{RGB}{52, 89, 149}
\definecolor{C2}{RGB}{251, 77, 61}
\definecolor{C3}{RGB}{3, 206, 164}
\definecolor{C4}{RGB}{202, 21, 81}
\hypersetup{colorlinks=true, linkcolor=BlueRome, citecolor=BlueRome, urlcolor=BlueRome}

\usepackage[normalem]{ulem}

\newcommand*{\Mpop}{\mathcal{M}}

\usepackage{epstopdf}
\usepackage{tikz}
\usepackage[export]{adjustbox}
\usepackage{thmtools}
\usepackage{thm-restate}
\usepackage{enumerate}

\usepackage{accents}

\usepackage{tikzit}

\tikzstyle{dot}=[inner sep=0.3mm, minimum width=2mm, minimum height=2mm, draw, shape=circle, font={\footnotesize}, tikzit fill=magenta]
\tikzstyle{white dot}=[dot, fill=white, text depth=-0.2mm, tikzit category=ZH-pf, draw=black]
\tikzstyle{white phase dot}=[minimum size=5mm, font={\footnotesize\boldmath}, shape=rectangle, rounded corners=2mm, inner sep=0.2mm, outer sep=-2mm, scale=0.8, tikzit shape=circle, draw=black, fill=white, tikzit category=ZH-pf, tikzit fill=white, tikzit draw=blue]
\tikzstyle{gray dot}=[dot, fill={rgb,255: red,180; green,180; blue,180}, text depth=-0.2mm, tikzit category=ZH-pf]
\tikzstyle{gray phase dot}=[white phase dot, tikzit shape=circle, tikzit draw=blue, fill={rgb,255: red,180; green,180; blue,180}, font={\footnotesize\boldmath}]
\tikzstyle{hadamard}=[fill=white, draw, inner sep=0.6mm, minimum height=1.5mm, minimum width=1.5mm, shape=rectangle, tikzit shape=rectangle, tikzit category=ZH-pf]
\tikzstyle{small hadamard}=[hadamard]
\tikzstyle{lambda}=[hadamard, fill={rgb,255: red,180; green,180; blue,180}, tikzit shape=rectangle]
\tikzstyle{halfscalar}=[star, fill=black, draw=black, minimum size=8pt, inner sep=0pt]
\tikzstyle{box}=[shape=rectangle, text height=1.5ex, text depth=0.25ex, yshift=0.2mm, fill=white, draw=black, minimum height=3mm, minimum width=5mm, font={\small}]
\tikzstyle{Z dot}=[inner sep=0mm, minimum size=2mm, shape=circle, draw=black, fill={zx_green}, tikzit fill=green]
\tikzstyle{Z phase dot}=[minimum size=5mm, font={\footnotesize\boldmath}, shape=rectangle, rounded corners=2mm, inner sep=0.2mm, outer sep=-2mm, scale=0.8, tikzit shape=circle, draw=black, fill={zx_green}, tikzit draw=blue, tikzit fill=green]
\tikzstyle{X dot}=[Z dot, shape=circle, draw=black, fill={zx_red}, tikzit fill=red]
\tikzstyle{X phase dot}=[Z phase dot, tikzit shape=circle, tikzit draw=blue, fill={zx_red}, font={\footnotesize\color{black}\boldmath}, tikzit fill=red]
\tikzstyle{H box}=[hadamard]
\tikzstyle{st}=[star, star points=5, fill=white, draw=black, inner sep=1.2pt, line width=1.2pt, tikzit fill=blue, tikzit draw=red, tikzit category=ZH-pf]
\tikzstyle{triangle}=[regular polygon, regular polygon sides=3, fill=white, draw=black, inner sep=0pt, minimum width=1em, tikzit draw=blue, tikzit category=ZH-pf, tikzit fill=cyan]
\tikzstyle{not}=[fill={rgb,255: red,180; green,180; blue,180}, draw=black, shape=circle, font={$\neg$}, dot]
\tikzstyle{vertex}=[inner sep=0mm, minimum size=1mm, shape=circle, draw=black, fill=black]
\tikzstyle{vertex set}=[inner sep=0mm, minimum size=1mm, shape=circle, draw=black, fill=white, font={\footnotesize\boldmath}]
\tikzstyle{wide point}=[fill=white, draw, shape=isosceles triangle, shape border rotate=-90, isosceles triangle stretches=true, inner sep=0pt, minimum width=1.5cm, minimum height=6.12mm, yshift=-0.0mm]
\tikzstyle{medium gray box}=[semilarge box, fill={rgb,255: red,180; green,180; blue,180}]
\tikzstyle{small box}=[rectangle, inline text, fill=white, draw, minimum height=5mm, yshift=-0.5mm, minimum width=5mm, font={\small}]
\tikzstyle{small gray box}=[small box, fill={rgb,255: red,180; green,180; blue,180}]
\tikzstyle{medium box}=[rectangle, inline text, fill=white, draw, minimum height=5mm, yshift=-0.5mm, minimum width=8mm, font={\small}]
\tikzstyle{ddot}=[line width=1.6pt, inner sep=0mm, minimum width=2.5mm, minimum height=2.5mm, draw, shape=circle]
\tikzstyle{dd white}=[ddot, fill=white, tikzit draw=green]
\tikzstyle{dd white phase}=[white phase dot, line width=1.6pt, tikzit draw=yellow]
\tikzstyle{dd gray}=[ddot, fill={rgb,255: red,180; green,180; blue,180}, tikzit draw=green]
\tikzstyle{dd gray phase}=[gray phase dot, line width=1.6pt, tikzit draw=yellow]

\tikzstyle{simple}=[-]
\tikzstyle{hadamard edge}=[-, dashed, dash pattern=on 2pt off 1pt, thick, draw=gray]
\tikzstyle{gray}=[-, draw={blue!60!white}, tikzit draw=blue]
\tikzstyle{blue}=[-, draw={blue!60!white}, tikzit draw=blue]
\tikzstyle{brace edge}=[-, tikzit draw=blue, decorate, decoration={brace,amplitude=1mm,raise=-1mm}]
\tikzstyle{diredge}=[->]
\tikzstyle{not edge}=[-, dashed, dash pattern=on 2pt off 1.5pt, thick, draw={rgb,255: red,255; green,68; blue,68}]
\tikzstyle{double edge}=[-, double, shorten <=-1mm, shorten >=-1mm, double distance=2pt]
\tikzstyle{boldedge}=[-, line width=1.6pt, shorten <=-0.17mm, shorten >=-0.17mm, tikzit draw=blue]

\newcommand*{\rey}{R\'enyi }

\newtheorem*{thm*}{Theorem}

\theoremstyle{remark}

\newcommand*{\ot}{\otimes}
\newcommand*{\nn}{\nonumber}

\newcommand*{\id}{\mathbb{1}}

\newcommand*{\mc}{\mathcal}

\newcommand*{\ex}{\mathrm{e}}

\DeclareMathOperator{\tr}{tr}

\hypersetup{
pdftitle={Magic Resources of the Heisenberg Picture},
pdfsubject={Many-body dynamics, quantum magic},
pdfauthor={Neil Dowling, Pavel Kos, Xhek Turkeshi},
pdfkeywords={Many-body quantum physics, quantum magic, quantum computation}
}

\begin{document}
\title[]{Magic Resources of the Heisenberg Picture} 

\author{Neil Dowling}
\email[]{ndowling@uni-koeln.de}
\affiliation{Institut f\"ur Theoretische Physik, Universit\"at zu K\"oln, Z\"ulpicher Strasse 77, 50937 K\"oln, Germany}
\affiliation{School of Physics \& Astronomy, Monash University, Clayton, VIC 3800, Australia}

\author{Pavel Kos}
\affiliation{Max-Planck-Institut f\"ur Quantenoptik, Hans-Kopfermann-Str. 1, 85748 Garching}

\author{ 
Xhek Turkeshi}
\affiliation{Institut f\"ur Theoretische Physik, Universit\"at zu K\"oln, Z\"ulpicher Strasse 77, 50937 K\"oln, Germany}

\begin{abstract}
We study a non-stabilizerness resource theory for operators, which is dual to that describing states. We identify that the stabilizer R\'enyi entropy analog in operator space is a good magic monotone satisfying the usual conditions while inheriting efficient computability properties and providing a tight lower bound to the minimum number of non-Clifford gates in a circuit. 
Operationally, this measure quantifies how well an operator can be approximated by one with only a few Pauli strings -- analogous to how entanglement entropy relates to tensor-network truncation.
A notable advantage of operator stabilizer entropies is their inherent locality, as captured by a Lieb-Robinson bound. This feature makes them particularly 
suited for studying local dynamical magic resource generation in many-body systems. 
We compute this quantity analytically in two distinct regimes.
First, we show that under random evolution, operator magic typically reaches near-maximal value for all R\'enyi indices, and we evaluate the Page correction. 
Second, harnessing both dual unitarity and ZX graphical calculus, we solve the operator stabilizer entropy for interacting integrable XXZ circuit, finding that it quickly saturates to a constant value. 
Overall, this measure sheds light on the structural properties of many-body non-stabilizerness generation and can inspire Clifford-assisted tensor network methods.
\end{abstract}

\maketitle
\emph{Introduction.---} 
While quantum computers are expected to surpass classical ones in solving particular tasks, many relevant quantum problems can also be efficiently simulated on classical computers. 
A distinctive framework where this is possible is Clifford operations acting on the class of stabilizer states, implementable with polynomial resources through the Gottesman-Knill theorem~\cite{gottesman1998,Aaronson2004}. 
This efficient simulability indicates the stabilizer framework's non-universal nature: alone, it cannot generate all possible quantum states, and non-stabilizer states or non-Clifford gates -- i.e., \emph{quantum magic resources} -- are required.
Quantifying magic resources for many-body systems is challenging, yet fundamental for its relevance in quantum technologies and applications~\cite{Zhou2020-cd,White2021Conformal,Leone2021quantumchaosis,Haferkamp_2022,Liu_2022,gu2024magicinduced,Ahmadi_2024,niroula2024phaset,catalano2024magicphase,smith2024nonstabrydber}.
While a zoo of geometric monotones exists~\cite{PhysRevA.59.141,Bravyi2019simulationofquantum,PhysRevLett.118.090501,Beverland_2020,Jiang2023,Bu_2024}, they are often impractical for many-body systems due to high computational complexity. 
A recent breakthrough was the development of the stabilizer R\'enyi entropy (SRE), a magic monotone~\cite{Leone2022stab,leone2024monotonesmagic} that provides  
a lower bound for several other relevant measures~\cite{PhysRevA.83.032317,PhysRevLett.118.090501,Beverland_2020,Xhek2023flatness,Jiang2023}. 
A key advantage of the SRE is its efficient computability in many practical scenarios, including matrix product states~\cite{Haug_2023mps,Lami2023,Tarabunga2024} and variational wavefunctions~\cite{Tarabunga2024magicingeneralized}. 
On the other hand, determining how magic {resources} are generated and propagated in many-body dynamics has remained a challenging problem.

\begin{figure}[t!]
    \centering
    \includegraphics[width=0.95\linewidth]{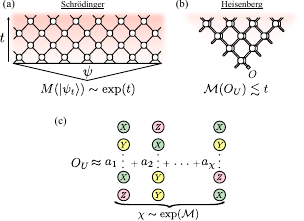}
    \caption{Schematic of the comparison of magic resources for states versus operators; cf. Eq.~\eqref{eq:fundamental}. (a) In the Schr\"odinger picture, magic resources $M$ tend to grow exponentially fast for a local circuit or dynamics. (b) 
    In the Heisenberg picture, {the growth of} magic resource $\mc{M}$ is bounded by a Lieb-Robinson light cone for an initially local operator $O$. (c) Operationally, any operator which can be well-approximated by an operator with only a polynomial number $\chi$ of Pauli coefficients must have {slowly growing operator magic resource: $\mathcal{M}\sim \mathcal{O}(\log(t))$}.
    }
    \label{fig:cartoon}
\end{figure}

Let us apply these ideas to the fundamental problem of computing  expectation values of observables
\begin{equation}
    \langle O\rangle_t = \mathrm{tr}[ O U_t \rho U_t^\dagger]\;,\label{eq:fundamental}
\end{equation}
describing the vast majority of quantum experiments and computations. 
In Eq.~\eqref{eq:fundamental}, $O$ is an operator of interest and $\rho$ is the initial density matrix, while $U_t$ is a unitary evolution. 
Simulating this value typically requires a choice: we can either evolve the quantum state $\rho_U = U_t \rho U_t^\dagger$ 
(Schr\"odinger picture), or evolve the operator $O_U = U^\dagger_t O U_t$ (Heisenberg picture); cf. Fig.~\ref{fig:cartoon}. 
{The latter approach is beneficial in many applications. 
For instance, in tensor network methods, one often finds an exponential improvement in the efficiency of computing Eq.~\eqref{eq:fundamental} in the Heisenberg picture when dealing with local operators, especially under integrable dynamics~\cite{Prosen2007,Prosen2007a,Dubail_2017,Jonay2018,Alba2019,Kos2020,Kos2020II,Alba2021}. 
This fact naturally raises the question: \textit{can a similar principle apply to magic resource theory? More specifically, how can we quantify the magic resource of an operator?}}

In this Letter, we introduce a class of magic monotones for operators and demonstrate their advantages over state-based measures. 
These monotones provide clear operational bounds on how well an operator can be approximated within the stabilizer formalism; see Fig.~\ref{fig:cartoon} (c). 
They are defined as the entropy of the squared Pauli coefficients of a Heisenberg operator, {generalizing magic resources to operator space.}
We call this monotone the \textit{operator stabilizer entropy} (OSE), and it inherits desirable properties from the SRE, such as: being efficiently computable using tensor networks~\cite{Haug_2023mps}, being a class of R\'enyi entropies, and providing a lower bound on the $T$-count $\tau(U)$ -- the minimum number of non-Clifford gates 
needed when $U$ is decomposed as Clifford unitaries plus single qubit gates $T = \exp(-i (\pi/8) \sigma_z)$. 
Beyond this, the OSE has several advantages over the stabilizer R\'enyi entropy. 
First, it exhibits a Lieb-Robinson light cone structure for local evolution, which constrains its growth to be linear, unlike the stabilizer R\'enyi entropy that is extensive even after short-times~\cite{turkeshi2024magicspreadingrandomquantum}.
Additionally, it equals the $T$-count for diagonal $T$-gate circuits, saturating the lower bound. 
Lastly, for long-time random evolution, they are typically maximal, up to a $\mathcal{O}(1)$ correction.

{We present evidence that operator magic resources can scale significantly more slowly than in the state case,} even when accounting for light-cone effects. 
To this end, we analytically compute the OSE in an interacting integrable dual-unitary XXZ circuit model for an arbitrary initially local operator across all R\'enyi indices and times. 
Using graphical ZX calculus techniques~\cite{aceto_interacting_2008,Coecke_2011,vandewetering2020zxcalc} adapted from~\cite{lopez2024exactsolutionlongrangestabilizer}, we show that it quickly saturates to a constant value.

This behavior contrasts sharply with the stabilizer R\'enyi entropy, which grows extensively with system size -- even after a single layer of the dynamics~\cite{lopez2024exactsolutionlongrangestabilizer}. 
{This suggests that in many-body quantum systems, tracking magic resources through the Heisenberg picture may offer significant computational advantages~\cite{Rakovszky2022,angrisani2024classically,Dowlin2024LOE-OSRE}.}

\emph{Operator Stabilizer Entropy and its Properties.---} 
{We denote $\mathcal{P}_N=\tilde{\mathcal{P}}_N/\{\pm i \id \}$ to be the set of Pauli strings~\cite{sarkar2019setscommutinganticommutingpaulis,Leone2022stab},} with $\tilde{\mathcal{P}}_N$ and  $\mathcal{C}_N$ respectively the Pauli and Clifford groups acting on $N$ qubits with total dimension $D=2^N$. 
Throughout, we use base-2 logarithms. 

{
{Consider some Hilbert-Schmidt normalized operator $O$, with $\tr[O^\dagger O]/D=1$---including, for instance, unitaries.} The operator stabilizer R\'enyi entropies (OSE) of $O$ are defined as 
\begin{equation}
    \begin{split}
        \Mpop^{(\alpha)}(O_U) &:=\frac{1}{1-\alpha} \log P^{(\alpha)}(O_U),\\ 
        P^{(\alpha)}(O_U)&:= \sum_{P \in \mathcal{P}_N}\left(\frac{1}{D}\tr[O_U P]\right)^{2\alpha} 
    \end{split}
    \label{eq:zio}
\end{equation}
}
where $\alpha \geq 0$, and $\alpha \in \{ 0,1,\infty\}$ are defined via their respective limits. 
The OSE is the R\'enyi entropy of the distribution $\{ \Pi_i := (({1}/{D})\tr[O_U P_i])^{2} \}$, with $P_i$ a Pauli string {labeled by some $i \in \{1,2,\dots,D^2\}$. $P^{(\alpha)}(O_U)$ are the \textit{operator stabilizer purities} which in turn can be used to define linear R\'enyi entropies,} $\Mpop^{(\alpha)}_\mathrm{lin}(O_U):=1-P^{(\alpha)}(O_U)$~\cite{leone2024monotonesmagic,Haug_2024,turkeshi2024magicspreadingrandomquantum}.

The OSE~\eqref{eq:zio} is a good magic monotone from the perspective of magic resource theory. In particular, it is {(i) \textit{faithful}, in that $\Mpop^{(\alpha)}(O_U) = 0 $ only for $O_U \in \mc{P}_N$,} (ii) \textit{stable under free operations}, in that $\Mpop^{(\alpha)}(C^\dagger O_U C) = \Mpop^{(\alpha)}(O_U) $ for $C \in \mc{C}_N 
   $, (iii) \textit{additive}, such that $\Mpop^{(\alpha)}( A_U \otimes B_V) = \Mpop^{(\alpha)}( A_U ) +\Mpop^{(\alpha)}( B_V) $, and (iv) \textit{bounded} as $0 \leq \Mpop^{(\alpha)}(O_U) \leq 2N$.
This upper bound can be saturated, e.g., for an operator that is a uniform superposition over all Pauli strings, or (approximately) for typical operators when $N\gg 1$, as discussed later. See App.~\ref{ap:OSEproperties} for proof of the above and other properties of the OSE.

Returning to the problem of computing evaluating Eq.~\eqref{eq:fundamental}, we recall that stabilizer simulation methods allow the computation of Pauli expectation values of stabilizer states with costs that are polynomial in $N$. 
Consider a resource-free (stabilizer) initial state $\rho = \ket{\psi}\bra{\psi} = C^\dagger \ket{0}\bra{0} C$ for $C \in \mathcal{C}_N$. 
In the Heisenberg picture, the number {of} such Pauli expectation values that need to be computed to exactly evaluate Eq.~\eqref{eq:fundamental} for some Hermitian operator $O_U$ is the \textit{rank $r$ of its Pauli decomposition},
\begin{equation}
    O_U = \sum_{i=1}^r a_i P_i,\label{eq:superposition}
\end{equation}
where $\Mpop^{(0)}(O_U) = \log(r)$ and {$a_i \in \mathbb{R}$}. 
In practice, many of the amplitudes may be small $|a_i|\ll 1$ and thus contribute negligibly to expectation values.

To exploit this fact, {one can} consider the truncated operator $\tilde{O}_U:=\sum_{i=1}^\chi a_i P_i$, where we retain only the largest $\chi<r$ amplitudes $a_i$. 
The resulting error in this approximation is the operator norm distance between $O_U$ and $\tilde{O}_U$, which bounds the error in computing expectation values for any initial $\rho$, 
\begin{equation}
    |\tr[O_U \rho] - \tr[\tilde{O}_U \rho]| \leq \| \sum_{i=\chi+1}^r a_i P_i \|_\infty =: \epsilon \;.\label{eq:thisss}
\end{equation}
This follows from H\"older's inequality with $\| X \|_\infty$ meaning the largest singular value of $X$. 
From Eq.~\eqref{eq:thisss}, we can study the efficiency of Pauli truncation through relating the scaling of OSE to $\epsilon$. {More specifically, if the OSE for $\alpha \geq 1$ scales extensively, $\mc{M}^{(1)}(O_U) \geq c N $ for some $c\in \mathbb{R}$, then for a given small $\epsilon$, the required rank of the Pauli decomposition of the truncated operator scales as $\chi \sim \mc{O} (\exp(N))$}. We prove this in App.~\ref{ap:operational}. 
Because of this exponential cost, 
one cannot efficiently compute expectation values for $O_U$ using {Pauli truncation} methods when OSE is large. 
We note that a slow scaling of the OSE for $\alpha\geq 1$ does not necessarily imply that $O_U$ can be efficiently approximated. 
{On the other hand, a sub-extensive $\mathcal{M}^{(0)}(O_U) \propto  \log(N)$ means the rank $r$ of the Pauli decomposition scales polynomially in system size $N$, implying efficient stabilizer implementations. OSE scaling therefore dictates the efficiency of Pauli truncation, analogous to the relation between entanglement entropies and tensor network methods~\cite{Verstraete2006,Schuch_2008}. 
Pauli truncation as described here is the general description of \textit{Pauli-path propagation}~\cite{angrisani2024classically,angrisani2025simulatin} and \textit{sparse Pauli dynamics}~\cite{Chan2024,begusic2024realtime} methods, while it is also closely related to \textit{dissipation-assisted operator evolution} (DAOE)~\cite{Rakovszky2022,lloyd2023ballisticdif,srivatsa2024prob}.} Further improvements may be made by integrating these methods with state-of-the-art stabilizer simulators~\cite{Bravyi2016,Bravyi2019simulationofquantum}, which achieve approximate sampling with scaling $\mc{O}(1.17^{\tau})$.

\textit{Comparison to Other Monotones.---} 
The OSE is a direct generalization of state SRE~\cite{Leone2022stab} to operator space. 
Moreover, optimizing $\Mpop^{(\alpha)}(O_U)$ over all initial operators $O$, one arrives at the \emph{magic power}, which can be used to lower-bound the circuit complexity of $U$~\cite{Bu_2024}. 
The OSE {of a Heisenberg-evolved resource free $O\in \mc{P}_N$, $O_U=U^\dagger O U$,} lower bounds the $T-$count $\tau(U)$, defined as the number of single $T-$gates together with Clifford gates required to
represent given circuit $U$
\begin{equation}
    \Mpop^{(\alpha)}(O_U) \leq \tau(U)\;. \label{eq:t-bound}
\end{equation}
This follows from a counting argument: 
a single $T-$gate at-most doubles the rank of the superposition \eqref{eq:superposition}, while Cliffords preserve it. Note that Eq.~\eqref{eq:t-bound} can be extended for generic operators $O$, where the bound has an additional term accounting for the initial magic resource of $O$.
In fact, the example of $U$ consisting of a tensor product of $T-$gates saturates this bound for $O\in \{ \sigma_x , \sigma_y\}^{\otimes N}$. 
This can be verified directly by computing the action of a single $T$-gate and applying the additivity of OSE.
Notably, the value is independent of the \rey index $\alpha$. {This fact stands in contrast to the SRE of tensor products of magic states $\ket{H}:=T \ket{+}$, which depends explicitly on $\alpha$~\cite{Leone2022stab}.} Beyond this, note also that a similar counting argument leads to the conclusion that the OSE approaches half the $T$-count $\Mpop^{(\alpha)}(O_U) \approx \tau(U)/2$ for random Clifford circuits interspersed with $T$-gates in the regime $N \gg \tau(U)$. See App.~\ref{ap:OSEproperties} for further discussion and proofs.

We can also compare the Schr\"odinger and Heisenberg picture stabilizer entropies in a more systematic way.  
For a given unitary {$U$}, we compare the average-case OSE and SRE over initial Pauli operators and stabilizer states respectively. The former is related to the SRE of the state representation of $U$~\cite{Leone2021choiSRE}, while {the latter} is termed \textit{nonstabilizing power}~\cite{Leone2022stab}.
Using results in Ref.~\cite{Leone2021choiSRE,Leone2022stab} we find,
\begin{equation}
    \mathbb{E}_{\ket{\psi} \in \mathrm{Stab}} \left({M}^{(2)}_{\mathrm{lin}}(U\ket{\psi}) \right) \leq \frac{1}{4}\mathbb{E}_{O \in \mc{P}_N} \left( {\mc{M}}^{(2)}_{\mathrm{lin}}(O_U) \right) \;. \label{eq:power}
\end{equation}
{This result is striking because both in the average case of Eq.~\eqref{eq:power} and in the particular case of $U=T^{\otimes \tau}$ for some integer $\tau$}, the Heisenberg picture magic measures provide 
a tighter bound on the $T-$count. {
Lastly, we note that the OSE also lower bounds the \emph{unitary stabilizer nullity}~\cite{Jiang2023}, as recently proven in Ref.~\cite{Dowlin2024LOE-OSRE}.}

\begin{figure}[t]
    \centering
    \includegraphics[width=0.95\linewidth]{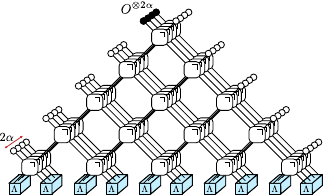}
    \caption{Tensor network diagram for the generalized Pauli purity, leading to the OSE \eqref{eq:zio}, for an initially local operator $O$ under brickwork circuit dynamics. Time goes from top to bottom, and each brick represents a doubled-picture two-site unitary, $U \otimes U^*$. At the bottom we have $\Lambda$ tensors from Eq.~\eqref{eq:Lambda}, and at the light cone edges white bullets representing vectorised identity $\ket{\circ} = \sum_i \ket{ii}$, connecting the copies $U$ to $U^*$. Black dots represent the initial local operator $O$, and the normalization is not shown.}
    \label{fig:OSE_tn}
\end{figure}

\textit{Efficient Computability and Light-cone.---} Another merit of the OSE is its efficient computability via tensor network methods, similar to the procedure laid out in Ref.~\cite{Haug_2023mps} for SRE. To clarify this point, we rewrite the stabilizer purity from Eq.~\eqref{eq:zio} as a single overlap in the $2\alpha-$replica space,
\begin{equation}
        P^{(\alpha)}(O_U)= \frac{1}{D^{2\alpha}}\tr[(O_U \otimes O_U^* )^{\otimes \alpha } \mathbb{\Lambda}] \label{eq:tn_form} 
\end{equation}
where we introduced the replica operator
\begin{equation}
    \mathbb{\Lambda}:=\sum_{P_1,\dots, P_N \in \mc{P}_1}\Big(( P_1 \otimes \dots \otimes P_N )\otimes ( P_1^* \otimes \dots \otimes P_N^* )\Big)^{\otimes  \alpha}. \label{eq:lambda}
\end{equation}
One can reshuffle indices of the above as {$ \mathbb{\Lambda} \leftarrow (\Lambda^{(\alpha)})^{\otimes n}$}, to arrive at a tensor network in replica space as a function of the single-site tensors 
\begin{equation}
     \Lambda^{(\alpha)} := \frac{1}{4} \sum_{P \in \mathcal{P}_1} (P \otimes P^*)^{\otimes \alpha}.
     \label{eq:Lambda}
\end{equation}
Fig.~\ref{fig:OSE_tn} more specifically depicts the OSE for a brickwork circuit Floquet dynamics on a one-dimensional spin-chain~\cite{Nahum2017,Nahum2018,Keyserlingk2018,amos2018,Khemani2018,Fisher2023}. Here, each layer of discrete evolution is on alternating next-neighbor pairs of spins, as in the first-order trotter approximation for a local Hamiltonian. 
Note that the support of an initially local operator $O_U$ grows by at-most two sites per time-step $t$. Then the extensive sum 
in Eq.~\eqref{eq:tn_form} reduces to only a sum over maximum length $2t$ Pauli strings, which are non-identity only within the support of $O_U$. This generalizes directly to quasi-local operators, local operators on multiple sites (where one instead needs only Pauli strings on the union of the support of each Heisenberg-evolved operator), and also for continuous evolution (up to an exponential error) in the form of a Lieb-Robinson bound. Therefore, for local dynamics and an initially local $O$, 
\begin{equation}
    \Mpop^{\alpha} (O_U) \leq t. \label{eq:lightCone}
\end{equation}
From the expression Eq.~\eqref{eq:tn_form}, one could adapt the techniques of Refs.~\cite{Haug_2023mps,Tarabunga2024} to efficiently compute the OSE when the bond dimension of the corresponding {matrix product operator} representation of $O_U$ grows slowly; often the case for integrable systems~\cite{Prosen2007,Prosen2007a,Dubail_2017,Jonay2018,Alba2019,Kos2020,Kos2020II,Alba2021}. 

{We note that the OSE can be constructed from measurements in the computational basis of a state encoding of the operator $O_U$. Namely, we show in App.~\ref{ap:measurement} that the OSE can be found by preparing the state $|{O_U}\rangle \! \rangle := (U \ot U^*) (O \ot \id) \ket{\phi^+}$ and performing a full measurement of the output in the computational basis, where $\ket{\phi^+}$ is a maximally entangled state across system/ancilla. They could therefore be measured experimentally in quantum devices when one has access both to ancilla qubits and backwards-in-time evolution, similar to protocols for measuring information scrambling~\cite{Landsman_2019,Mi2021,Green_2022,liang2024observatio}.}

\textit{Operator stabilizer entropy of typical observables.---} It is instructive to compute the {typical OSE for} $O_U = U O U^\dagger$, obtained evolving a Pauli string $O \in \mc{P}_N$ under a Haar {sampled} unitary, $U \in \mathbb{H}$. $\mathbb{H}$ is the unique, {unitarily invariant measure over the unitary group}, and is approximately reproduced from random brickwork circuit{s} after an exponential number of layers~\cite{Haferkamp2022}.
{In doing so, we shed light on the typical} stationary values of operator magic monotones at late times for chaotic evolutions. 

First we consider the average Pauli purities $P^{(\alpha)}$, requiring the computation of $2\alpha$ Haar moments 
\begin{equation}
    \overline{P^{(\alpha)}}:= 
    \mathbb{E}_{\mathbb{H}}\left[\sum_{P\in \mathcal{P}_N} \left(\frac{\mathrm{tr}(O_U P)}{D}\right)^{2\alpha} \right]\;, \label{eq:p1}
\end{equation}
{where for ease of notation we drop the explicit dependence on $O_U$.}
These can be evaluated exactly via Weingarten calculus~\cite{weingarten_asymptotic_1978,collins_integration_2006,puchala_symbolic_2017,Roberts2017-en}, for instance resulting in
\begin{equation}
    \overline{P^{(2)}} \equiv 
    \mathbb{E}_{\mathbb{H}}\left[\sum_{P\in \mathcal{P}_N} \left(\frac{\mathrm{tr}(O_U P)}{D}\right)^{4} \right] = \frac{3 (D^2-8)}{(D^2(D^2-9))},\label{eq:p2}
\end{equation}
for $\alpha=2$ while other expressions for $\alpha\le 5$, together with further details, can be found in App.~\ref{ap:haar}. 
Eq.~\eqref{eq:p2} allows us to bound $\overline{\Mpop^{(2)}}$ via Jensen's inequality. For example, from Eq.~\eqref{eq:t-bound} and $N=2$ two qubits we find that $2.22 \leq \overline{\Mpop^{(2)}} \leq \tau(U)$. We can therefore be assured that to reproduce Haar random dynamics on two qubits, on average more than two $T-$gates are required. This is consistent with the $\mathcal{O}(N)$ scaling found in~\cite{Leone2021quantumchaosis}.

In the scaling limit $N \gg 1$, we can further compute the behavior of the moments for any $\alpha$ using the asymptotics of the Weingarten symbols~\cite{collins_integration_2006}, obtaining 
\begin{equation}
    \overline{\Mpop^{(\alpha)}}\simeq 2 N 
    + \frac{\log[(2\alpha-1)!!]}{1-\alpha} + O \left(\frac{1}{D}\right)\;. \label{eq:oseAsymp}
\end{equation}
The leading term for any $\alpha \geq 2$ is maximal, 
meaning that observables in the late time are approximately equally distributed over all Pauli strings. {This result should be contrasted with the SRE for Haar random states, which depends on R\'enyi index $\alpha$ and is equal to zero as $\alpha \to \infty$.} We note that the subleading (negative) correction exactly matches that of the filtered stabilizer entropy for Haar states in Ref.~\cite{turkeshi2023paulispectrummagictypical}, suggesting a universality in this page-curve-like correction.

Key to the derivation of Eq.~\eqref{eq:oseAsymp} is the asymptotic self-averaging of $\Mpop^{(\alpha)}$. The fluctuations of the moments $P^{(\alpha)}$ are exponentially suppressed in $N$, ensuring that the difference in $\overline{\Mpop^{(\alpha)}}$ and $(1-\alpha)^{-1}\log \overline{P^{(\alpha)}}$ is small. Finally, for any $\alpha \geq 2$ the OSE exhibits strong typicality for random evolution $U$: non-negligible deviations from the average values Eqs.~\eqref{eq:p2}-\eqref{eq:oseAsymp} are exponentially suppressed in $D$. Therefore, the distribution of Pauli weights $\{ \Pi_i \}$ will tend to be approximately maximally mixed (and so the OSE will be maximal for all $\alpha$), for any \textit{individual} sample $U \in \mathbb{H}$; see App.~\ref{ap:typicality}.

\textit{Dual Unitary XXZ Circuit Dynamics.---} We will now compute the OSE for an interacting-integrable local circuit model, with full details of the calculation to be found in App.~\ref{ap:DU}. Consider an infinite one-dimension qubit spin chain. The dual unitary XXZ model consists of layers of two-site next-neighbor unitary gates, each  
given by
\begin{equation}
    U = \exp\left(- i \left( \frac{\pi}{4} (\sigma_x\otimes \sigma_x+\sigma_y\otimes \sigma_y) +\left(J+\frac{\pi}{4}\right)\sigma_z\otimes \sigma_z  \right) \right), \label{eq:two-site}
\end{equation}
with $0 \leq J \leq \pi/4$, and arranged in a brickwork pattern (as in Fig.~\ref{fig:OSE_tn}). For $J=0$ this is a circuit of SWAP {gates}, while for $J\neq 0$ it is an interacting integrable model. The maps \eqref{eq:two-site} are unitary in both the space and time directions; a class which turns out to be particularly amenable to analytic computations~\cite{Bertini2019exact,Claeys2020,Bertini2021random}. 
We evaluate the OSE for arbitrary system size $N$ and circuit depth (time steps) $t$ for an initial local unitary operator,
\begin{equation}
    O= a_x \sigma_x + a_y \sigma_y+ a_z \sigma_z \label{eq:arbOp}
\end{equation}
where from unitarity, $a_x^2 + a_y^2 + a_z^2 = 1$. We find that {
\begin{align}
     \Mpop^{(\alpha)}(O_U)= \frac{1}{1-\alpha} \Big( \log \big(& a_z^{2\alpha} + (a_x^{2\alpha} +a_y^{2\alpha})\label{eq:XXZfinal} \\
     &\times \left( \cos^{2 \alpha }(2 J) + \sin^{2\alpha }(2 J)\right)^t \big)  \Big). \nn
\end{align}}
For $a_z \neq 0$, Eq.~\eqref{eq:XXZfinal} grows fast
before saturating to a constant. We can also apply the replica trick to Eq.~\eqref{eq:XXZfinal} to determine $\Mpop^{(1)}(O_U)$. Interestingly, we find that this value increases linearly with $t$, and the maximal rate $ (a_x^2 + a_y^2)$ is achieved for $J=\pi/8$.

The scaling we find in Eq.~\eqref{eq:XXZfinal} is surprising, as one might expect that magic resources would scale linearly as the light cone (and hence support of $O_U$) increases. In contrast, the SRE density was found in Ref.~\cite{lopez2024exactsolutionlongrangestabilizer} to be near-maximal after a single time step, with an expression similar to Eq.~\eqref{eq:XXZfinal}. 
. 
Moreover, the scaling of Eq.~\eqref{eq:XXZfinal} is similar to that found for the local operator entanglement of the same model, computed in Ref.~\cite{Kos2020II} for the $2-$R\'enyi 
entropy.

\textit{Discussion.---}
We have developed a resource theory of operator non-stabilizerness as quantified by the OSE.
A defining feature of this monotone is that it naturally embodies locality{, leading to magic resources propagating at-fastest linearly in time for a local operator.} In contrast, recent work has shown that the stabilizer R\'enyi entropies anti-concentrate exponentially fast for chaotic dynamics~\cite{turkeshi2024magicspreadingrandomquantum}, {while growing} linearly in depth for doped Clifford circuits~\cite{Haug_2024,haug2024probingquantumcomplexityuniversal}. 
Determining relationships between magic resources in the Schr\"odinger versus Heisenberg pictures and extracting a locality principle also for the former requires further investigation.

As exemplified by the exact expression for the XXZ dual unitary model, the OSE may also be amenable to exact computations of magic monotones in other solvable models~\cite{sachdevgapless1993,Prosen_2016,Bertini2019exact,wang2020solvable,kos2021thermalization,Klobas2021,yu2023hierarchical,kos2023circuits,claeys2024operatordynamicsentanglementspacetime}. This could shed light on the interplay of magic resources and many-body phenomena such as phase transitions~\cite{White2021Conformal,niroula2024phaset,catalano2024magicphase}, eigenstate thermalization~\cite{Deutsch1991,Srednicki,Rigol2008,Foini2019,Pappalardi2022freeETH,fava2023designsfreeprobability}, and information scrambling~\cite{Leone2022stab,garcia_resource_2023,Oliviero2024,gu2024magicinduced,Ahmadi_2024}.

Regarding this latter concept, out-of-time order correlators are a key metric~\cite{Shenker_Stanford_2014,Maldacena_Shenker_Stanford_2016,Swingle2018}, and have been shown to necessarily witness operator entanglement growth~\cite{Hosur2016,Styliaris2021,dowling2023scrambling}. The OSE is in fact closely related to operator entanglement~\cite{Dowlin2024LOE-OSRE}, providing an avenue for uncovering interplay between entanglement and magic resources~\cite{gu2024magicinduced}.

Finally, from a practical point of view, there has been recent progress towards the goal of combining stabilizer and tensor-network methods for more efficient simulation~\cite{
lami2024quantumstatedesignsclifford,lami2024learningstabilizergroupmatrix,masotllima2024,paviglianiti2024estimatingnonstabilizernessdynamicssimulating,mello2024hybridstabilizermatrixproduct,qian2024augmentingdensitym,qian2024augmentedtimedependent,nakhl2024stabilizer,Gu_2024}. {In light of our results and the remarkable efficiency of Heisenberg picture tensor network methods, understanding the behavior of operator stabilizer resources may lead to innovations in classical simulation.}

\begin{acknowledgments}
    \emph{Acknowledgments.---}ND thanks Gregory White and Kavan Modi for useful discussions and collaboration on related projects. PK thanks Jordi Montana Lopez and Ignacio Cirac for discussions on related topics. XT thanks Piotr Sierant and Emanuele Tirrito for discussions on related topics. ND is supported by an Australian Government Research Training Program Scholarship and the Monash Graduate Excellence Scholarship. PK acknowledges financial support from the Alexander von Humboldt Foundation. XT acknowledges support from DFG under Germany's Excellence Strategy – Cluster of Excellence Matter and Light for Quantum Computing (ML4Q) EXC 2004/1 – 390534769, and DFG Collaborative Research Center (CRC) 183 Project No. 277101999 - project B01. 
\end{acknowledgments}

%

%
\onecolumngrid
\section*{Appendix}

\appendix

\tableofcontents

\section{Properties of Operator Stabilizer Entropies}\label{ap:OSEproperties}
In this section, we prove the following key properties of the OSE: 
\begin{enumerate}
    \item[1.] OSE satisfies the properties of a R\'enyi entropy.
    \item[2.] OSE is a unitary magic monotone.
    \item[3.] {If the $\alpha \geq 1$ OSE scales linearly for an operator $O_U$, $\Mpop^{(1)}(O_U) \geq c N$ for some $c\in \mathbb{R}$, then via Pauli truncation $O_U$ cannot be well-approximated by an operator with $\mathrm{poly}(N)$ Pauli strings.}
    \item[4.] {The OSE of $O_U$ lower-bounds the $T-$count of $U$;} this bound is saturated for $U=T^{\otimes \tau}$ for some positive integer $\tau$.
    \item[5.] The linear OSE averaged over initial Pauli operators, upper bounds the linear stabilizer R\'enyi entropy (SRE) (averaged over initial stabilizer states) by a factor of $1/4$, up to exponentially small corrections in $N$.
    \item[6.] For $t$-doped Clifford circuits, OSE is approximately \textit{half its upper bound}, i.e., $\tau(U)/2$, in the large system size limit $N \gg \tau$.   
    \item[7.] OSE is measurable in an experimental setting, through an ancilla space.
\end{enumerate}

\subsection{OSE is a \rey Entropy}
{Without loss of generality, we consider the numbering $\mathcal{P}_N=\{P_i\}_{i=1,2,\dots, D^2}$.} 
We are required to show that $\Pi_i := \big( ({1}/{D})\tr[O_U P_i]\big)^{2} = \big( ({1}/{D}^2)\tr[O_U P_i] \tr[O_U P_i]^* \big) $ is a probability. Clearly, $\Pi_i \geq 0$, as it is the absolute value of a complex number. Using that $\sum_{i=1}^{D^2} P_i \otimes P_i^\dagger = D \mathbb{S}$ where $\mathbb{S}=\sum_{i,j=0}^{D-1}(|i\rangle\ot|j\rangle)(\langle j|\ot\langle i|)$ is the SWAP unitary, 
\begin{equation}
    \sum_{i=1}^{D^2} \Pi_i = \sum_{i=1}^{D^2} \left( \frac{1}{D}\tr[O_U P_i]\right)^{2}= \frac{1}{D^2 } \sum_{i=1}^{D^2} \tr[(O_U \otimes O_U^\dagger)(P_i \otimes P^\dagger_i)] = \frac{D}{D^2 } \tr[(O_U \otimes O_U^\dagger) \mathbb{S}] = \frac{1}{D } \tr[O_U\cdot O_U^\dagger]=1,
\end{equation}
where the final equality {follows from the normalization of $O$, as
\begin{equation}
    O\cdot O^\dagger=\sum_i |a_i|^2 \mathbb{1} + \sum_{i,j} a_i a_j^* P_i P_j =  \mathbb{1} + \sum_{i,j} a_i a_j^* P_i P_j,
\end{equation}
and $P=P_iP_j \in \mc{P}\backslash \{ \mathbb{I}\}$ are traceless.} Therefore, as each $\Pi_i \geq 0$ and $\sum_i \Pi_i = 1$, $\{\Pi_i\}_{i=1,\dots,D^2}$ is a probability distribution, hence the OSE \eqref{eq:zio} is equal to $\alpha$-R\'enyi entropies.

\subsection{OSE is a magic monotone}
As a valid magic monotone, we show the following OSE properties hold.
\begin{enumerate}[(i)]
    \item \textit{Faithfulness}: {$\Mpop^{(\alpha)}(O_U)\ge0$ and $ \Mpop^{(\alpha)}(O_U)=0\iff  O_U \in \mc{P}_N$.}
    \item \textit{Stability under free operations}: $\Mpop^{(\alpha)}(C^\dagger O_U C) = \Mpop^{(\alpha)}(O_U)$ for any Clifford unitary $C\in \mathcal{C}_N$. 
    \item \textit{Additivity}: $\Mpop^{(\alpha)}(A_U \otimes B_V) = \Mpop^{(\alpha)}(A_U) + \Mpop^{(\alpha)}(B_V)$.
\end{enumerate}
We will consider statements in terms of $\log P^{(\alpha)}(O_U)$, as the proofs for $\Mpop^{(\alpha)}(O_U)$ follows directly. 
    \begin{enumerate}[(i)]
        \item {First, we note that as it is a R\'enyi entropy, $\Mpop^{(\alpha)}(O_U)\geq 0$, and it is maximized for the uniform distribution, $\Mpop^{(\alpha)}(O)\leq \log(D^2)=2N$. 
        Now, consider a Pauli operator $O_U =: P_j \in \mathcal{P}_N$ for some $j\in \{1,\dots,D^2\}$,
        \begin{align}
            \log P^{(\alpha)}(O_U) &=  \log \sum_{i=1}^{D^2} \left(\frac{1}{D}\tr[P_j P_i]\right)^{2\alpha} \\
            &=\log \sum_{i=1}^{D^2} (\delta_{ij})^{2\alpha} =0,
        \end{align}
        where we have used that $\mathrm{tr}[P_i P_j] = \delta_{ij} D$. 
        Therefore it follows that $\Mpop^{(\alpha)}(O_U) = 0$ for any $\alpha$. Now let us assume the converse: $\Mpop^{\alpha}(O_U)=0$. Since the argument of the logarithm is positive, 
        \begin{equation}
            P^{(\alpha)}(O_U)=1.
        \end{equation}
        Expanding $O_U = \sum_{i} a_{i} P_i$, again using that $\mathrm{tr}[P_i P_j] = \delta_{ij} D$, we have
        \begin{equation}
            \sum_{i=1}^{D^2} a_{i}^{2\alpha}=1, \label{eq:sum_cond}
        \end{equation}
        for all positive integers $\alpha$. Given that the sum is over a finite number of terms and that Eq.~\eqref{eq:sum_cond} should be valid for any $\alpha$, taking the limit $\alpha \to \infty$ of both sides leads to the conclusion that a single amplitude $a_j=1$, while the rest are zero: $\{ |a_i|^2 \} = \{ \delta_{ij} \}$ for some $j$. In other words, $O_U$ is a Pauli string as required.}
        \item For $C \in \mathcal{C}_N$ take $O_U \to C^\dagger O_U C $, then
        \begin{align}
              P^{(\alpha)}(C^\dagger O_U C) &=  \sum_{P_i \in \mathcal{P}_N} \left(\frac{1}{D}\tr[C^\dagger O_U C P_i]\right)^{2\alpha} \nn \\
              &= \sum_{P_i' \in \mathcal{P}_N} \left(\frac{1}{D}\tr[O_U P_i']\right)^{2\alpha} \label{eq:partii} =  P^{(\alpha)}(O_U),
        \end{align}
        {where we have used the fact that a Pauli string transforms under a Clifford operation into another Pauli string up to a phase $\pm1$, i.e., $P' = \pm C P C^\dagger$ with $P'\in \mathcal{P}_N$. The square in Eq.~\eqref{eq:partii} simplifies the $\pm 1$ for any integer $\alpha$, leaving $P^{(\alpha)}(O_U)$ unchanged.} Consequently, it holds $\Mpop^{\alpha}(O_U) = \Mpop^{\alpha}(C^\dagger O_U C)$ as required.
        \item Consider any product Heisenberg operator $A_U \otimes B_V \in \mc{B}(\mc{H}_a \otimes \mc{H}_b) $ on the combined Hilbert space $\mc{H}_a \otimes \mc{H}_b$, with $\dim (\mc{H}_{a}) = D_a = 2^{N_a}$ and $\dim (\mc{H}_{b}) = D_b = 2^{N_b}$. We define the Pauli strings acting on $\mathcal{H}_a$ ($\mathcal{H}_b$) by $\mathcal{P}_{N_a}$ ($\mathcal{P}_{N_b}$); then
        \begin{align}
            \log P^{(\alpha)}(A_U \otimes B_V) &= \log \sum_{P \in \mathcal{P}_N} \left(\frac{1}{D}\tr[A_U \otimes B_V P]\right)^{2\alpha}\nn  \\
            &= \log \sum_{P_a \in \mathcal{P}_{N_a} ,P_b \in \mathcal{P}_{N_b}} \left(\frac{1}{D_a D_b}\tr[(A_U \otimes B_V) (P_a \otimes P_b) ]\right)^{2\alpha} \label{eq:partiii} \\
            &= \log \sum_{P_a \in \mathcal{P}_{N_a} } \left(\frac{1}{D_a}\tr[A_U  P_a ]\right)^{2\alpha} +\log  \sum_{P_b \in \mathcal{P}_{N_b}}  \left(\frac{1}{D_b} \tr[ B_V P_b ]\right)^{2\alpha} \nn\\
            &= \log P^{(\alpha)}(A_U)+\log P^{(\alpha)}(B_V). \nn
        \end{align}
       Here, we have used the fact that all Pauli strings are product operators in order to split up the summation of Pauli strings into two separate sums. 
    \end{enumerate}
The above proofs for (i) and (ii) generalize directly to the linear OSE, which, however, is multiplicative rather than additive.

\subsection{OSE Scaling and the Efficiency of Pauli Truncation}  \label{ap:operational}
{
Consider a Hilbert-Schmidt normalized Heisenberg operator $O_U$ and arbitrary initial state $\rho$. We would like to bound, in terms of the OSE of $O_U$, the error in expectation values 
\begin{equation}
    \Delta := |\tr[(O_U - \tilde{O}_U)\rho]|, \label{eq:delta}
\end{equation}
where $\tilde{O}_U $ is an approximate operator for $O_U$, to be defined below. $O_U$ admits a Pauli decomposition (cf. Eq.~\eqref{eq:superposition} from the Main Text)
\begin{equation}
    O_U = \sum_{i=1}^{r} a_i P_i, \label{eq:pauli_decomp}
\end{equation}
where $1\leq r \leq D^2 $ is an integer, $ a_i \in \mathbb{R}\backslash \{ 0 \}$, the operator is Hilbert-Schmidt normalized as $\sum_i |a_i|^2 =1$, 
and the ordering $i=1,2,\dots,r$ is fixed to have $|a_1| \geq |a_2| \geq \dots \geq |a_r|$. 
We define Pauli truncation to be a simulation algorithm where, after Heisenberg evolution, only $\chi$ Pauli strings with the largest coefficients in the decomposition \eqref{eq:pauli_decomp} are kept, resulting in the truncated operator
\begin{equation}
    \tilde{O}_U =  \sum_{i=1}^{\chi} a_i P_i. \label{eq:truncd}
\end{equation}

Without loss of generality, we assume there exists an index $0<\chi = \chi(\delta)<r$ for a given cutoff $\delta\ge 0$ such that $|a_\chi|> \delta > |a_{\chi+1}|$~\footnote{Note that this construction and the following analysis can be directly generalized to other truncation strategies, beyond choosing a Pauli decomposition of $O_U$---in this case, the OSE could not be the relevant simulation complexity resource.}. (The cases $\chi=0$ and $\chi=r$ are trivial and correspond, respectively, to full truncation $\tilde{O}_U=0$, and no truncation $\tilde{O}_U=O_U$.) We note that, in general, Pauli truncation results in $\tilde{O}_U$ not being unitary, even if $O_U$ is (and that $\|\tilde{O}_U \|_\infty \neq 1$), analogous to the case for operator tensor network truncation (matrix product operator) methods~\cite{Hartmann2009}. Nonetheless, it is still meaningful to compute expectation values of $\tilde{O}_U$, as Hermicity is preserved under truncation. 

We further assume that it is efficient to compute expectation values of operators with only a polynomial (in $N$) number of Pauli coefficients, and leave open the particular method of truncation. Pauli truncation as we describe here is a general description of the \textit{Pauli-path propagation} technique of Refs.~\cite{angrisani2024classically,angrisani2025simulatin}, and \textit{sparse Pauli dynamics} method of Refs.~\cite{Chan2024,begusic2024realtime}. It is also closely related to the  \textit{dissipation-assisted operator evolution} (DAOE) method of Refs.~\cite{Rakovszky2022,lloyd2023ballisticdif,srivatsa2024prob}, with the caveat that two-point correlations are generally considered there, cf. Eq.~\eqref{eq:delta}. 

We say that if one can well-approximate $O_U$ by $\tilde{O}_U$ through retaining only $\chi \sim \mathrm{poly}(N)$ Pauli strings, then Pauli truncation is efficient. By well-approximate, we formally mean that 
\begin{equation}
    \| O_U - \tilde{O}_U \|_\infty \leq \epsilon \label{eq:eps}
\end{equation}
for some small $\epsilon$, where $\| X\|_p $ refers to the Schatten $p$-norm of $X$, with $\| X\|_\infty$ being the largest singular value of $X$. The condition Eq.~\eqref{eq:eps} directly implies that expectation values according to any state $\rho$ are at worst $\epsilon$-approximated by $\tilde{O}_U$, as from Eq.\eqref{eq:delta},
\begin{equation}
     |\tr[(O-\tilde{O})\rho]|  \leq \| O_U - \tilde{O}_U \|_\infty \| \rho \|_1 = \| O_U - \tilde{O}_U \|_\infty. \label{eq:delta2}
\end{equation}
Here we have used H\"older's inequality, and that $\rho$ is positive semi-definite with unit trace. \\

We will now prove (i) that an extensive $1$-R\'enyi OSE, $\Mpop^{(1)}(O_U) = \mc{O}(N)$, means that $O_U$ cannot be approximated using Pauli truncation, and (ii) that a sub-extensive $0$-R\'enyi OSE, $\Mpop^{(0)}(O_U) = \mc{O}( \log(N))$, then $O_U$ can be well-approximated using Pauli truncation.\\

First, we note that through the standard relations between Schatten $p$-norms, 
\begin{align}
    \| O_U - \tilde{O}_U  \|_\infty &\geq D^{-1/2} \| O_U - \tilde{O}_U  \|_2 \\
    &=D^{-1/2} \|  \sum_{i=\chi+1}^r a_i P_i \|_2 \\
    &= D^{-1/2} \sqrt{\sum_{i,j =\chi+1}^r a_i a_j \tr[P_i P_j] } \\
    &= \sqrt{\sum_{i=\chi+1}^r a_i^2}. \label{eq:tracedist1}
\end{align}
Here, we have used that $\tr[P_i P_j] = \delta_{ij} \tr[\id] = D \delta_{ij}$. The next step is to notice that the final line \eqref{eq:tracedist1} is the trace norm distance between two pure states. We will follow a similar argument to those made with respect to the simulability of states using tensor network methods; see Refs.~\cite{Verstraete2006,Schuch_2008}. We define the two (normalized) pure states 
\begin{align}
    &| O_U \rangle \! \rangle := (O_U \otimes \mathbb{1})|\phi^+\rangle,  \quad \text{and} \nn \\
    &| \tilde{O}_U \rangle \! \rangle := \frac{1}{\sqrt{\sum_{i=1}^\chi a_i^2}}(\tilde{O}_U \otimes \mathbb{1})|\phi^+\rangle = \frac{1}{n}(\tilde{O}_U \otimes \mathbb{1})|\phi^+\rangle, \label{eq:CJI}
\end{align}
where $|\phi^+\rangle := (1/\sqrt{D}) \sum_i \ket{i} \otimes \ket{ i }$ is a maximally entangled state across two copies of Hilbert space. $| O_U \rangle \! \rangle $ is the pure quantum state corresponding to $O_U$ through the Choi–Jamio\l kowski isomorphism, while $| \tilde{O}_U \rangle \! \rangle $ is that for the normalized $ (1/n))\tilde{O}_U$. For brevity, we write the normalization as $n:= {\sqrt{\sum_{i=1}^\chi a_i^2}}$. Note that we do not consider $\tilde{O}_U$ to be normalized in the method of Pauli truncation, and the normalized state $| \tilde{O}_U \rangle \! \rangle$ is an auxiliary object for the purpose of the present proof. It is immediate to verify that the Hilbert-Schmidt normalization enforced earlier is sufficient to guarantee that $| O_U \rangle \! \rangle$ is a normalized pure state (similarly so for $| \tilde{O}_U \rangle \! \rangle$)
\begin{align}
   \langle \! \langle O_U| O_U \rangle \! \rangle &= \langle \phi^+ |(O_U \otimes \mathbb{1})^\dagger (O_U \otimes \mathbb{1})|\phi^+\rangle \\
   &= \frac{1}{D}\tr[O_U^\dagger O_U] =1. 
\end{align}
Using the definitions Eq.~\eqref{eq:CJI}, we have that Eq.~\eqref{eq:tracedist1} can be rewritten as 
\begin{align}
    \frac{1}{2}\| |{O_U}\rrangle \llangle O_U |  - |{\tilde{O}_U}\rrangle \llangle {\tilde{O}_U} | \|_1 &= \sqrt{1 - \llangle O_U|\tilde{O}_U\rrangle^2}, \\
    &=  \sqrt{1 - \frac{1}{D^2} \left( \frac{1}{n}\tr[\tilde{O}_U^{\dagger } \tilde{O}_U ]\right)^2 } \\
    &= \sqrt{1 - \frac{1}{D^2} \left( \frac{\sum_{i=1}^\chi a_i^2}{n}\tr[\id]\right)^2 }\\
    &=\sqrt{1 - n^2 } = \sqrt{\sum_{i=\chi+1}^r a_i^2}, \label{eq:truncation}
\end{align}
where we first have used the equivalence between trace distance and fidelity for pure states~\cite{wilde_2017}, and then in the final line that from the normalization of $O_U$, $1=\sum_{i=1}^r a_i^2 = \sum_{i=1}^\chi a_i^2+\sum_{i=\chi+1}^r a_i^2 = n^2 + \sum_{i=\chi+1}^r a_i^2$.  

The strategy now will be to lower-bound Eq.~\eqref{eq:truncation} in terms of a the $1$-R\'enyi OSE of $O_U$. Recall the {Fannes–Audenaert inequality {for any positive semi-definite $\rho$ and $\sigma$}~\cite{wilde_2017},}
\begin{equation}
    |S(\rho) - S(\sigma) | \leq T \log(D) + h(T)\;.
\end{equation}
Here, $S:=-\mathrm{tr}(\rho\log \rho)$ is the von Neumann entropy, {$h(T):= - T\log T - (1-T)\log (1-T)\leq 1$ is the binary entropy,} $D$ is the dimension of $\rho$, and $T := \frac{1}{2}\|\rho - \sigma \|_1$ is the trace distance. Define $\$(\cdot) := \sum_{i=1}^{D^2} |P_i\rrangle\llangle P_i| (\cdot)|P_i\rrangle\llangle P_i| $ as the dephasing channel with respect to the basis of Choi states of Pauli matrices. This channel is completely positive trace-preserving (CPTP), and so
\begin{align}
    |S(\$(|O_U\rrangle\llangle{O_U}|)) - S(\$(|{\tilde{O}_U}\rrangle\llangle{\tilde{O}_U}|)) | &\leq \frac{1}{2} \| \$(|{O_U}\rrangle\llangle{O_U}|) - \$(|{\tilde{O}_U}\rrangle\llangle{\tilde{O}_U}|)\|_{1}\log(D^2) + 1 \nn\\
    &\leq \frac{1}{2} \| |{O_U}\rrangle\llangle{O_U}| - \|{\tilde{O}_U}\rrangle\llangle{\tilde{O}_U}| \|_{1}\log(D^2) + 1.\label{eq:cptp}
\end{align}
Here, we have used the monotonicity of distance measures under CPTP maps. We notice that $S(\$(\ket{O_U}\!\rangle\langle\!\bra{O_U}))$ is exactly the Shannon entropy of the square Pauli coefficients of the operator $O_U$, equal to the $\alpha=1$ OSE $\Mpop^{(1)}(O_U)$. This can be seen as $\$(\ket{O_U}\!\rangle\langle\!\bra{O_U})$ is a diagonal density matrix in the Pauli basis, with components $\{ | a_i |^2\}_{i=1}^{r}$. We will also use that truncation may not increase the entropy of a distribution, even after renormalization by $n$,
\begin{equation}
    \Mpop^{(1)}(O_U) \geq \Mpop^{(1)}(\frac{1}{n}\tilde{O}_U).
\end{equation}
Moreover, we can bound $\Mpop^{(1)}((1/n)\tilde{O}_U)$ by the Pauli rank of $\tilde{O}_U$, 
\begin{equation}
    \Mpop^{(1)}(\frac{1}{n}\tilde{O}_U) \leq \log(\chi).
\end{equation}
Putting this together with Eqs.~\eqref{eq:eps}, \eqref{eq:tracedist1}, \eqref{eq:truncation}, and \eqref{eq:cptp}, we finally arrive at 
\begin{align}
    \epsilon &\geq \| O_U - \tilde{O}_U  \|_\infty \\
    &\geq \frac{1}{2} \| |{O_U}\rrangle\llangle{O_U}| - \|{\tilde{O}_U}\rrangle\llangle{\tilde{O}_U}| \|_{1} \\
    &\geq \frac{1}{2 N}(\Mpop^{(1)}(O_U) - \log(\chi) -1). \label{eq:finalTrunc}
\end{align}
Now we assume that for the Pauli truncation protocol, one requires some sufficiently small error $\epsilon$, but that the $1$-R\'enyi OSE is extensive, $\Mpop^{(1)}(O_U) \geq c N$ for some $c \in \mathbb{R}$. Then from Eq.~\eqref{eq:finalTrunc} we know that 
\begin{equation}
    \chi \geq \exp(N(c-2\epsilon) -1).
\end{equation}
Therefore, under the assumption of a sufficiently small $\epsilon$, one needs to retain an exponential number of Pauli strings in $\tilde{O}$, and so this system with extensive OSE cannot be simulated efficiently using Pauli truncation.\\

Now we will make the simpler argument that if the $0$-R\'enyi OSE is logarithmic in $N$, then Pauli truncation can be done efficiently. This follows trivially from the relation that 
\begin{equation}
    \Mpop^{(0)}(O_U) = \log(r),
\end{equation}
such that the initial operator only has $\mathrm{poly}(N)$ Pauli strings, and so none of them need to be truncated to exactly represent $O_U$ with a polynomial number of Pauli strings. \\

} 

We also note that the above proofs are independent of whether $O_U$ is initially resource-free (i.e. initially a Pauli string), as we care only about its scaling with time and/or system size.

\subsection{OSE bounds $T-$count} \label{ap:t-count}
We here give proof that OSE bounds the $T$-count, denoted here $\tau(U)$, for doped Clifford circuits. Consider an arbitrary initial $O$ and $O_{U_t}=U_t^\dagger O U_t$ decomposed in the Pauli basis,
\begin{equation}
    O_{U_t} = \sum_i^{r_t} a_i P_i.\label{eq:superposition1}
\end{equation}
Here, we take the time $t$ to be discrete, with each step $t \to t+1$ denoting a layer of circuit that contains exactly one $T-$gate and arbitrary Clifford gates. Then, every $T-$gate can at most double the size of the superposition $r_{t+1}\leq 2r_t$. This is because $T^\dagger P T$ results in, at most, a superposition of two Pauli strings (as it acts only on a single qubit and $T$ does not commute locally with only an $X$ or $Y$ Pauli), for each element of the sum in Eq.~\eqref{eq:superposition1}.
Intermittent Cliffords, on the other hand, preserve the number of elements $r$. 
Therefore, after $\tau(U)$ $T-$gates the number of terms in the sum Eq.~\eqref{eq:superposition1} is at most $r=2^{\tau \Mpop^{(0)}(O)}$, recalling that $2^{\Mpop^{(0)}(O)}$ is the initial rank of Pauli superposition. The highest-entropy value of the coefficients $\{|a_i|^2\}$ is the uniform distribution, $|a_i|= 1/\sqrt{r}$ for each $1 \leq i \leq r$, which results in the bound 
\begin{equation}
    \Mpop^{(\alpha)}(O_U) \leq \tau(U) + \Mpop^{(0)}(O). \label{eq:t-bound1}
\end{equation}
Here, in contrast to the result presented in Eq.~\eqref{eq:t-bound}, we have an additive contribution when the initial operator is not a Pauli string, which is equal to the logarithm of the rank of its Pauli decomposition.  

\subsection{Relation between Average State and Operator Stabilizer R\'enyi entropies}\label{ap:power}
{Using a result for the explicit expression for the $2-$linear nonstabilizer power $\mathbb{E}_{\ket{\psi} \in \mathrm{Stab}} \left({M}^{(2)}_{\mathrm{lin}}(U\ket{\psi}) \right)$ from Ref.~\cite{Leone2022stab} (see Eq.~(S.43) therein), we can substitute in the average linear OSE over initial Pauli operators, $\mathbb{E}_{O \in \mc{P}_N} \left( {\mc{M}}^{(2)}_{\mathrm{lin}}(O_U) \right)$, to arrive at Eq.~\eqref{eq:power}
\begin{align}
     \mathbb{E}_{\ket{\psi} \in \mathrm{Stab}} \left({M}^{(2)}_{\mathrm{lin}}(U\ket{\psi}) \right) &\leq \frac{D+3}{4(D+4)}\mathbb{E}_{O \in \mc{P}_N} \left( {\mc{M}}^{(2)}_{\mathrm{lin}}(O_U) \right) - \frac{15}{D+4} \\
     &\leq \frac{1}{4}\mathbb{E}_{O \in \mc{P}_N} \left( {\mc{M}}^{(2)}_{\mathrm{lin}}(O_U) \right).
\end{align}}

\subsection{OSE for generic $T-$doped Clifford circuits}
Consider layers of deep, random Clifford circuits interspersed with $T-$gates on random bits. This is called the $T-$doped Clifford ensemble. 

We will make an informal argument that for large systems, in the average case, the OSE will tend towards 
\begin{equation}
    \Mpop^{(\alpha)}(O_U) \overset{N \gg 1}{\approx} \tau(U)/2. \label{eq:t-dopedConjecture}
\end{equation}
This can be understood as follows: every random Clifford maps a Pauli string to a random Pauli string. Then, if a $T-$gate does not commute with a Pauli string, it creates a uniform superposition of two Pauli strings. This will happen for half of all random Pauli strings, as this only depends on whether the local site where $T$ acts is $\sigma_x$ or $\sigma_y$. If the system is very large, $N \gg \tau(U)$, we arrive directly at Eq.~\eqref{eq:t-dopedConjecture} as all Pauli strings resulting from applying a $T-$gate will almost surely be independent of other Pauli strings in the superposition of $O_U$, from a counting argument. 

\subsection{Experimental Measurement Protocol} \label{ap:measurement}
The OSE can be experimentally measured adapting several existing methods~\cite{Landsman_2019,Mi2021,Green_2022,liang2024observatio}. 
Here we briefly present a protocol based on the Choi–Jamio\l kowski isomorphism (CJI)~\cite{Mele_2024}. Namely, an operator may be encoded in a quantum state through its action on half of a maximally entangled bell state $\ket{\phi^+}=\sum_{n=0}^{D-1}|n\rangle\ot|n\rangle/\sqrt{D}$,
    \begin{equation}
        | O \rangle \! \rangle := (O \otimes \id) |{\phi^+}\rangle.
    \end{equation}
    The Heisenberg evolution of $O$ is then implemented through the unitary $U \ot U^*$, as 
    \begin{equation}
        | O_U \rangle \! \rangle = (O_U \otimes \id) |{\phi^+}\rangle = (U \ot U^*)(O \otimes \id) |{\phi^+}\rangle.
    \end{equation}
    In a quantum device, one can then prepare $| O_U \rangle \! \rangle$ using an ancilla space and access to backwards-in-time evolution $U^*$, with the OSE involving the computation of
    \begin{equation}
        \frac{1}{D^2}|\mathrm{tr}[O_U P_i]|^2 =  |\langle \! \langle O | P_i  \rangle \! \rangle|^2 
    \end{equation}
    for several Pauli strings $P_i$.
    We remark that the Choi states of the Pauli strings $|P_i\rangle \! \rangle$ correspond to computational basis states in the doubled space. 
    Individual computational basis overlaps are implemented via standard methods~\cite{Huang_2020}. 
    In general, the number of $P_i$ we must compute scales as $O(\exp(\mathcal{M}))$. 
    In particular, at late time $\mathcal{M}\propto N$: 
 hence the same quantum computational complexity holds for stabilizer R\'enyi entropy, and other state magic resources~\cite{niroula2024phaset}, as for the OSE. 
    However, for short times or slowly growing OSE, the light cone structure drastically reduce the sampling complexity to the evaluation of $O(\exp(t))$ Pauli strings. 
    In contrast, SRE grows extensively after just one time step, requiring $O(\exp(N))$ resources regardless, cf. for instance~\cite{turkeshi2024magicspreadingrandomquantum}.
    A detailed study on the experimental implementation and on the sampling efficiency for the OSE presents an important avenue for future research.

\section{Haar random evolution} \label{ap:haar}
This section details the closed-form expression of the operator stabilizer entropy for an initial Pauli string evolved under a deep Haar random circuit.
We then discuss the concentration of measure properties and self-averaging results for this setup.

\subsection{OSE of Typical Operators} \label{eq:haar_calc}
We first detail the derivation of the exact expressions for the Haar averaged Pauli purities, given for $\alpha=2$ in Eq.~\eqref{eq:p2} and for $\alpha\leq 5$ below in Eq.~\eqref{eq:alpha5}, and their resultant OSE. 
From Eq.~\eqref{eq:p1}, the key calculation is 
\begin{equation}
    \mathbb{E}_{\mathbb{H}}[ \mathrm{tr}(O_U P)^{\beta}]=\sum_{\pi,\sigma} \mathrm{Wg}_{\pi,\sigma} \mathrm{tr}(O^{\otimes \beta} T_\pi)\mathrm{tr}(P^{\otimes \beta} T_\sigma)\;,\label{eq:wein1}
\end{equation}
where $\beta$ is a positive integer, $\mathrm{Wg}_{\pi,\sigma}$ are the Weingarten symbols and $\pi,\sigma\in S_\beta$ are permutations of $\beta$ elements, and $T_\pi$ are the representation of the permutation operators in the space of $\beta$-replicas~\cite{collins_integration_2006,Mele_2024,turkeshi2024error}. Performing the double sum in Eq.~\eqref{eq:wein1} over the permutations is unfeasible beyond $\alpha=3$, due to the highly complex $\mathcal{O}([(2\alpha)!]^2)$ computational costs. However, since $P^{\otimes \beta}$ is invariant under permutations, their expectation with $T_\sigma$ is fixed solely by the cycle structure of the permutation $\sigma$, denoted $\lambda_\sigma$. 
Recalling that $\lambda_\sigma\vdash \beta$ is an integer partition of $\beta$, we have $\mathrm{tr}(P^{\otimes\beta} T_\sigma) = \prod_{c\in \lambda_\sigma}\mathrm{tr}(P^c)$.
Additionally, since $P^2=\mathbb{1}$, we have that $\mathrm{tr}(P^c)= \delta_{P,\mathbb{1}} D$ if $c$ is odd and $\mathrm{tr}(P^c)= D$ if $c$ is even. This implies that $\mathrm{tr}(P^{\otimes\beta} T_\sigma) = D^{\ell(\lambda_\sigma)} \delta_{\lambda_\sigma}(P)$, where $\ell(\lambda)$ is the length of the integer partition and $\delta_{\lambda_\sigma}(P) = \delta_{P,\mathbb{1}}$ if any $c\in {\lambda_\sigma}$ is odd and $\delta_{\lambda_\sigma}(P)=1$ otherwise.

For the operator stabilizer purity, we consider $\beta=2\alpha$ and sum Eq.~\eqref{eq:wein1} over all Pauli strings $P$. 
Let us divide the permutation group in a part that contains only cycles with even length $S_{2\alpha}^+$, and permutations that contains at least one odd length cycle $S_{2\alpha}^-$. For instance $(12)(34)\in S_{4}^+$, while $(123)(4)\in S_{4}^-$. 
We have
\begin{equation}
    \begin{split}
        \overline{P^{(\alpha)}}=& \sum_{\pi \in S^+_{2\alpha}}\sum_{\sigma\in S^+_{2\alpha}} \Big(D^{\ell(\lambda_\pi)+\ell(\lambda_\sigma) +2-2\alpha }\mathrm{Wg}_{\pi,\sigma}\Big)\\
     &+\sum_{\pi \in S^+_{2\alpha}}\sum_{\sigma\in S^-_{2\alpha}} \Big(D^{\ell(\lambda_\pi)+\ell(\lambda_\sigma) -2\alpha }\mathrm{Wg}_{\pi,\sigma}\Big).
    \end{split}
\end{equation}
Since $O$ is a non-identity Pauli string, the sum over $\pi$ has also been restricted to only even length cycles. This expression is further simplified by the fact that these sums depend only on the cycle structure. In particular
\begin{equation}
    \overline{P^{(\alpha)}} = \sum_{\pi \in S^+_{2\alpha}}\sum_{\lambda \vdash 2\alpha} D^{\ell(\lambda_\pi)+\ell(\lambda) -2\alpha} D^{2 \delta_\lambda(O)}   a_\lambda \mathrm{Wg}_{\pi,\sigma(\lambda)}, 
    \label{eq:ziooo1}
\end{equation}
where $\lambda \vdash 2\alpha$ are integer partitions of $2\alpha$, $\sigma(\lambda)$ the permutation fixed by the standard Young tableau of $\lambda$, $a_\lambda$ the number of permutations with cycle structure fixed by $\lambda$. 
This expression is exponentially more efficient, since it requires $\mathcal{O}((2\alpha)! p(2\alpha))$ resources, with $p(n)\sim \exp(\pi \sqrt{2n/3})$ is the Euler partition function. 
This enables us to deter-
mine the closed-form expressions for $\alpha\le 5$, 
\begin{equation}
    \begin{split}
        \overline{P^{(2)}}&=\frac{3 \left(D^2-8\right)}{D^2 \left(D^2-9\right)}\\
    \overline{P^{(3)}}&=\frac{15 \left(D^6-33 D^4+216 D^2-256\right)}{D^4 \left(D^6-35 D^4+259 D^2-225\right)}\\
    \overline{P^{(4)}}&=\frac{105 \left(D^8-81 D^6+1776 D^4-10432 D^2+15360\right)}{D^6 \left(D^8-84 D^6+1974 D^4-12916 D^2+11025\right)} \\
    \overline{P^{(5)}}&=\frac{945 \left(D^{12}-170 D^{10}+9657 D^8-224080 D^6+2199488 D^4-8985600 D^2+12386304\right)}{D^8 \left(D^2-9\right)^2 \left(D^8-156 D^6+7374 D^4-106444 D^2+99225\right)}\label{eq:alpha5}
    \end{split}
\end{equation}
We note that, in the scaling limit, the general form is 
\begin{equation}
    \overline{P^{(\alpha)}} = \frac{(2\alpha-1)!!}{D^{2(\alpha-1)}} \frac{R(D)}{Q(D)}\;,
\end{equation}
with the algebraic function $R(D)/Q(D)$ being the ratio of two polynomials in $D$ such that, for $N\gg 1$, $R(D)/Q(D) = 1 + \mathcal{O}(1/D)$.


We will now justify analytically this asymptotic random scaling as $N\gg 1$, Eq.~\eqref{eq:oseAsymp}. 
The key ingredient is the large $D$ expansion of the Weingarten symbol $\mathrm{Wg}_{\pi,\sigma}\simeq D^{-2\alpha}\delta_{\pi,\sigma} + \mathcal{O}(1/D)$~\cite{weingarten_asymptotic_1978}. 
Inserting this expression in Eq.~\eqref{eq:ziooo1} leads at leading order to 
\begin{equation}
    \overline{P^{(\alpha)}} \simeq \sum_{2\lambda \vdash \alpha }a_\lambda\frac{D^{2\ell({\lambda})+2}}{D^{4\alpha}},\label{eq:asss}
\end{equation}
where we denoted $2\lambda$ the integer partitions of $2\alpha$ such that any $c\in \lambda$ is even. 
Up to corrections $\mathcal{O}(1/D)$, Eq.~\eqref{eq:asss} is fixed by the maximal value of $\ell(\tilde\lambda)=\alpha$ which occurs when all $\tilde\lambda_i=1$. In that case, $a_\lambda= (2\alpha-1)!!$, where $n!!$ denotes double factorial of $n$, and the final result is 
\begin{equation}
    \overline{P^{(\alpha)}} \simeq \frac{(2\alpha-1)!!}{D^{2\alpha-2}}+\mathcal{O}(1/D)\;.
    \label{eq:resmom}
\end{equation}
From Eq.~\eqref{eq:resmom}, we recover the scaling limit of the OSE in Eq.~\eqref{eq:oseAsymp}. We are able to go from the moments Eq.~\eqref{eq:resmom} to the average OSE $\overline{\Mpop^{(\alpha)}}$ due to Haar concentration of measure~\cite{turkeshi2024error}, with the suppression of deviations from the average as $N \gg 1$. In the following section we supply a rigorous proof for this concentration of measure.


\subsection{Typicality of the Average Case} \label{ap:typicality}
It is important to ask: What can we learn from $\overline{P^{(\alpha)}}$ about the average-case OSE, $\overline{M_\alpha}$? 
In this section, we first present a simple heuristic argument describing the self-averaging of these quantities in the scaling limit. Afterward, we present rigorous proof using a concentration of measured arguments. 

Let us compute the relative fluctuations for $P^{(2)}$, namely
\begin{equation}
    \mathcal{F}(P^{(2)}) = \sqrt{\frac{\overline{(P^{(2)})^2}-\overline{P^{(2)}}^2}{\overline{P^{(2)}}^2}}.
\end{equation}
We need to evaluate
\begin{equation}
   \overline{(P^{(2)})^2} = \mathbb{E}\left[\sum_{P_1,P_2} \frac{\mathrm{tr}(O_U P_1)^4\mathrm{tr}(O_U P_2)^4}{D^8}\right]\;.
\end{equation}
Consider first the case $P_1=P_2$. This is equivalent to computing $\overline{P^{(4)}}\simeq 945/D^6$. However, this term is subleading compared to $\overline{P^{(2)}}^2\simeq 9/D^4$. Let us now focus on $P_1\neq P_2$. 
Following the same steps as for the scaling limit detailed in App.~\ref{eq:haar_calc}, we have
\begin{equation}
    \overline{(P^{(2)})^2} \simeq \sum_{P_1\neq P_2}\sum_{\sigma \in S_8 } \frac{D^{2\lambda_\sigma}}{D^{16}} \delta_{\sigma^{-1}}(O)\delta_{\sigma}(P_1P_2) + \mathcal{O}(1/D^6)\;.
\end{equation}
The maximal term occurs when $\lambda$ couple pairs and has length $4$. However, since $\mathrm{tr}(P_1 P_2) = 0$, the only non-trivial contributions come from $\sigma\in S_4\times S_4$. In the scaling limit, this leads to 
\begin{equation}
    \overline{(P^{(2)})^2} \simeq  \frac{9}{D^4} + \mathcal{O}(1/D^6).
\end{equation}
This contribution cancels out, and one is left with $\mathcal{F}\simeq 1/D$, meaning the relative fluctuations are exponentially suppressed in system size.

We now prove that the average case is typical through a standard application of Levy's Lemma via concentration of measure. We first formulate the informal statement as a precise mathematical theorem.

\begin{thm*}
    For any $\epsilon >0$ and for a randomly sampled $U$ according to the (global) Haar measure $U \sim \mathbb{H}$,
 \begin{equation}
    \mathrm{Pr}_{U\sim \mathbb{H}} \left\{|P^{(\alpha)}(U^\dagger O U)- \braket{P^{(\alpha)}(U^\dagger O U)}_{\mathbb{H}}|\geq \epsilon \right\} \leq \exp \left(-\frac{D \epsilon^2 }{64 (\alpha+1)^2} \right).
\end{equation}
\end{thm*}
\begin{proof}
    We apply Levy's Lemma, which states that for $U$ sampled according to the Haar measure $\mathbb{H}$, $f: \mathbb{U}_d \to \mathbb{R}$ a Lipschitz continuous function with Lipschitz constant $K$, and $\epsilon >0$ then
        \begin{equation}
            \mathrm{Pr}_{U\sim \mathbb{H}} \left\{|f(U) - \braket{f(U)}_{\mathbb{H}}|\geq \epsilon \right\} \leq \exp \left(-\frac{D \epsilon^2 }{4 K^2} \right)  
        \end{equation}
        where $K$ is defined such that for all $U,V\in \mathbb{U}_D$ 
        \begin{equation}
            |f(U) - f(V) |\leq K \|U-V \|_2.
        \end{equation}
    The task is to determine a Lipschitz constant $K$ for the function  $f(U) = \sum_{P \in \mathcal{P}} (\frac{1}{D}\tr[O_U P])^{2\alpha} $. 
    
    We will first take $\alpha\geq 2$ to be an even integer and then use the result to also provide a Lipschitz constant for odd $\alpha$. We have that 
    \begin{align}
        |f(U) - f(V) | &= | \sum_{P \in \mathcal{P}} (\frac{1}{D}\tr[U^\dagger O UP])^{2\alpha} - \sum_{P \in \mathcal{P}} (\frac{1}{D}\tr[V^\dagger O V P])^{2\alpha} |\\
        &=\frac{1}{D^{2\alpha -2}}  | \tr[ (U^\dagger O U)^{\otimes  2 \alpha}  (\Lambda^{(\alpha)})^{\otimes n} - (V^\dagger O V)^{ \otimes 2\alpha}  (\Lambda^{(\alpha)})^{\otimes n} ] | \\
        &\leq \frac{1}{D^{2\alpha-2 }} \|(\Lambda^{(\alpha)})^{\otimes n} \|_1  \| (U^\dagger O U)^{\otimes 2 \alpha}  - (V^\dagger O V)^{ \otimes 2\alpha}    \|_\infty \label{eq:levy1}
    \end{align}
    where we have applied the relation, $|\tr[AB]| \leq \|A\|_1 \| B\|_\infty$. Recall the definition Eq.~\eqref{eq:lambda}, up to reshuffling of indices equal to $(\Lambda^{(\alpha)})^{\otimes n} = (1/D^2) (\sum_{P \in \mathcal{P}_1} ( P \otimes P^* )^{\alpha})^{\otimes n}$. Using this, we can directly evaluate the first norm in the expression Eq.~\eqref{eq:levy1} for even $\alpha$, 
    \begin{align}
        \|(\Lambda^{(\alpha)})^{\otimes n} \|_1 &=  \tr[ \sqrt{ \Lambda^{(\alpha)})^{\otimes n}  }]   =  \tr[  (\Lambda^{(\alpha)})^{\otimes n} ]  \\
        &= \frac{1}{D^2} \tr[  (\sum_{P \in \mathcal{P}_1} ( P \otimes P^* )^{ \alpha })^{\otimes n}] = \frac{1}{D^2} \tr[ \id^{\otimes  2 \alpha n} ] = D^{2 \alpha-2},
    \end{align}
    where we have used that for even $\alpha$, $(\Lambda^{(\alpha)})^{\otimes n} $ is a normalized projector and so $\sqrt{(\Lambda^{(\alpha)})^{\otimes n}} = \sqrt{((\Lambda^{(\alpha)})^{\otimes n})^2} = (\Lambda^{(\alpha)})^{\otimes n}$. We note that for odd $\alpha$, we cannot make the same argument as $(\Lambda^{(\alpha)})^{\otimes n} $ is not a projector (and is instead unitary Hermitian)~\cite{Gross_2021,Haug_2024}. In that case, one finds instead that $\|(\Lambda^{(\alpha)})^{\otimes n} \|_1 =D^{2 \alpha-1} $, which is not small enough to find a Lipschitz constant that sees concentration of measure using the present method. We will instead handle odd $\alpha$ separately at the end. We have also used the fact that all Pauli strings are traceless, except for the identity. 
    
    Continuing for even integer $\alpha$, for the second norm in Eq.~\eqref{eq:levy1} we apply the triangle inequality,
    \begin{align}
        |f(U) - f(V) | &\leq \frac{D^{2 \alpha-2}}{D^{2 \alpha-2}} \|(U^\dagger)^{\otimes 2\alpha}  O^{\otimes 2\alpha}  (U^{\otimes 2\alpha} -V^{\otimes 2\alpha} )- ((V^\dagger)^{\otimes 2\alpha} -(U^\dagger)^{\otimes 2\alpha} ) O^{\otimes 2\alpha}  V^{\otimes 2\alpha} \|_\infty \\
        &\leq \|(U^\dagger)^{\otimes 2\alpha}  O^{\otimes 2\alpha}  (U^{\otimes 2\alpha} -V^{\otimes 2\alpha} )\|_\infty  + \| ((V^\dagger)^{\otimes 2\alpha} -(U^\dagger)^{\otimes 2\alpha} ) O^{\otimes 2\alpha}  V^{\otimes 2\alpha} \|_\infty  \\
        &= 2 \|U^{\otimes 2\alpha}-V^{\otimes 2\alpha} \|_\infty . \label{eq:levy2}
    \end{align}
    We have used that Schatten $p-$norms are unitarily invariant and that $\| A \| = \| A^\dagger\| $. Now we will use the following identity,
    \begin{equation}
        \|U^{\otimes 2\alpha}-V^{\otimes 2\alpha} \|_\infty \leq 2 \alpha \|U-V \|_\infty. \label{eq:levy3}
    \end{equation} 
    This relation can be derived by iteratively applying the following, 
    \begin{align}
        \| A\otimes A^* - B \otimes B^*\| &=\| (A\otimes A^* - A\otimes B^*) - (B \otimes B^* -A\otimes B^*) \| \\
        &\leq \| (A\otimes A^* - A\otimes B^*)\| + \| (B \otimes B^* -A\otimes B^*)\| \\
        &= \| A \| \| A^* - B^*\| +  \| B  -A \| \| B^* \| \\
        &= \| A - B\| (\| A \| + \| B \|). \label{eq:triangle}
    \end{align}
    where we have again utilized the triangle inequality, that $\| X \otimes Y\| = \| X\| \| Y \|$. An equivalent relation applies when replacing $A^*$ ($B^*$) with $A$ ($B$). Applying that unitaries have unit operator norm, $\| U \|_\infty = 1 $, we arrive at Eq.~\eqref{eq:levy3}. 
    
    Now we simply use that $\| A \|_\infty \leq  \|A \|_2 $ and Eq.~\eqref{eq:levy3}, such that from Eq.~\eqref{eq:levy2} we have 
    \begin{equation}
        |f(U) - f(V) | \leq 4 \alpha \|U -V \|_2. \label{eq:evenK}
    \end{equation}
    We can, therefore, bound the Lipschitz constant as $K\leq 4 \alpha$. 
    For odd $\alpha$, we will now use the above derived upper bounds for the even $\alpha-1$ and $\alpha +1$.
    In particular, for a given sampling of $U$ and $V$, then 
    \begin{equation}
        \Delta f(\alpha) := |f(U) - f(V) | = |\sum p_i^\alpha - \sum q_j^\alpha |,
    \end{equation}
    where $p_i = ({1}/{D}\tr[U^\dagger O UP])^{2}$ is a probability distribution. Now, depending on the properties of the distributions $\{ p_i\}$ and $\{ q_i\}$, either $\Delta f(\alpha-1) \leq \Delta f(\alpha) \leq \Delta f(\alpha+1) $, or $\Delta f(\alpha-1) \geq \Delta f(\alpha) \geq \Delta f(\alpha+1) $. We can rest assured that one of these cases is true, given the monotonic property of any single element $p_i^{\alpha}$ with $\alpha$. Therefore, we have in all cases that $\Delta f(\alpha)$ is bounded by the largest upper bound out of $\Delta f(\alpha-1)$ and $\Delta f(\alpha+1)$, so from Eq.~\eqref{eq:evenK},
    \begin{equation}
        \Delta f(\alpha) \leq \max \{ 4 (\alpha-1),4 (\alpha+1)\} \|U -V \|_2 \leq 4 (\alpha+1)\|U -V \|_2.
    \end{equation}
    Therefore, for any $\alpha$, we can choose a Lipschitz constant as $K =  4 (\alpha+1)$. According to Levy's Lemma, the operator stabilizer purities exhibit strong typicality. 
\end{proof}

\section{Computation of the OSE for XXZ Dual Unitary Circuits} \label{ap:DU}
In this section, we will exactly evaluate the OSE for all R\'enyi entropies in the dual unitary XXZ model, cumulating in the result of Eq.~\eqref{eq:XXZfinal} and analysis thereof.

\subsection{ZX Caclulus}
We will leverage the tools of ZX calculus. In this formalism, one can, in principle, perform any quantum mechanics calculations entirely graphically. It is algebraically complete, and in practice, it is particularly useful to work in this representation when an expression predominantly contains Cliffords and a small number of non-Cliffords. We will provide a short summary of the tools we require, following Refs.~\cite{vandewetering2020zxcalc,lopez2024exactsolutionlongrangestabilizer} closely. 

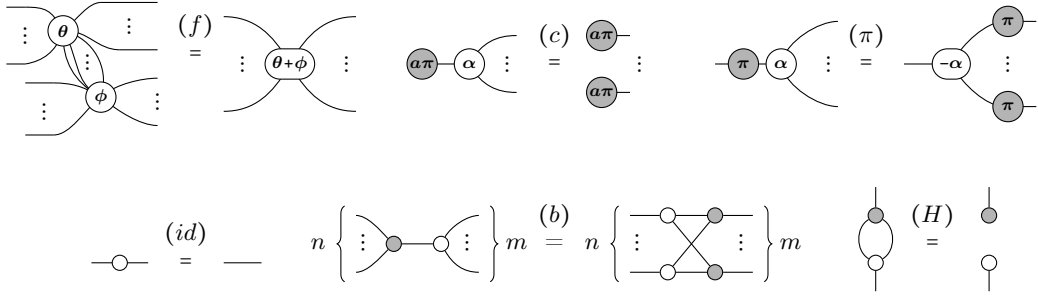
\begin{figure}
    \centering
    \begin{tikzpicture}
	\begin{pgfonlayer}{nodelayer}
		\node [style=white phase dot] (0) at (-10.5, 1.375) {$\phi$};
		\node [style=none] (1) at (-9, 1.875) {};
		\node [style=none] (3) at (-11.75, 0.375) {};
		\node [style=none] (4) at (-9, 0.625) {};
		\node [style=none, rotate=90] (7) at (-12, 1) {...};
		\node [style=none, rotate=90] (8) at (-9, 1.25) {...};
		\node [style=white phase dot] (10) at (-11.5, 3.125) {$\theta$};
		\node [style=none] (11) at (-10.5, 3.875) {};
		\node [style=none] (13) at (-10, 2.625) {};
		\node [style=none, rotate=90] (14) at (-9.75, 3.25) {...};
		\node [style=none, rotate=90] (17) at (-12.5, 3) {...};
		\node [style=none] (18) at (-9, 2.625) {};
		\node [style=none] (19) at (-9, 3.875) {};
		\node [style=none] (22) at (-12.5, 0.375) {};
		\node [style=none] (24) at (-8, 2.5) {$=$};
		\node [style=none, rotate=90] (25) at (-10.85, 2.3) {...};
		\node [style=none] (26) at (-7.25, 3.5) {};
		\node [style=none, rotate=90] (28) at (-6.75, 2.25) {...};
		\node [style=none] (29) at (-3.75, 1) {};
		\node [style=none, rotate=90] (30) at (-4, 2.25) {...};
		\node [style=white phase dot] (32) at (-5.5, 2.25) {$\ \theta\!+\!\phi\ $};
		\node [style=none] (33) at (-3.75, 3.5) {};
		\node [style=none] (34) at (-7.25, 1) {};
		\node [style=none] (35) at (-8, 3.25) {$(f)$};
		\node [style=white phase dot] (36) at (12, 2.25) {$\ -\alpha\ $};
		\node [style=none] (37) at (14.25, 3.375) {};
		\node [style=none] (38) at (9.75, 2.25) {$=$};
		\node [style=none] (39) at (10.75, 2.25) {};
		\node [style=none] (40) at (14.25, 1.125) {};
		\node [style=gray phase dot] (42) at (13.5, 1.125) {$\pi$};
		\node [style=gray phase dot] (43) at (6.5, 2.25) {$\pi$};
		\node [style=none] (44) at (9, 3.375) {};
		\node [style=white phase dot] (45) at (7.5, 2.25) {$\alpha$};
		\node [style=none, rotate=90] (46) at (8.75, 2.25) {...};
		\node [style=none, rotate=90] (48) at (13.5, 2.25) {...};
		\node [style=none] (50) at (5.75, 2.25) {};
		\node [style=none] (51) at (9, 1.125) {};
		\node [style=gray phase dot] (52) at (13.5, 3.375) {$\pi$};
		\node [style=none] (53) at (9.75, 3) {$(\pi)$};
		\node [style=none, rotate=90] (55) at (3.75, 2.25) {...};
		\node [style=white phase dot] (57) at (-0.75, 2.25) {$\alpha$};
		\node [style=none, rotate=90] (63) at (0.25, 2.25) {...};
		\node [style=none] (66) at (1.5, 3) {$(c)$};
		\node [style=white dot] (87) at (-10, -3) {};
		\node [style=none] (88) at (-10.75, -3) {};
		\node [style=none] (89) at (-9.25, -3) {};
		\node [style=none] (90) at (-7.25, -3) {};
		\node [style=none] (91) at (-6.25, -3) {};
		\node [style=none] (92) at (-8.25, -2.25) {$(id)$};
		\node [style=none] (93) at (-8.25, -3) {$=$};
		\node [style=gray dot] (131) at (-2.75, -2.5) {};
		\node [style=none] (132) at (-3.75, -1.75) {};
		\node [style=none] (133) at (-3.75, -3.25) {};
		\node [style=white dot] (134) at (-1.5, -2.5) {};
		\node [style=none] (135) at (-0.5, -1.75) {};
		\node [style=none] (136) at (-0.5, -3.25) {};
		\node [style=none] (137) at (1.5, -2.5) {=};
		\node [style=none] (138) at (3.5, -1.75) {};
		\node [style=none] (139) at (3.5, -3.25) {};
		\node [style=none] (140) at (6.75, -1.75) {};
		\node [style=none] (141) at (6.75, -3.25) {};
		\node [style=white dot] (142) at (4.5, -1.75) {};
		\node [style=white dot] (143) at (4.5, -3.25) {};
		\node [style=gray dot] (144) at (5.75, -1.75) {};
		\node [style=gray dot] (145) at (5.75, -3.25) {};
		\node [style=none, rotate=90] (146) at (-3.5, -2.45) {$\cdots$};
		\node [style=none, rotate=90] (147) at (-0.75, -2.45) {$\cdots$};
		\node [style=none] (148) at (-4.25, -3.5) {};
		\node [style=none] (149) at (-4.25, -1.5) {};
		\node [style=none] (150) at (-4.75, -2.5) {$n$};
		\node [style=none] (151) at (0, -1.5) {};
		\node [style=none] (152) at (0, -3.5) {};
		\node [style=none] (153) at (0.5, -2.5) {$m$};
		\node [style=none, rotate=90] (154) at (3.75, -2.45) {$\cdots$};
		\node [style=none, rotate=90] (155) at (6.5, -2.45) {$\cdots$};
		\node [style=none] (156) at (3, -3.5) {};
		\node [style=none] (157) at (3, -1.5) {};
		\node [style=none] (158) at (2.5, -2.5) {$n$};
		\node [style=none] (159) at (7.25, -1.5) {};
		\node [style=none] (160) at (7.25, -3.5) {};
		\node [style=none] (161) at (7.75, -2.5) {$m$};
		\node [style=none] (162) at (1.5, -1.75) {$(b)$};
		\node [style=gray dot] (168) at (10, -1.75) {};
		\node [style=white dot] (169) at (10, -3) {};
		\node [style=none] (170) at (10, -1) {};
		\node [style=none] (171) at (10, -3.75) {};
		\node [style=none] (172) at (11.5, -2.5) {$=$};
		\node [style=none] (173) at (11.5, -1.75) {$(H)$};
		\node [style=gray dot] (174) at (13, -1.75) {};
		\node [style=none] (175) at (13, -1) {};
		\node [style=white dot] (176) at (13, -3) {};
		\node [style=none] (177) at (13, -3.75) {};
		\node [style=none] (178) at (-13, 3.75) {};
		\node [style=none] (179) at (-13, 2.25) {};
		\node [style=none] (180) at (-12.5, 2.25) {};
		\node [style=none] (181) at (-12.5, 3.75) {};
		\node [style=none] (182) at (-11.75, 1.75) {};
		\node [style=none] (183) at (-12.5, 1.75) {};
		\node [style=gray phase dot] (184) at (-2, 2.25) {$a \pi$};
		\node [style=gray phase dot] (185) at (2.75, 3) {$a \pi$};
		\node [style=gray phase dot] (186) at (2.75, 1.5) {$a \pi$};
		\node [style=none] (187) at (3.5, 3) {};
		\node [style=none] (188) at (3.5, 1.5) {};
		\node [style=none] (189) at (0.5, 3) {};
		\node [style=none] (190) at (0.5, 1.5) {};
		\node [style=none] (191) at (1.5, 2.25) {$=$};
	\end{pgfonlayer}
	\begin{pgfonlayer}{edgelayer}
		\draw [in=-141, out=0, looseness=0.75] (3.center) to (0);
		\draw [in=180, out=-39, looseness=0.75] (0) to (4.center);
		\draw [in=180, out=39, looseness=0.75] (0) to (1.center);
		\draw [in=180, out=0] (10) to (13.center);
		\draw [in=180, out=39, looseness=0.75] (10) to (11.center);
		\draw [bend left=45] (10) to (0);
		\draw (11.center) to (19.center);
		\draw (13.center) to (18.center);
		\draw (22.center) to (3.center);
		\draw [bend left] (0) to (10);
		\draw [in=-120, out=0] (34.center) to (32);
		\draw [in=180, out=-60] (32) to (29.center);
		\draw [in=180, out=60] (32) to (33.center);
		\draw [in=120, out=0] (26.center) to (32);
		\draw [in=180, out=-75, looseness=0.75] (45) to (51.center);
		\draw [in=180, out=75, looseness=0.75] (45) to (44.center);
		\draw [in=180, out=0, looseness=0.50] (43) to (45);
		\draw (50.center) to (43);
		\draw [in=-60, out=180, looseness=0.75] (42) to (36);
		\draw [in=0, out=180, looseness=0.75] (36) to (39.center);
		\draw [in=60, out=180, looseness=0.75] (52) to (36);
		\draw (37.center) to (52);
		\draw (40.center) to (42);
		\draw (88.center) to (89.center);
		\draw (90.center) to (91.center);
		\draw [in=150, out=-60] (10) to (0);
		\draw [in=-135, out=0, looseness=0.75] (133.center) to (131);
		\draw [in=0, out=135, looseness=0.75] (131) to (132.center);
		\draw (131) to (134);
		\draw [in=180, out=60, looseness=0.75] (134) to (135.center);
		\draw [in=-180, out=-60] (134) to (136.center);
		\draw (139.center) to (143);
		\draw (138.center) to (142);
		\draw (142) to (145);
		\draw (143) to (144);
		\draw (142) to (144);
		\draw (143) to (145);
		\draw (144) to (140.center);
		\draw (145) to (141.center);
		\draw [style=brace edge] (148.center) to (149.center);
		\draw [style=brace edge] (151.center) to (152.center);
		\draw [style=brace edge] (156.center) to (157.center);
		\draw [style=brace edge] (159.center) to (160.center);
		\draw [bend left=60] (168) to (169);
		\draw [bend right=60] (168) to (169);
		\draw (169) to (171.center);
		\draw (170.center) to (168);
		\draw (175.center) to (174);
		\draw (176) to (177.center);
		\draw [bend left] (10) to (180.center);
		\draw (179.center) to (180.center);
		\draw (181.center) to (178.center);
		\draw [bend left, looseness=1.25] (181.center) to (10);
		\draw [bend left=15] (182.center) to (0);
		\draw (182.center) to (183.center);
		\draw (184) to (57);
		\draw (188.center) to (186);
		\draw (187.center) to (185);
		\draw [bend right] (57) to (190.center);
		\draw [bend left] (57) to (189.center);
	\end{pgfonlayer}
\end{tikzpicture}
    \caption{A summary of graphical rewrite rules in ZX calculus. These are: $(f)$usion of spiders, the $(c)$opy rule, ($\pi$)-commutation, $(id)$entity, $(b)$ialbegra rule, and the $(H)$opf rule. The Hopf rule can be derived from the others but turns out to be particularly useful.  }
    \label{fig:zx-rules}
\end{figure}

The basic objects in ZX calculus are Z and X spiders, defined respectively as 
\begin{equation}
    \begin{tikzpicture}
	\begin{pgfonlayer}{nodelayer}
		\node [style=white dot] (0) at (0, 0) {$\theta$};
		\node [style=none, rotate=90] (1) at (1, 0) {...};
		\node [style=none, rotate=90] (2) at (-1, 0) {...};
		\node [style=none] (3) at (1.5, 1.5) {};
		\node [style=none] (4) at (1.5, 1.25) {};
		\node [style=none] (5) at (1.5, 1) {};
		\node [style=none] (6) at (1.5, -0.5) {};
		\node [style=none] (7) at (1.5, -1) {};
		\node [style=none] (8) at (-1.5, 1.25) {};
		\node [style=none] (9) at (-1.5, 0.75) {};
		\node [style=none] (10) at (-1.5, -0.75) {};
		\node [style=none] (11) at (-1.5, -1.25) {};
		\node [style=none] (12) at (1.5, 0.75) {};
		\node [style=none] (14) at (-1.75, -1.5) {};
		\node [style=none] (15) at (-1.75, 1.5) {};
		\node [style=none] (16) at (1.75, 1.75) {};
		\node [style=none] (17) at (1.75, -1.25) {};
		\node [style=none] (18) at (-2.25, 0) {$m$};
		\node [style=none] (19) at (2.25, 0.25) {$n$};
	\end{pgfonlayer}
	\begin{pgfonlayer}{edgelayer}
		\draw [style=simple, bend left, looseness=0.75] (8.center) to (0);
		\draw [style=simple, bend left=15] (9.center) to (0);
		\draw [style=simple, bend right] (10.center) to (0);
		\draw [style=simple, bend right=45, looseness=0.75] (11.center) to (0);
		\draw [style=simple, bend left=45] (0) to (3.center);
		\draw [style=simple, bend left] (0) to (4.center);
		\draw [style=simple, bend left] (0) to (12.center);
		\draw [style=simple, bend right=15] (0) to (6.center);
		\draw [style=simple, bend right] (0) to (7.center);
		\draw [style=brace edge] (14.center) to (15.center);
		\draw [style=brace edge] (16.center) to (17.center);
	\end{pgfonlayer}
\end{tikzpicture}:= \ket{0}^{\otimes m}\bra{0}^{\otimes n}+ \ex^{i \theta}\ket{1}^{\otimes m}\bra{1}^{\otimes n}, \qquad \begin{tikzpicture}
	\begin{pgfonlayer}{nodelayer}
		\node [style=white dot] (0) at (0, 0) {$\theta$};
		\node [style=none, rotate=90] (1) at (1, 0) {...};
		\node [style=none, rotate=90] (2) at (-1, 0) {...};
		\node [style=none] (3) at (1.5, 1.5) {};
		\node [style=none] (4) at (1.5, 1.25) {};
		\node [style=none] (5) at (1.5, 1) {};
		\node [style=none] (6) at (1.5, -0.5) {};
		\node [style=none] (7) at (1.5, -1) {};
		\node [style=none] (8) at (-1.5, 1.25) {};
		\node [style=none] (9) at (-1.5, 0.75) {};
		\node [style=none] (10) at (-1.5, -0.75) {};
		\node [style=none] (11) at (-1.5, -1.25) {};
		\node [style=none] (12) at (1.5, 0.75) {};
		\node [style=gray dot] (13) at (0, 0) {$\theta$};
		\node [style=none] (14) at (-1.75, -1.5) {};
		\node [style=none] (15) at (-1.75, 1.5) {};
		\node [style=none] (16) at (1.75, 1.75) {};
		\node [style=none] (17) at (1.75, -1.25) {};
		\node [style=none] (18) at (-2.25, 0) {$m$};
		\node [style=none] (19) at (2.25, 0.25) {$n$};
	\end{pgfonlayer}
	\begin{pgfonlayer}{edgelayer}
		\draw [style=simple, bend left, looseness=0.75] (8.center) to (0);
		\draw [style=simple, bend left=15] (9.center) to (0);
		\draw [style=simple, bend right] (10.center) to (0);
		\draw [style=simple, bend right=45, looseness=0.75] (11.center) to (0);
		\draw [style=simple, bend left=45] (0) to (3.center);
		\draw [style=simple, bend left] (0) to (4.center);
		\draw [style=simple, bend left] (0) to (12.center);
		\draw [style=simple, bend right=15] (0) to (6.center);
		\draw [style=simple, bend right] (0) to (7.center);
		\draw [style=brace edge] (14.center) to (15.center);
		\draw [style=brace edge] (16.center) to (17.center);
	\end{pgfonlayer}
\end{tikzpicture}:=\ket{+}^{\otimes m}\bra{+}^{\otimes n}+ \ex^{i \theta}\ket{-}^{\otimes m}\bra{-}^{\otimes n}.
\end{equation}
If $\theta=0$, we omit it in the diagram. Remarkably, one can rewrite any quantum circuit in terms of these fundamental units. Then, it is possible to prove a number of rewrite rules, as detailed in Fig.~\ref{fig:zx-rules}. ZX can be seen as an extension to regular tensor network expressions, where we include Pauli commutation relations in the form of the additional rules of Fig.~\ref{fig:zx-rules}. As such, the directionality of the interior wires does not matter, and if there are no open wires, then the diagram corresponds to a constant. We also point out that the Pauli matrices graphically are 
\begin{equation}
    \sigma_x = \begin{tikzpicture}
	\begin{pgfonlayer}{nodelayer}
		\node [style=none] (0) at (-1, 0) {};
		\node [style=none] (1) at (1, 0) {};
		\node [style=gray dot] (2) at (0, 0) {$\pi$};
	\end{pgfonlayer}
	\begin{pgfonlayer}{edgelayer}
		\draw (0.center) to (1.center);
	\end{pgfonlayer}
\end{tikzpicture}
, \quad \sigma_z =\begin{tikzpicture}
	\begin{pgfonlayer}{nodelayer}
		\node [style=none] (0) at (-1, 0) {};
		\node [style=none] (1) at (1, 0) {};
		\node [style=white dot] (2) at (0, 0) {$\pi$};
	\end{pgfonlayer}
	\begin{pgfonlayer}{edgelayer}
		\draw (0.center) to (1.center);
	\end{pgfonlayer}
\end{tikzpicture}
, \quad \sigma_y = -i\left( \begin{tikzpicture}
	\begin{pgfonlayer}{nodelayer}
		\node [style=none] (0) at (-1, 0) {};
		\node [style=none] (1) at (0.75, 0) {};
		\node [style=white dot] (2) at (-0.5, 0) {$\pi$};
		\node [style=gray dot] (3) at (0.25, 0) {$\pi$};
	\end{pgfonlayer}
	\begin{pgfonlayer}{edgelayer}
		\draw (0.center) to (1.center);
	\end{pgfonlayer}
\end{tikzpicture}
 \right).
\end{equation}

\subsection{Dual Unitary XXZ Model}
Layers of the dual unitary XXZ circuit involve $2-$body gates which in alternating layers act on even ($\dots,(0,1),(2,3),\dots$) and odd ($\dots,(1,2),(3,4),\dots$) labeled sites respectively, 
\begin{equation}
    U_{e/o} = (\mathbb{S} \cdot \exp(-i J \sigma_z \otimes \sigma_z ))^{\otimes N/2},
    \label{eq:ziobno}
\end{equation}
where $\mathbb{S}$ is the two-qubit SWAP gate, and where we can take the number of sites $N \to \infty$. Applying a layer of this evolution to a local operator $O = \sigma_x^{(j)}$ (a single site $\sigma_x$ operator on the $j^{th}$ site, with implicit identities elsewhere), {each two-body gate $\exp(-i J \sigma_z \otimes \sigma_z )$} which has overlapping support with $\sigma_x$ anti-commutes with this gate, and all other gates commute (due to independent support). Then the SWAP part just translates the operator $\sigma_x$ to a different site (assume $j$ is even without loss of generality):
\begin{equation}
    \mathbb{S}^{\otimes N/2} \exp(-i \sum_{i\, \text{even}}J \sigma_z^{(i)} \otimes \sigma_z^{(i+1)} )) \sigma_x^{(j)} = \sigma_x^{(j+1)}  \exp(-i J\sum_{i\, \text{even}} (-1)^{\delta_{ji}} \sigma_z^{(i)} \otimes \sigma_z^{(i+1)} )).
\end{equation}
The equivalent occurs for the odd layers. Then after $(t/2)-$steps with Floquet evolution operator $U_t= (U_oU_e)^{t/2}$, commuting $\sigma_x^{(j)}$ through $U_t$ flips the sign of the $ZZ$ gate, and otherwise the gate cancels due to $U$ vs $U^\dagger $ 
\begin{equation}
    O_U = U_t^\dagger \sigma_x^{(j)} U_t = \sigma_x^{(j+t)} \exp(-i 2J \sum_{i=0}^{t-1} \sigma_z^{(t+j)} \otimes \sigma_z^{(j+i)}). \label{eq:diag_commuted}
\end{equation}
Here, notice that $O_U$ now has support on $t$ qubits; cf. the full lightcone of $2t$ sites. This is because the other $\exp(-i J \sigma_z \otimes \sigma_z )$ gates commute with each other after implementing the SWAPs.

\subsection{Proof for Local Pauli Observables}
Recall the tensor network expression for the OSE (Fig.~\ref{fig:OSE_tn} and Eq.~\eqref{eq:tn_form}), and moreover that from the the local operator's light-cone, its nontrivial support after $t$ steps is on only $2t$ sites,
\begin{equation}
    \frac{1}{2^{2 N \alpha}}\tr[(O_U \otimes O_U^* )^{\otimes \alpha } (\Lambda^{(\alpha)})^{\otimes n}] = \frac{1}{2^{2 t \alpha}}\tr[(O_U \otimes O_U^* )^{\otimes \alpha } (\Lambda^{(\alpha)})^{\otimes t}] . \label{eq:tn_form1}
\end{equation}
Following a similar argument to \cite{lopez2024exactsolutionlongrangestabilizer}, we decompose
\begin{equation}
    \Lambda^{(\alpha )} = \sum_{P \in \mathcal{P}_1} (P \otimes P^*)^{\otimes \alpha} =4  \Lambda^{(\alpha )}_x \Lambda^{(\alpha )}_z,
\end{equation}
with 
\begin{align}
    &\Lambda^{(\alpha )}_x= \frac{1}{2}(\sigma_0^{\ot 2 \alpha} +  \sigma_x^{\ot 2 \alpha}), \text{ and}\\
    &\Lambda^{(\alpha )}_z=\frac{1}{2}(\sigma_0^{\ot 2 \alpha} +  \sigma_z^{\ot 2 \alpha}). 
\end{align}
We can rewrite these simply using the $ZX$ calculus notation, 
\begin{equation}
    \Lambda^{(\alpha )}_z=\begin{tikzpicture}
	\begin{pgfonlayer}{nodelayer}
		\node [style=gray dot] (0) at (0, 0) {};
		\node [style=white dot] (1) at (2, 2) {};
		\node [style=white dot] (2) at (2, 1) {};
		\node [style=white dot] (3) at (2, -1) {};
		\node [style=white dot] (4) at (2, -2) {};
		\node [style=none] (5) at (2.5, 1) {};
		\node [style=none] (6) at (1.5, 1) {};
		\node [style=none] (7) at (1.5, 2) {};
		\node [style=none] (8) at (2.5, 2) {};
		\node [style=none] (9) at (2.5, 2) {};
		\node [style=none] (10) at (1.5, -1) {};
		\node [style=none] (11) at (2.5, -1) {};
		\node [style=none] (12) at (1.5, -2) {};
		\node [style=none] (13) at (2.5, -2) {};
		\node [style=none, rotate=90] (14) at (2, 0) {...};
	\end{pgfonlayer}
	\begin{pgfonlayer}{edgelayer}
		\draw [style=simple] (0) to (1);
		\draw [style=simple] (0) to (2);
		\draw [style=simple] (0) to (3);
		\draw [style=simple] (0) to (4);
		\draw (7.center) to (1);
		\draw (1) to (9.center);
		\draw (6.center) to (2);
		\draw (2) to (5.center);
		\draw (10.center) to (3);
		\draw (3) to (11.center);
		\draw (12.center) to (4);
		\draw (4) to (13.center);
	\end{pgfonlayer}
\end{tikzpicture}, \quad \Lambda^{(\alpha )}_x= \begin{tikzpicture}
	\begin{pgfonlayer}{nodelayer}
		\node [style=none] (5) at (2.5, 1) {};
		\node [style=none] (6) at (1.5, 1) {};
		\node [style=none] (7) at (1.5, 2) {};
		\node [style=none] (8) at (2.5, 2) {};
		\node [style=none] (9) at (2.5, 2) {};
		\node [style=none] (10) at (1.5, -1) {};
		\node [style=none] (11) at (2.5, -1) {};
		\node [style=none] (12) at (1.5, -2) {};
		\node [style=none] (13) at (2.5, -2) {};
		\node [style=none, rotate=90] (14) at (2, 0) {...};
		\node [style=white dot] (15) at (0, 0) {};
		\node [style=gray dot] (16) at (2, 2) {};
		\node [style=gray dot] (17) at (2, 1) {};
		\node [style=gray dot] (18) at (2, -1) {};
		\node [style=gray dot] (19) at (2, -2) {};
		\node [style=gray dot] (20) at (2, -1) {};
	\end{pgfonlayer}
	\begin{pgfonlayer}{edgelayer}
		\draw [style=simple] (11.center) to (20);
		\draw [style=simple] (20) to (10.center);
		\draw [style=simple] (19) to (12.center);
		\draw [style=simple] (19) to (13.center);
		\draw [style=simple] (17) to (5.center);
		\draw [style=simple] (16) to (9.center);
		\draw [style=simple] (16) to (7.center);
		\draw [style=simple] (17) to (6.center);
		\draw [style=simple] (20) to (15);
		\draw [style=simple] (17) to (15);
		\draw [style=simple] (16) to (15);
		\draw [style=simple] (19) to (15);
	\end{pgfonlayer}
\end{tikzpicture}, \label{eq:zx_lambda}
\end{equation}
where there are $2\alpha$ horizontal wires ($\Lambda^{(\alpha )}$ acts on $2\alpha$ replica spaces of a single qubit). In the above, and for the rest of the proof, we ignore any normalizations. Then, given the final expression will be valid for any $\alpha \geq 1$ and any parameter $J$, we can find the correct normalization at the end given that $\{ \Pi_i \}$ is a probability distribution,
\begin{equation}
    \frac{1}{2^{2 t}}\tr[(O_U \otimes O_U^* ) (\Lambda^{(1)})^{\otimes t}] \overset{!}{=}1,
\end{equation}
and that the magic is zero for $J=0$, as this corresponds to a circuit of SWAPs (which is clearly Clifford). 
Further, the two-site $ZZ$ rotation gate in Eq.~\eqref{eq:diag_commuted} also admits a simple expression in the ZX representation~\cite{lopez2024exactsolutionlongrangestabilizer},
\begin{equation}
    \exp(-i 2J \sigma_z^{(a)} \otimes \sigma_z^{(b)}) = \begin{tikzpicture}
	\begin{pgfonlayer}{nodelayer}
		\node [style=white dot] (0) at (0, 1) {};
		\node [style=white dot] (1) at (0, -1) {};
		\node [style=gray dot] (2) at (0, 0) {};
		\node [style=white dot] (3) at (1, 0) {4J};
		\node [style=none] (4) at (-2, 1) {};
		\node [style=none] (5) at (2, 1) {};
		\node [style=none] (6) at (-2, -1) {};
		\node [style=none] (7) at (2, -1) {};
	\end{pgfonlayer}
	\begin{pgfonlayer}{edgelayer}
		\draw [style=simple] (5.center) to (0);
		\draw [style=simple] (0) to (4.center);
		\draw [style=simple] (0) to (2);
		\draw [style=simple] (2) to (3);
		\draw [style=simple] (2) to (1);
		\draw [style=simple] (6.center) to (1);
		\draw [style=simple] (1) to (7.center);
	\end{pgfonlayer}
\end{tikzpicture}. \label{eq:ZZ}
\end{equation}
Using this, the ZX diagram for Eq.~\eqref{eq:diag_commuted} is 
\begin{equation}
    O_U= \begin{tikzpicture}
	\begin{pgfonlayer}{nodelayer}
		\node [style=white dot] (15) at (-2, 2.75) {};
		\node [style=white dot] (16) at (-2, 0.75) {};
		\node [style=gray dot] (17) at (-2, 1.75) {};
		\node [style=white dot] (18) at (-1, 1.75) {${\theta}$};
		\node [style=none] (21) at (-4, 0.75) {};
		\node [style=none] (22) at (4.5, 0.75) {};
		\node [style=white dot] (23) at (0, 2.75) {};
		\node [style=white dot] (24) at (0, -0.75) {};
		\node [style=gray dot] (25) at (0, 1.75) {};
		\node [style=white dot] (26) at (1, 1.75) {${\theta}$};
		\node [style=none] (28) at (4.5, 2.75) {};
		\node [style=none] (29) at (-4, -0.75) {};
		\node [style=none] (30) at (4.5, -0.75) {};
		\node [style=none, rotate=90] (31) at (3.5, -1.5) {...};
		\node [style=white dot] (32) at (3.5, 2.75) {};
		\node [style=white dot] (33) at (3.5, -2.5) {};
		\node [style=gray dot] (34) at (3.5, 1.75) {};
		\node [style=white dot] (35) at (4.5, 1.75) {${\theta}$};
		\node [style=none] (36) at (-4, -2.5) {};
		\node [style=none] (37) at (4.5, -2.5) {};
		\node [style=none] (38) at (2.5, 1.75) {...};
		\node [style=gray dot] (39) at (-3, 2.75) {$\pi$};
		\node [style=none] (40) at (-4, 2.75) {};
		\node [style=none] (41) at (3.5, -2) {};
		\node [style=none] (42) at (3.5, -1) {};
	\end{pgfonlayer}
	\begin{pgfonlayer}{edgelayer}
		\draw [style=simple] (15) to (17);
		\draw [style=simple] (17) to (18);
		\draw [style=simple] (17) to (16);
		\draw [style=simple] (21.center) to (16);
		\draw [style=simple] (16) to (22.center);
		\draw [style=simple] (28.center) to (23);
		\draw [style=simple] (23) to (25);
		\draw [style=simple] (25) to (26);
		\draw [style=simple] (25) to (24);
		\draw [style=simple] (29.center) to (24);
		\draw [style=simple] (24) to (30.center);
		\draw [style=simple] (15) to (23);
		\draw [style=simple] (32) to (34);
		\draw [style=simple] (34) to (35);
		\draw [style=simple] (36.center) to (33);
		\draw [style=simple] (33) to (37.center);
		\draw [style=simple] (39) to (15);
		\draw [style=simple] (40.center) to (39);
		\draw [style=simple] (34) to (42.center);
		\draw [style=simple] (41.center) to (33);
	\end{pgfonlayer}
\end{tikzpicture}, \label{eq:ZZa}
\end{equation}
where the horizontal wires are the Hilbert spaces of qubits $t,t-1,\dots,0$ from top to bottom, and for brevity we write $\theta := 4J$, and $\bar{\theta}:=-4J$. $O_U^*$ has the same diagram, but one replaces the angle $\theta$ with its negative, $\overline{\theta}$. Substituting these expressions \eqref{eq:zx_lambda}-\eqref{eq:ZZa} into Eq.~\eqref{eq:tn_form1}, we have the full ZX diagram for the OSE 
\begin{equation}
    \tr[(O_U \otimes O_U^* )^{\otimes \alpha } (\Lambda^{(\alpha)})^{\otimes t}] = \tr \left[ \begin{tikzpicture}
	\begin{pgfonlayer}{nodelayer}
		\node [style=white dot] (15) at (-1.75, 7) {};
		\node [style=white dot] (16) at (-1.75, 5) {};
		\node [style=gray dot] (17) at (-1.75, 6) {};
		\node [style=white dot] (18) at (-0.75, 6) {$\theta$};
		\node [style=white dot] (23) at (0.25, 7) {};
		\node [style=white dot] (24) at (0.25, 3.5) {};
		\node [style=gray dot] (25) at (0.25, 6) {};
		\node [style=white dot] (26) at (1.25, 6) {$\theta$};
		\node [style=none, rotate=90] (31) at (3.75, 2.75) {...};
		\node [style=white dot] (32) at (3.75, 7) {};
		\node [style=white dot] (33) at (3.75, 1.75) {};
		\node [style=gray dot] (34) at (3.75, 6) {};
		\node [style=white dot] (35) at (4.75, 6) {$\theta$};
		\node [style=none] (38) at (2.75, 6) {...};
		\node [style=gray dot] (39) at (-2.75, 7) {$\pi$};
		\node [style=none] (41) at (3.75, 2.25) {};
		\node [style=none] (42) at (3.75, 3.25) {};
		\node [style=white dot] (43) at (-1.75, 0.5) {};
		\node [style=white dot] (44) at (-1.75, -1.5) {};
		\node [style=gray dot] (45) at (-1.75, -0.5) {};
		\node [style=white dot] (46) at (-0.75, -0.5) {$\overline{\theta}$};
		\node [style=white dot] (49) at (0.25, 0.5) {};
		\node [style=white dot] (50) at (0.25, -3) {};
		\node [style=gray dot] (51) at (0.25, -0.5) {};
		\node [style=white dot] (52) at (1.25, -0.5) {$\overline{\theta}$};
		\node [style=none, rotate=90] (56) at (3.75, -3.75) {...};
		\node [style=white dot] (57) at (3.75, 0.5) {};
		\node [style=white dot] (58) at (3.75, -4.75) {};
		\node [style=gray dot] (59) at (3.75, -0.5) {};
		\node [style=white dot] (60) at (4.75, -0.5) {$\overline{\theta}$};
		\node [style=none] (63) at (2.75, -0.5) {...};
		\node [style=gray dot] (64) at (-2.75, 0.5) {$\pi$};
		\node [style=none] (66) at (3.75, -4.25) {};
		\node [style=none] (67) at (3.75, -3.25) {};
		\node [style=none] (84) at (-4.5, 7) {};
		\node [style=none] (85) at (-4.5, 5) {};
		\node [style=none] (86) at (-4.5, 3.5) {};
		\node [style=none] (87) at (-4.5, 1.75) {};
		\node [style=none] (88) at (-4.5, 0.5) {};
		\node [style=none] (89) at (-4.5, -1.5) {};
		\node [style=none] (90) at (-4.5, -3) {};
		\node [style=gray dot] (91) at (-3.5, 7) {};
		\node [style=gray dot] (92) at (-3.5, 5) {};
		\node [style=gray dot] (93) at (-3.5, 3.5) {};
		\node [style=gray dot] (94) at (-3.5, 1.75) {};
		\node [style=gray dot] (95) at (-3.5, 0.5) {};
		\node [style=gray dot] (96) at (-3.5, -1.5) {};
		\node [style=gray dot] (97) at (-3.5, -3) {};
		\node [style=gray dot] (98) at (-3.5, -4.75) {};
		\node [style=none] (99) at (5.5, 7) {};
		\node [style=none] (100) at (5.5, 1.75) {};
		\node [style=none] (101) at (5.5, 0.5) {};
		\node [style=none] (102) at (-4.5, -4.75) {};
		\node [style=none] (103) at (5.5, -4.75) {};
		\node [style=white dot] (104) at (4.75, 7) {};
		\node [style=white dot] (105) at (4.75, 1.75) {};
		\node [style=white dot] (106) at (4.75, 0.5) {};
		\node [style=white dot] (107) at (4.75, -3) {};
		\node [style=white dot] (108) at (4.75, -4.75) {};
		\node [style=none] (109) at (5.5, -3) {};
		\node [style=none] (110) at (5.5, -1.5) {};
		\node [style=none] (111) at (5.5, 3.5) {};
		\node [style=white dot] (112) at (4.75, -1.5) {};
		\node [style=white dot] (113) at (4.75, 3.5) {};
		\node [style=white dot] (114) at (4.75, 5) {};
		\node [style=none] (115) at (5.5, 5) {};
		\node [style=white dot] (182) at (-6.5, 3) {};
		\node [style=white dot] (183) at (-6.5, 1) {};
		\node [style=white dot] (184) at (-6.5, -1) {};
		\node [style=white dot] (185) at (-6.5, -3) {};
		\node [style=gray dot] (186) at (7.5, 3) {};
		\node [style=gray dot] (187) at (7.5, 1) {};
		\node [style=gray dot] (188) at (7.5, -1) {};
		\node [style=gray dot] (189) at (7.5, -3) {};
		\node [style=none, rotate=90] (190) at (-0.5, -5.75) {...};
		\node [style=none] (191) at (1.5, -5.75) {($2\alpha$ copies)};
		\node [style=none, rotate=90] (192) at (7.5, -4.75) {...};
		\node [style=none, rotate=90] (193) at (-6.5, -4.75) {...};
		\node [style=none] (194) at (-6, 2.25) {};
		\node [style=none] (195) at (-6.25, 1.75) {};
		\node [style=none] (196) at (-6, 0.25) {};
		\node [style=none] (197) at (-6.25, -0.25) {};
		\node [style=none] (198) at (-6, -1.75) {};
		\node [style=none] (199) at (-6.25, -2.25) {};
		\node [style=none] (200) at (-6, -3.75) {};
		\node [style=none] (201) at (-6.25, -4) {};
		\node [style=none] (202) at (7, 2.25) {};
		\node [style=none] (203) at (7.25, 2) {};
		\node [style=none] (204) at (7, 0.25) {};
		\node [style=none] (205) at (7.25, 0) {};
		\node [style=none] (206) at (7, -1.75) {};
		\node [style=none] (207) at (7.25, -2) {};
		\node [style=none] (208) at (7, -3.75) {};
		\node [style=none] (209) at (7.25, -4) {};
	\end{pgfonlayer}
	\begin{pgfonlayer}{edgelayer}
		\draw [style=simple] (15) to (17);
		\draw [style=simple] (17) to (18);
		\draw [style=simple] (17) to (16);
		\draw [style=simple] (23) to (25);
		\draw [style=simple] (25) to (26);
		\draw [style=simple] (25) to (24);
		\draw [style=simple] (15) to (23);
		\draw [style=simple] (32) to (34);
		\draw [style=simple] (34) to (35);
		\draw [style=simple] (39) to (15);
		\draw [style=simple] (34) to (42.center);
		\draw [style=simple] (41.center) to (33);
		\draw [style=simple] (43) to (45);
		\draw [style=simple, in=180, out=0] (45) to (46);
		\draw [style=simple] (45) to (44);
		\draw [style=simple] (49) to (51);
		\draw [style=simple] (51) to (52);
		\draw [style=simple] (51) to (50);
		\draw [style=simple] (43) to (49);
		\draw [style=simple] (57) to (59);
		\draw [style=simple] (59) to (60);
		\draw [style=simple] (64) to (43);
		\draw [style=simple] (59) to (67.center);
		\draw [style=simple] (66.center) to (58);
		\draw [style=simple] (49) to (57);
		\draw [style=simple] (23) to (32);
		\draw [style=simple] (92) to (16);
		\draw [style=simple] (85.center) to (92);
		\draw [style=simple] (84.center) to (15);
		\draw [style=simple] (86.center) to (24);
		\draw [style=simple] (87.center) to (33);
		\draw [style=simple] (96) to (44);
		\draw [style=simple] (89.center) to (96);
		\draw [style=simple] (88.center) to (95);
		\draw [style=simple] (90.center) to (50);
		\draw [style=simple] (102.center) to (58);
		\draw [style=simple] (58) to (103.center);
		\draw [style=simple] (57) to (101.center);
		\draw [style=simple] (33) to (100.center);
		\draw [style=simple] (32) to (99.center);
		\draw [style=simple] (44) to (110.center);
		\draw [style=simple] (50) to (109.center);
		\draw [style=simple] (24) to (111.center);
		\draw [style=simple] (95) to (64);
		\draw [style=simple] (16) to (115.center);
		\draw [style=simple] (182) to (91);
		\draw [style=simple] (182) to (95);
		\draw [style=simple] (183) to (92);
		\draw [style=simple] (183) to (96);
		\draw [style=simple] (184) to (93);
		\draw [style=simple] (184) to (97);
		\draw [style=simple] (185) to (94);
		\draw [style=simple] (185) to (98);
		\draw [style=simple] (189) to (108);
		\draw [style=simple] (105) to (189);
		\draw [style=simple] (113) to (188);
		\draw [style=simple] (188) to (107);
		\draw [style=simple] (112) to (187);
		\draw [style=simple] (187) to (114);
		\draw [style=simple] (104) to (186);
		\draw [style=simple] (186) to (106);
		\draw [style=simple] (182) to (194.center);
		\draw [style=simple] (182) to (195.center);
		\draw [style=simple] (183) to (196.center);
		\draw [style=simple] (183) to (197.center);
		\draw [style=simple] (184) to (198.center);
		\draw [style=simple] (184) to (199.center);
		\draw [style=simple] (185) to (200.center);
		\draw [style=simple] (185) to (201.center);
		\draw [style=simple] (202.center) to (186);
		\draw [style=simple] (186) to (203.center);
		\draw [style=simple] (187) to (204.center);
		\draw [style=simple] (187) to (205.center);
		\draw [style=simple] (188) to (206.center);
		\draw [style=simple] (188) to (207.center);
		\draw [style=simple] (189) to (208.center);
		\draw [style=simple] (189) to (209.center);
	\end{pgfonlayer}
\end{tikzpicture} \right]. \label{eq:full_tn}
\end{equation}
We apply tools of ZX calculus to simplify this expression significantly. 

First we consider a single qubit replica space $0 \leq i < t$ (other than site where $\sigma_x$ acts), as well as its connecting spiders $\Lambda_{x/z}^{(\alpha)}$,
\begin{equation}
    \begin{split}
        \begin{tikzpicture}
	\begin{pgfonlayer}{nodelayer}
		\node [style=none] (38) at (0, 1.25) {...};
		\node [style=none] (63) at (0, -2) {...};
		\node [style=white dot] (210) at (0, 2.75) {};
		\node [style=gray dot] (211) at (-4.25, 1.25) {};
		\node [style=gray dot] (212) at (-2, 1.25) {};
		\node [style=gray dot] (213) at (2, 1.25) {};
		\node [style=gray dot] (214) at (4, 1.25) {};
		\node [style=white dot] (215) at (-4.25, -1.25) {};
		\node [style=white dot] (216) at (-2, -1.25) {};
		\node [style=white dot] (217) at (2, -1.25) {};
		\node [style=white dot] (218) at (4, -1.25) {};
		\node [style=gray dot] (219) at (0, -2.75) {};
		\node [style=none] (226) at (0, 1.75) {$2\alpha$};
		\node [style=none] (227) at (0, -1.5) {$2\alpha$};
		\node [style=white dot] (228) at (-4.5, -0.25) {};
		\node [style=white dot] (229) at (-2.25, -0.25) {};
		\node [style=white dot] (230) at (1.75, -0.25) {};
		\node [style=white dot] (231) at (3.75, -0.25) {};
		\node [style=gray dot] (235) at (-5.25, -0.5) {};
		\node [style=white dot] (237) at (-5.25, -1.25) {$\theta$};
		\node [style=gray dot] (238) at (-3, -0.5) {};
		\node [style=white dot] (240) at (-3, -1.25) {$\bar{\theta}$};
		\node [style=gray dot] (241) at (3, -0.5) {};
		\node [style=white dot] (243) at (3, -1.25) {$\bar{\theta}$};
		\node [style=gray dot] (244) at (1, -0.5) {};
		\node [style=white dot] (246) at (1, -1.25) {$\theta$};
		\node [style=none] (247) at (0.5, -0.75) {};
		\node [style=none] (248) at (2.5, -0.75) {};
		\node [style=none] (249) at (-3.5, -0.75) {};
		\node [style=none] (250) at (-5.75, -0.75) {};
	\end{pgfonlayer}
	\begin{pgfonlayer}{edgelayer}
		\draw [style=simple] (219) to (217);
		\draw [style=simple] (219) to (218);
		\draw [style=simple] (216) to (219);
		\draw [style=simple] (215) to (219);
		\draw [style=simple] (210) to (211);
		\draw [style=simple] (210) to (212);
		\draw [style=simple] (210) to (213);
		\draw [style=simple] (210) to (214);
		\draw [style=simple, bend right=15, looseness=1.25] (211) to (215);
		\draw [style=simple, bend left=15] (211) to (215);
		\draw [style=simple, bend right=15] (212) to (216);
		\draw [style=simple, bend left=15] (212) to (216);
		\draw [style=simple, bend right=15] (213) to (217);
		\draw [style=simple, bend left=15] (213) to (217);
		\draw [style=simple, bend right=15] (214) to (218);
		\draw [style=simple, bend left=15] (214) to (218);
		\draw [style=simple] (235) to (228);
		\draw [style=simple] (235) to (237);
		\draw [style=simple] (238) to (240);
		\draw [style=simple] (238) to (229);
		\draw [style=simple] (241) to (243);
		\draw [style=simple] (241) to (231);
		\draw [style=simple] (244) to (246);
		\draw [style=simple] (230) to (244);
		\draw [style=not edge] (235) to (250.center);
		\draw [style=not edge] (238) to (249.center);
		\draw [style=not edge] (244) to (247.center);
		\draw [style=not edge] (241) to (248.center);
	\end{pgfonlayer}
\end{tikzpicture} &\overset{(f)}{=} \begin{tikzpicture}
	\begin{pgfonlayer}{nodelayer}
		\node [style=none] (38) at (0, 1.25) {...};
		\node [style=none] (63) at (0, -2) {...};
		\node [style=white dot] (210) at (0, 2.75) {};
		\node [style=gray dot] (211) at (-4.25, 1.25) {};
		\node [style=gray dot] (212) at (-2, 1.25) {};
		\node [style=gray dot] (213) at (2, 1.25) {};
		\node [style=gray dot] (214) at (4, 1.25) {};
		\node [style=white dot] (215) at (-4.25, -1.25) {};
		\node [style=white dot] (216) at (-2, -1.25) {};
		\node [style=white dot] (217) at (2, -1.25) {};
		\node [style=white dot] (218) at (4, -1.25) {};
		\node [style=gray dot] (219) at (0, -2.75) {};
		\node [style=none] (226) at (0, 1.75) {$2\alpha$};
		\node [style=none] (227) at (0, -1.5) {$2\alpha$};
		\node [style=white dot] (228) at (-2.75, -1.75) {};
		\node [style=white dot] (229) at (-1, -2) {};
		\node [style=white dot] (230) at (1, -2) {};
		\node [style=white dot] (231) at (2.75, -1.75) {};
		\node [style=gray dot] (235) at (-3.5, -2) {};
		\node [style=white dot] (237) at (-3.5, -2.75) {$\theta$};
		\node [style=gray dot] (238) at (-2, -2.5) {};
		\node [style=white dot] (240) at (-2, -3.25) {$\bar{\theta}$};
		\node [style=gray dot] (241) at (3, -2.5) {};
		\node [style=white dot] (243) at (3, -3.25) {$\bar{\theta}$};
		\node [style=gray dot] (244) at (1.25, -2.75) {};
		\node [style=white dot] (246) at (1.25, -3.5) {$\theta$};
		\node [style=none] (247) at (0.75, -3) {};
		\node [style=none] (248) at (2.5, -2.75) {};
		\node [style=none] (249) at (-2.5, -2.75) {};
		\node [style=none] (250) at (-4, -2.25) {};
	\end{pgfonlayer}
	\begin{pgfonlayer}{edgelayer}
		\draw [style=simple] (219) to (217);
		\draw [style=simple] (219) to (218);
		\draw [style=simple] (216) to (219);
		\draw [style=simple] (215) to (219);
		\draw [style=simple] (210) to (211);
		\draw [style=simple] (210) to (212);
		\draw [style=simple] (210) to (213);
		\draw [style=simple] (210) to (214);
		\draw [style=simple, bend right=15, looseness=1.25] (211) to (215);
		\draw [style=simple, bend left=15] (211) to (215);
		\draw [style=simple, bend right=15] (212) to (216);
		\draw [style=simple, bend left=15] (212) to (216);
		\draw [style=simple, bend right=15] (213) to (217);
		\draw [style=simple, bend left=15] (213) to (217);
		\draw [style=simple, bend right=15] (214) to (218);
		\draw [style=simple, bend left=15] (214) to (218);
		\draw [style=simple] (235) to (228);
		\draw [style=simple] (235) to (237);
		\draw [style=simple] (238) to (240);
		\draw [style=simple] (238) to (229);
		\draw [style=simple] (241) to (243);
		\draw [style=simple] (241) to (231);
		\draw [style=simple] (244) to (246);
		\draw [style=simple] (230) to (244);
		\draw [style=not edge] (235) to (250.center);
		\draw [style=not edge] (238) to (249.center);
		\draw [style=not edge] (244) to (247.center);
		\draw [style=not edge] (241) to (248.center);
	\end{pgfonlayer}
\end{tikzpicture}\overset{(H)}{=}  \begin{tikzpicture}
	\begin{pgfonlayer}{nodelayer}
		\node [style=none] (38) at (0, 0.75) {...};
		\node [style=none] (63) at (0, -1.25) {...};
		\node [style=white dot] (210) at (0, 2.25) {};
		\node [style=gray dot] (211) at (-2, 0.75) {};
		\node [style=gray dot] (212) at (-1.25, 0.75) {};
		\node [style=gray dot] (213) at (1, 0.75) {};
		\node [style=gray dot] (214) at (1.75, 0.75) {};
		\node [style=white dot] (215) at (-4.25, -0.5) {};
		\node [style=white dot] (216) at (-2, -0.5) {};
		\node [style=white dot] (217) at (2, -0.5) {};
		\node [style=white dot] (218) at (4, -0.5) {};
		\node [style=gray dot] (219) at (0, -2) {};
		\node [style=none] (226) at (0, 1.25) {$2\alpha$};
		\node [style=none] (227) at (0, -0.75) {$2\alpha$};
		\node [style=white dot] (228) at (-2.75, -1) {};
		\node [style=white dot] (229) at (-1, -1.25) {};
		\node [style=white dot] (230) at (1, -1.25) {};
		\node [style=white dot] (231) at (2.75, -1) {};
		\node [style=gray dot] (235) at (-3.5, -1.25) {};
		\node [style=white dot] (237) at (-3.5, -2) {$\theta$};
		\node [style=gray dot] (238) at (-2, -1.75) {};
		\node [style=white dot] (240) at (-2, -2.5) {$\bar{\theta}$};
		\node [style=gray dot] (241) at (3, -1.75) {};
		\node [style=white dot] (243) at (3, -2.5) {$\bar{\theta}$};
		\node [style=gray dot] (244) at (1.25, -2) {};
		\node [style=white dot] (246) at (1.25, -2.75) {$\theta$};
		\node [style=none] (247) at (0.75, -2.25) {};
		\node [style=none] (248) at (2.5, -2) {};
		\node [style=none] (249) at (-2.5, -2) {};
		\node [style=none] (250) at (-4, -1.5) {};
	\end{pgfonlayer}
	\begin{pgfonlayer}{edgelayer}
		\draw [style=simple] (219) to (217);
		\draw [style=simple] (219) to (218);
		\draw [style=simple] (216) to (219);
		\draw [style=simple] (215) to (219);
		\draw [style=simple] (210) to (211);
		\draw [style=simple] (210) to (212);
		\draw [style=simple] (210) to (213);
		\draw [style=simple] (210) to (214);
		\draw [style=simple] (235) to (228);
		\draw [style=simple] (235) to (237);
		\draw [style=simple] (238) to (240);
		\draw [style=simple] (238) to (229);
		\draw [style=simple] (241) to (243);
		\draw [style=simple] (241) to (231);
		\draw [style=simple] (244) to (246);
		\draw [style=simple] (230) to (244);
		\draw [style=not edge] (235) to (250.center);
		\draw [style=not edge] (238) to (249.center);
		\draw [style=not edge] (244) to (247.center);
		\draw [style=not edge] (241) to (248.center);
	\end{pgfonlayer}
\end{tikzpicture}  \\
   &\overset{(c),(f)}{=}   \left(\begin{tikzpicture}
	\begin{pgfonlayer}{nodelayer}
		\node [style=none] (38) at (0.25, 1.75) {...};
		\node [style=gray dot] (211) at (-2, 1.5) {};
		\node [style=gray dot] (212) at (-1.25, 1.5) {};
		\node [style=gray dot] (213) at (1.75, 1.5) {};
		\node [style=gray dot] (219) at (3.25, 0.5) {};
		\node [style=none] (226) at (0.25, 2.25) {$(2\alpha-1)$};
		\node [style=none] (227) at (0.25, -1) {$2\alpha$};
		\node [style=white dot] (228) at (-0.25, 0.75) {};
		\node [style=white dot] (229) at (-0.25, -0.25) {};
		\node [style=white dot] (230) at (-0.25, -2) {};
		\node [style=white dot] (231) at (-0.25, -3) {};
		\node [style=gray dot] (235) at (-2.25, 0.75) {};
		\node [style=white dot] (237) at (-1.5, 0.25) {$\theta$};
		\node [style=gray dot] (238) at (-2.25, -0.25) {};
		\node [style=white dot] (240) at (-1.5, -0.75) {$\bar{\theta}$};
		\node [style=gray dot] (241) at (-2.25, -3) {};
		\node [style=white dot] (243) at (-1.5, -3.5) {$\bar{\theta}$};
		\node [style=gray dot] (244) at (-2.25, -2) {};
		\node [style=white dot] (246) at (-1.5, -2.5) {$\theta$};
		\node [style=none] (247) at (-3.25, -2) {};
		\node [style=none] (248) at (-3.25, -3) {};
		\node [style=none] (249) at (-3.25, -0.25) {};
		\node [style=none] (250) at (-3.25, 0.75) {};
		\node [style=gray dot] (251) at (-2, 2.75) {};
		\node [style=gray dot] (252) at (-1.25, 2.75) {};
		\node [style=gray dot] (253) at (1.75, 2.75) {};
		\node [style=none, rotate=90] (254) at (-0.25, -1) {...};
	\end{pgfonlayer}
	\begin{pgfonlayer}{edgelayer}
		\draw [style=simple] (235) to (228);
		\draw [style=simple] (235) to (237);
		\draw [style=simple] (238) to (240);
		\draw [style=simple] (238) to (229);
		\draw [style=simple] (241) to (243);
		\draw [style=simple] (241) to (231);
		\draw [style=simple] (244) to (246);
		\draw [style=simple] (230) to (244);
		\draw [style=not edge] (235) to (250.center);
		\draw [style=not edge] (238) to (249.center);
		\draw [style=not edge] (244) to (247.center);
		\draw [style=not edge] (241) to (248.center);
		\draw [style=simple] (251) to (211);
		\draw [style=simple] (252) to (212);
		\draw [style=simple] (253) to (213);
		\draw [style=simple] (219) to (229);
		\draw [style=simple] (219) to (228);
		\draw [style=simple] (219) to (230);
		\draw [style=simple] (219) to (231);
	\end{pgfonlayer}
\end{tikzpicture} \right) \overset{(id),(f)}{=}  \begin{tikzpicture}
	\begin{pgfonlayer}{nodelayer}
		\node [style=gray dot] (219) at (0.5, 0) {};
		\node [style=none] (227) at (2, 0) {$2\alpha$};
		\node [style=white dot] (237) at (1.5, 1.5) {$\theta$};
		\node [style=white dot] (240) at (1.5, 0.75) {$\bar{\theta}$};
		\node [style=white dot] (243) at (1.5, -1.5) {$\bar{\theta}$};
		\node [style=white dot] (246) at (1.5, -0.75) {$\theta$};
		\node [style=none] (247) at (-0.75, -0.75) {};
		\node [style=none] (248) at (-0.75, -1.5) {};
		\node [style=none] (249) at (-0.75, 0.75) {};
		\node [style=none] (250) at (-0.75, 1.5) {};
		\node [style=none, rotate=90] (254) at (1.5, 0) {...};
		\node [style=none] (255) at (-1.75, 1.5) {};
		\node [style=none] (256) at (-1.75, 0.75) {};
		\node [style=none] (257) at (-1.75, -0.75) {};
		\node [style=none] (258) at (-1.75, -1.5) {};
	\end{pgfonlayer}
	\begin{pgfonlayer}{edgelayer}
		\draw [style=simple] (237) to (219);
		\draw [style=simple] (219) to (240);
		\draw [style=simple] (219) to (246);
		\draw [style=simple] (219) to (243);
		\draw [style=simple] (249.center) to (219);
		\draw [style=simple] (250.center) to (219);
		\draw [style=simple] (247.center) to (219);
		\draw [style=simple] (248.center) to (219);
		\draw [style=not edge] (255.center) to (250.center);
		\draw [style=not edge] (256.center) to (249.center);
		\draw [style=not edge] (257.center) to (247.center);
		\draw [style=not edge] (258.center) to (248.center);
	\end{pgfonlayer}
\end{tikzpicture} \label{eq:tn_3}
    \end{split}
\end{equation}
Here, we have used dashed (red) wires to indicate that this is a component of a larger diagram. 

Similarly, for the single qubit replica space of site $i=t$ (the site where $\sigma_x$ acts),
\begin{equation}
    \begin{split}
        \begin{tikzpicture}
	\begin{pgfonlayer}{nodelayer}
		\node [style=none] (38) at (0, 1.25) {...};
		\node [style=none] (63) at (0, -2) {...};
		\node [style=white dot] (210) at (0, 2.75) {};
		\node [style=gray dot] (211) at (-3.25, 1.25) {};
		\node [style=gray dot] (212) at (-1.75, 1.25) {};
		\node [style=gray dot] (213) at (1.75, 1.25) {};
		\node [style=gray dot] (214) at (3.25, 1.25) {};
		\node [style=white dot] (215) at (-3.25, -1.25) {};
		\node [style=white dot] (216) at (-1.75, -1.25) {};
		\node [style=white dot] (217) at (1.75, -1.25) {};
		\node [style=white dot] (218) at (3.25, -1.25) {};
		\node [style=gray dot] (219) at (0, -2.75) {};
		\node [style=none] (226) at (0, 1.75) {$2\alpha$};
		\node [style=none] (227) at (0, -1.5) {$2\alpha$};
		\node [style=white dot] (228) at (-3.5, -0.75) {};
		\node [style=white dot] (229) at (-2, -0.75) {};
		\node [style=white dot] (230) at (1.5, -0.75) {};
		\node [style=white dot] (231) at (3, -0.75) {};
		\node [style=none] (232) at (-4, -1) {};
		\node [style=none] (233) at (-2.75, -1) {};
		\node [style=none] (234) at (0.75, -1) {};
		\node [style=none] (235) at (2.25, -1) {};
		\node [style=gray dot] (236) at (-3.75, 0) {$\pi$};
		\node [style=gray dot] (237) at (1.25, 0) {$\pi$};
		\node [style=gray dot] (238) at (2.75, 0) {$\pi$};
		\node [style=gray dot] (239) at (-2.25, 0) {$\pi$};
	\end{pgfonlayer}
	\begin{pgfonlayer}{edgelayer}
		\draw [style=simple] (219) to (217);
		\draw [style=simple] (219) to (218);
		\draw [style=simple] (216) to (219);
		\draw [style=simple] (215) to (219);
		\draw [style=simple] (210) to (211);
		\draw [style=simple] (210) to (212);
		\draw [style=simple] (210) to (213);
		\draw [style=simple] (210) to (214);
		\draw [style=simple, bend right, looseness=1.25] (211) to (215);
		\draw [style=simple, bend left] (211) to (215);
		\draw [style=simple, bend right, looseness=1.25] (212) to (216);
		\draw [style=simple, bend left] (212) to (216);
		\draw [style=simple, bend right, looseness=1.25] (213) to (217);
		\draw [style=simple, bend left] (213) to (217);
		\draw [style=simple, bend right, looseness=1.25] (214) to (218);
		\draw [style=simple, bend left] (214) to (218);
		\draw [style=not edge] (232.center) to (228);
		\draw [style=not edge] (233.center) to (229);
		\draw [style=not edge] (234.center) to (230);
		\draw [style=not edge] (235.center) to (231);
	\end{pgfonlayer}
\end{tikzpicture} &\overset{(f)}{=}  \begin{tikzpicture}
	\begin{pgfonlayer}{nodelayer}
		\node [style=none] (38) at (0, 1.25) {...};
		\node [style=none] (63) at (0, -2) {...};
		\node [style=white dot] (210) at (0, 2.75) {};
		\node [style=gray dot] (211) at (-3.25, 1.25) {};
		\node [style=gray dot] (212) at (-1.75, 1.25) {};
		\node [style=gray dot] (213) at (1.75, 1.25) {};
		\node [style=gray dot] (214) at (3.25, 1.25) {};
		\node [style=white dot] (215) at (-3.25, -1.25) {};
		\node [style=white dot] (216) at (-1.75, -1.25) {};
		\node [style=white dot] (217) at (1.75, -1.25) {};
		\node [style=white dot] (218) at (3.25, -1.25) {};
		\node [style=gray dot] (219) at (0, -2.75) {};
		\node [style=none] (226) at (0, 1.75) {$2\alpha$};
		\node [style=none] (227) at (0, -1.5) {$2\alpha$};
		\node [style=white dot] (228) at (-2.25, -1.75) {};
		\node [style=white dot] (229) at (-1.25, -1.75) {};
		\node [style=white dot] (230) at (1.25, -1.75) {};
		\node [style=white dot] (231) at (2.25, -1.75) {};
		\node [style=none] (232) at (-2.75, -1.75) {};
		\node [style=none] (233) at (-1.75, -1.75) {};
		\node [style=none] (234) at (0.75, -1.75) {};
		\node [style=none] (235) at (1.75, -1.75) {};
		\node [style=gray dot] (236) at (-2.25, 1.75) {$\pi$};
		\node [style=gray dot] (237) at (1.25, 1.75) {$\pi$};
		\node [style=gray dot] (238) at (2.25, 1.75) {$\pi$};
		\node [style=gray dot] (239) at (-1.25, 1.75) {$\pi$};
	\end{pgfonlayer}
	\begin{pgfonlayer}{edgelayer}
		\draw [style=simple] (219) to (217);
		\draw [style=simple] (219) to (218);
		\draw [style=simple] (216) to (219);
		\draw [style=simple] (215) to (219);
		\draw [style=simple] (210) to (211);
		\draw [style=simple] (210) to (212);
		\draw [style=simple] (210) to (213);
		\draw [style=simple] (210) to (214);
		\draw [style=simple, bend right, looseness=1.25] (211) to (215);
		\draw [style=simple, bend left] (211) to (215);
		\draw [style=simple, bend right, looseness=1.25] (212) to (216);
		\draw [style=simple, bend left] (212) to (216);
		\draw [style=simple, bend right, looseness=1.25] (213) to (217);
		\draw [style=simple, bend left] (213) to (217);
		\draw [style=simple, bend right, looseness=1.25] (214) to (218);
		\draw [style=simple, bend left] (214) to (218);
		\draw [style=not edge] (232.center) to (228);
		\draw [style=not edge] (233.center) to (229);
		\draw [style=not edge] (234.center) to (230);
		\draw [style=not edge] (235.center) to (231);
	\end{pgfonlayer}
\end{tikzpicture}\overset{(H),(f)}{=} 
 \begin{tikzpicture}
	\begin{pgfonlayer}{nodelayer}
		\node [style=none] (38) at (0, 0.75) {...};
		\node [style=none] (63) at (0, -1.25) {...};
		\node [style=white dot] (210) at (0, 2.25) {};
		\node [style=gray dot] (211) at (-2, 0.75) {$\pi$};
		\node [style=gray dot] (212) at (-1, 0.75) {$\pi$};
		\node [style=gray dot] (213) at (1, 0.75) {$\pi$};
		\node [style=gray dot] (214) at (2, 0.75) {$\pi$};
		\node [style=white dot] (215) at (-3.25, -0.5) {};
		\node [style=white dot] (216) at (-1.75, -0.5) {};
		\node [style=white dot] (217) at (1.75, -0.5) {};
		\node [style=white dot] (218) at (3.25, -0.5) {};
		\node [style=gray dot] (219) at (0, -2) {};
		\node [style=none] (226) at (0, 1.25) {$2\alpha$};
		\node [style=none] (227) at (0, -0.75) {$2\alpha$};
		\node [style=white dot] (228) at (-2.25, -1) {};
		\node [style=white dot] (229) at (-1.25, -1) {};
		\node [style=white dot] (230) at (1.25, -1) {};
		\node [style=white dot] (231) at (2.25, -1) {};
		\node [style=none] (232) at (-2.75, -1) {};
		\node [style=none] (233) at (-1.75, -1) {};
		\node [style=none] (234) at (0.75, -1) {};
		\node [style=none] (235) at (1.75, -1) {};
	\end{pgfonlayer}
	\begin{pgfonlayer}{edgelayer}
		\draw [style=simple] (219) to (217);
		\draw [style=simple] (219) to (218);
		\draw [style=simple] (216) to (219);
		\draw [style=simple] (215) to (219);
		\draw [style=simple] (210) to (211);
		\draw [style=simple] (210) to (212);
		\draw [style=simple] (210) to (213);
		\draw [style=simple] (210) to (214);
		\draw [style=not edge] (232.center) to (228);
		\draw [style=not edge] (233.center) to (229);
		\draw [style=not edge] (234.center) to (230);
		\draw [style=not edge] (235.center) to (231);
	\end{pgfonlayer}
\end{tikzpicture} \\
        &\overset{(f),(c)}{=} \left( \begin{tikzpicture}
	\begin{pgfonlayer}{nodelayer}
		\node [style=none] (38) at (1, 1.75) {...};
		\node [style=gray dot] (211) at (-1.5, 1.75) {$\pi$};
		\node [style=gray dot] (212) at (-0.5, 1.75) {$\pi$};
		\node [style=gray dot] (213) at (2.5, 1.75) {$\pi$};
		\node [style=none] (226) at (1, 2.25) {$(2\alpha-1)$};
		\node [style=gray dot] (258) at (2.75, 0.5) {};
		\node [style=none] (259) at (-0.25, -1) {$2\alpha$};
		\node [style=white dot] (260) at (-0.75, 0.75) {};
		\node [style=white dot] (261) at (-0.75, -0.25) {};
		\node [style=white dot] (262) at (-0.75, -2) {};
		\node [style=white dot] (263) at (-0.75, -3) {};
		\node [style=none, rotate=90] (264) at (-0.75, -1) {...};
		\node [style=none] (269) at (-1.75, 0.75) {};
		\node [style=none] (270) at (-1.75, -0.25) {};
		\node [style=none] (271) at (-1.75, -2) {};
		\node [style=none] (272) at (-1.75, -3) {};
		\node [style=gray dot] (273) at (2.5, 3) {$\pi$};
		\node [style=gray dot] (274) at (-0.5, 3) {$\pi$};
		\node [style=gray dot] (275) at (-1.5, 3) {$\pi$};
	\end{pgfonlayer}
	\begin{pgfonlayer}{edgelayer}
		\draw [style=simple] (258) to (261);
		\draw [style=simple] (258) to (260);
		\draw [style=simple] (258) to (262);
		\draw [style=simple] (258) to (263);
		\draw [style=not edge] (269.center) to (260);
		\draw [style=not edge] (270.center) to (261);
		\draw [style=not edge] (271.center) to (262);
		\draw [style=not edge] (272.center) to (263);
		\draw (275) to (211);
		\draw (274) to (212);
		\draw (273) to (213);
	\end{pgfonlayer}
\end{tikzpicture} \right) {=} \begin{tikzpicture}
	\begin{pgfonlayer}{nodelayer}
		\node [style=gray dot] (258) at (2.5, 0) {};
		\node [style=none] (259) at (0.25, 0) {$2\alpha$};
		\node [style=white dot] (260) at (-0.25, 1.75) {};
		\node [style=white dot] (261) at (-0.25, 0.75) {};
		\node [style=white dot] (262) at (-0.25, -1) {};
		\node [style=white dot] (263) at (-0.25, -2) {};
		\node [style=none, rotate=90] (264) at (-0.25, 0) {...};
		\node [style=none] (269) at (-1.25, 1.75) {};
		\node [style=none] (270) at (-1.25, 0.75) {};
		\node [style=none] (271) at (-1.25, -1) {};
		\node [style=none] (272) at (-1.25, -2) {};
	\end{pgfonlayer}
	\begin{pgfonlayer}{edgelayer}
		\draw [style=simple] (258) to (261);
		\draw [style=simple] (258) to (260);
		\draw [style=simple] (258) to (262);
		\draw [style=simple] (258) to (263);
		\draw [style=not edge] (269.center) to (260);
		\draw [style=not edge] (270.center) to (261);
		\draw [style=not edge] (271.center) to (262);
		\draw [style=not edge] (272.center) to (263);
	\end{pgfonlayer}
\end{tikzpicture}. \label{eq:tn_4}
    \end{split}
\end{equation}
The dashed (red) line here indicates the rest of the diagram of Eq.~\eqref{eq:full_tn}; i.e., $t$ copies of Eq.~\eqref{eq:tn_3}. Substituting the simplifications \eqref{eq:tn_3} ($t-1$ copies on the left) and \eqref{eq:tn_4} (right) into Eq.~\eqref{eq:full_tn},
\begin{equation}
    \begin{split}
        \tr[(O_U \otimes O_U^* )^{\otimes \alpha } (\Lambda^{(\alpha)})^{\otimes t}] &= \begin{tikzpicture}
	\begin{pgfonlayer}{nodelayer}
		\node [style=gray dot] (258) at (4, -0.25) {};
		\node [style=none] (259) at (2, -0.25) {$2\alpha$};
		\node [style=white dot] (260) at (1.25, 1.75) {};
		\node [style=white dot] (261) at (1.25, 0.75) {};
		\node [style=white dot] (262) at (1.25, -1.25) {};
		\node [style=white dot] (263) at (1.25, -2.25) {};
		\node [style=none, rotate=90] (264) at (1.25, -0.25) {...};
		\node [style=gray dot] (280) at (-1.25, 3.5) {};
		\node [style=gray dot] (281) at (-1.25, 1.5) {};
		\node [style=gray dot] (282) at (-1.25, -1) {};
		\node [style=gray dot] (283) at (-1.25, -3) {};
		\node [style=none] (284) at (-1, 0.25) {$t$};
		\node [style=none, rotate=90] (285) at (-1.25, 0.25) {...};
		\node [style=white dot] (286) at (-1, 4.75) {$\theta$};
		\node [style=white dot] (288) at (-1.75, 4.75) {$\bar{\theta}$};
		\node [style=white dot] (289) at (-2.25, 4.25) {$\theta$};
		\node [style=white dot] (290) at (-2.5, 3.5) {$\bar{\theta}$};
		\node [style=white dot] (291) at (-2.5, 2.5) {$\theta$};
		\node [style=white dot] (292) at (-2.5, 2) {$\bar{\theta}$};
		\node [style=white dot] (293) at (-2.5, 1.5) {$\theta$};
		\node [style=white dot] (294) at (-2.5, 1) {$\bar{\theta}$};
		\node [style=white dot] (295) at (-2.5, -0.5) {$\theta$};
		\node [style=white dot] (296) at (-2.5, -1) {$\bar{\theta}$};
		\node [style=white dot] (297) at (-2.5, -1.5) {$\theta$};
		\node [style=white dot] (298) at (-2.5, -2) {$\bar{\theta}$};
		\node [style=white dot] (299) at (-2.5, -3) {$\theta$};
		\node [style=white dot] (300) at (-2.25, -3.75) {$\bar{\theta}$};
		\node [style=white dot] (301) at (-1.75, -4.25) {$\theta$};
		\node [style=white dot] (302) at (-1, -4.25) {$\bar{\theta}$};
		\node [style=none] (303) at (-3.75, 1.75) {$2\alpha$};
		\node [style=none] (304) at (-3.25, 2.75) {};
		\node [style=none] (305) at (-3.25, 0.75) {};
		\node [style=none] (306) at (-3.75, -1.25) {$2\alpha$};
		\node [style=none] (307) at (-3.25, -0.25) {};
		\node [style=none] (308) at (-3.25, -2.25) {};
		\node [style=none] (309) at (-3, -4.75) {$2\alpha$};
		\node [style=none] (310) at (-3.25, -3.5) {};
		\node [style=none] (311) at (-1.75, -5.25) {};
		\node [style=none] (312) at (-3.25, 4.75) {$2\alpha$};
		\node [style=none] (313) at (-2, 5.5) {};
		\node [style=none] (314) at (-3.25, 3.75) {};
	\end{pgfonlayer}
	\begin{pgfonlayer}{edgelayer}
		\draw [style=simple] (258) to (261);
		\draw [style=simple] (258) to (260);
		\draw [style=simple] (258) to (262);
		\draw [style=simple] (258) to (263);
		\draw [style=simple] (280) to (260);
		\draw [style=simple] (280) to (261);
		\draw [style=simple] (280) to (262);
		\draw [style=simple] (280) to (263);
		\draw [style=simple] (281) to (260);
		\draw [style=simple] (281) to (261);
		\draw [style=simple] (281) to (262);
		\draw [style=simple] (281) to (263);
		\draw [style=simple] (282) to (260);
		\draw [style=simple] (282) to (261);
		\draw [style=simple] (282) to (262);
		\draw [style=simple] (282) to (263);
		\draw [style=simple] (283) to (260);
		\draw [style=simple] (283) to (261);
		\draw [style=simple] (283) to (262);
		\draw [style=simple] (283) to (263);
		\draw [style=simple] (286) to (280);
		\draw [style=simple] (288) to (280);
		\draw [style=simple] (289) to (280);
		\draw [style=simple] (290) to (280);
		\draw [style=simple] (291) to (281);
		\draw [style=simple] (292) to (281);
		\draw [style=simple] (293) to (281);
		\draw [style=simple] (294) to (281);
		\draw [style=simple] (282) to (295);
		\draw [style=simple] (282) to (296);
		\draw [style=simple] (282) to (297);
		\draw [style=simple] (282) to (298);
		\draw [style=simple] (283) to (299);
		\draw [style=simple] (283) to (300);
		\draw [style=simple] (283) to (301);
		\draw [style=simple] (283) to (302);
		\draw [style=brace edge] (305.center) to (304.center);
		\draw [style=brace edge] (308.center) to (307.center);
		\draw [style=brace edge] (311.center) to (310.center);
		\draw [style=brace edge] (314.center) to (313.center);
	\end{pgfonlayer}
\end{tikzpicture} \overset{(b)}{=}  \begin{tikzpicture}
	\begin{pgfonlayer}{nodelayer}
		\node [style=gray dot] (305) at (4.25, 0) {};
		\node [style=white dot] (306) at (1.25, 0) {};
		\node [style=none] (307) at (3, 0) {$2\alpha$};
		\node [style=none, rotate=90] (308) at (2.5, 0) {...};
		\node [style=gray dot] (309) at (0, 0) {};
		\node [style=gray dot] (352) at (1.25, 0) {};
		\node [style=white dot] (353) at (0, 0) {};
		\node [style=gray dot] (354) at (-1.5, 3.25) {};
		\node [style=gray dot] (355) at (-1.5, 1.25) {};
		\node [style=gray dot] (356) at (-1.5, -1.25) {};
		\node [style=gray dot] (357) at (-1.5, -3.25) {};
		\node [style=none] (358) at (-1.25, 0) {$t$};
		\node [style=none, rotate=90] (359) at (-1.5, 0) {...};
		\node [style=white dot] (360) at (-1.25, 4.5) {$\theta$};
		\node [style=white dot] (361) at (-2, 4.5) {$\bar{\theta}$};
		\node [style=white dot] (362) at (-2.5, 4) {$\theta$};
		\node [style=white dot] (363) at (-2.75, 3.25) {$\bar{\theta}$};
		\node [style=white dot] (364) at (-2.75, 2.25) {$\theta$};
		\node [style=white dot] (365) at (-2.75, 1.75) {$\bar{\theta}$};
		\node [style=white dot] (366) at (-2.75, 1.25) {$\theta$};
		\node [style=white dot] (367) at (-2.75, 0.75) {$\bar{\theta}$};
		\node [style=white dot] (368) at (-2.75, -0.75) {$\theta$};
		\node [style=white dot] (369) at (-2.75, -1.25) {$\bar{\theta}$};
		\node [style=white dot] (370) at (-2.75, -1.75) {$\theta$};
		\node [style=white dot] (371) at (-2.75, -2.25) {$\bar{\theta}$};
		\node [style=white dot] (372) at (-2.75, -3.25) {$\theta$};
		\node [style=white dot] (373) at (-2.5, -4) {$\bar{\theta}$};
		\node [style=white dot] (374) at (-2, -4.5) {$\theta$};
		\node [style=white dot] (375) at (-1.25, -4.5) {$\bar{\theta}$};
		\node [style=none] (376) at (-4, 1.5) {$2\alpha$};
		\node [style=none] (377) at (-3.5, 2.5) {};
		\node [style=none] (378) at (-3.5, 0.5) {};
		\node [style=none] (379) at (-4, -1.5) {$2\alpha$};
		\node [style=none] (380) at (-3.5, -0.5) {};
		\node [style=none] (381) at (-3.5, -2.5) {};
		\node [style=none] (382) at (-3.25, -5) {$2\alpha$};
		\node [style=none] (383) at (-3.5, -3.75) {};
		\node [style=none] (384) at (-2, -5.5) {};
		\node [style=none] (385) at (-3.5, 4.5) {$2\alpha$};
		\node [style=none] (386) at (-2.25, 5.25) {};
		\node [style=none] (387) at (-3.5, 3.5) {};
	\end{pgfonlayer}
	\begin{pgfonlayer}{edgelayer}
		\draw [style=simple, bend left=75, looseness=1.50] (306) to (305);
		\draw [style=simple, in=255, out=-75, looseness=1.50] (306) to (305);
		\draw [style=simple, in=135, out=45, looseness=1.25] (306) to (305);
		\draw [style=simple, bend right=45, looseness=1.25] (306) to (305);
		\draw [style=simple] (306) to (309);
		\draw [style=simple] (360) to (354);
		\draw [style=simple] (361) to (354);
		\draw [style=simple] (362) to (354);
		\draw [style=simple] (363) to (354);
		\draw [style=simple] (364) to (355);
		\draw [style=simple] (365) to (355);
		\draw [style=simple] (366) to (355);
		\draw [style=simple] (367) to (355);
		\draw [style=simple] (356) to (368);
		\draw [style=simple] (356) to (369);
		\draw [style=simple] (356) to (370);
		\draw [style=simple] (356) to (371);
		\draw [style=simple] (357) to (372);
		\draw [style=simple] (357) to (373);
		\draw [style=simple] (357) to (374);
		\draw [style=simple] (357) to (375);
		\draw [style=simple] (354) to (353);
		\draw [style=simple] (355) to (353);
		\draw [style=simple] (356) to (353);
		\draw [style=simple] (357) to (353);
		\draw [style=brace edge] (378.center) to (377.center);
		\draw [style=brace edge] (381.center) to (380.center);
		\draw [style=brace edge] (384.center) to (383.center);
		\draw [style=brace edge] (387.center) to (386.center);
	\end{pgfonlayer}
\end{tikzpicture}  \\
        &\overset{(f),(c)}{=}  \left( \begin{tikzpicture}
	\begin{pgfonlayer}{nodelayer}
		\node [style=gray dot] (355) at (0.75, 0) {};
		\node [style=white dot] (364) at (-0.5, 1.25) {$\theta$};
		\node [style=white dot] (365) at (-0.5, 0.5) {$\overline{\theta}$};
		\node [style=white dot] (366) at (-0.5, -0.25) {$\theta$};
		\node [style=white dot] (367) at (-0.5, -1) {$\overline{\theta}$};
		\node [style=none] (376) at (-1.75, 0) {$2\alpha$};
		\node [style=none] (377) at (-1.25, 1.5) {};
		\node [style=none] (378) at (-1.25, -1.25) {};
	\end{pgfonlayer}
	\begin{pgfonlayer}{edgelayer}
		\draw [style=simple] (364) to (355);
		\draw [style=simple] (365) to (355);
		\draw [style=simple] (366) to (355);
		\draw [style=simple] (367) to (355);
		\draw [style=brace edge] (378.center) to (377.center);
	\end{pgfonlayer}
\end{tikzpicture} \right)^t = \left( \begin{tikzpicture}
	\begin{pgfonlayer}{nodelayer}
		\node [style=gray dot] (355) at (0.75, 0.25) {};
		\node [style=white dot] (364) at (-0.5, 1) {$\theta$};
		\node [style=white dot] (365) at (2, 1) {$\overline{\theta}$};
		\node [style=white dot] (366) at (-0.5, -0.5) {$\theta$};
		\node [style=white dot] (367) at (2, -0.5) {$\overline{\theta}$};
		\node [style=none] (376) at (-1.5, 0.25) {$\alpha$};
		\node [style=none] (377) at (-1.25, 1.25) {};
		\node [style=none] (378) at (-1.25, -0.75) {};
		\node [style=none, rotate=90] (379) at (-0.5, 0.25) {...};
		\node [style=none, rotate=90] (380) at (2, 0.25) {...};
	\end{pgfonlayer}
	\begin{pgfonlayer}{edgelayer}
		\draw [style=simple] (364) to (355);
		\draw [style=simple] (365) to (355);
		\draw [style=simple] (366) to (355);
		\draw [style=simple] (367) to (355);
		\draw [style=brace edge] (378.center) to (377.center);
	\end{pgfonlayer}
\end{tikzpicture} \right)^t \\
        &=\left( \left( \bra{0} + \ex^{i\theta } \bra{1}\right)^{\otimes \alpha} \left( \ket{-}\bra{- }^{\otimes \alpha} + \ket{+}\bra{ +}^{\otimes \alpha} \right) \left( \ket{0} + \ex^{-i\theta } \ket{1}\right)^{\otimes \alpha} \right)^t \label{eq:bigone}
    \end{split}.
\end{equation}
We can clearly simplify this final expression if we define the state $\ket{\psi_{\theta}}:=\ket{0} + \ex^{-i\theta } \ket{1}$, 
\begin{equation}
    \frac{1}{2^{2 t \alpha}}\tr[(O_U \otimes O_U^* )^{\otimes \alpha } (\Lambda^{(\alpha)})^{\otimes t}] = |\braket{\psi_{\theta}|-}|^{2 \alpha } +  |\braket{\psi_{\theta}|+}|^{2 \alpha },
\end{equation}
where we remind the reader that we are still ignoring normalizations. Then, 
\begin{equation}
    |\braket{\psi_{\theta}|\pm}|^{2} \propto  1 \pm \frac{1}{2}(\ex^{i \theta}+\ex^{-i \theta}) = 1 \pm \cos(\theta),
\end{equation}
and so 
\begin{align}
    |\braket{\psi_{\theta}|-}|^{2 \alpha } +  |\braket{\psi_{\theta}|+}|^{2 \alpha } &= (1 + \cos(\theta))^\alpha + (1 - \cos(\theta))^\alpha \\
    &=\sum^{\alpha}_{k=0} \binom{\alpha}{k} \cos^k(\theta) + (-1)^k \cos^k(\theta) = 2 \sum^{\lfloor  \alpha/2 \rfloor}_{k=0} \binom{\alpha}{2k} \cos^k(\theta) \\
    &= \sum^{\lfloor  \alpha/2 \rfloor}_{k=0} \binom{\alpha}{2k} \cos^{2k}(4J) = \cos^{2 \alpha}(2 J) + \sin^{2 \alpha}( 2 J),
\end{align}
where we have subbed back in the definition of $\theta = 4J$. 
To determine the multiplicative normalization $N$, we know that for $\alpha = 1$ and for any $t$
\begin{equation}
    \frac{1}{N}  ( \cos^{2 }(2 J) + \sin^{2 }(2 J) )^t \overset{!}{=} 1.
\end{equation}
This means that $N = a^{f(\alpha)}$ with $f(1) = 0$. We also know that magic is zero for a SWAP circuit ($J=0$) for any $\alpha \geq 2$ and $t$,
\begin{equation}
    \frac{1}{a^{f(\alpha)}} ( ( \cos^{2 \alpha }(0) + \sin^{2\alpha }(0) ))^t \overset{!}{=} 1,
\end{equation}
which means that ${a^{f(\alpha)}} = 1$. This infinite set of conditions can only be satisfied for $f(\alpha) = 0$. We have also verified numerically that the normalization is correct for small \rey index $\alpha $ and depth $t$.
So finally, the OSE for this model is 
\begin{equation}
    \Mpop^{(\alpha)}( U_t^\dagger \sigma_x^{(j)} U_t )  = \frac{t}{1-\alpha} \log \left(  \cos^{2 \alpha }(2 J) + \sin^{2\alpha }(2 J) \right), \label{eq:finalXXZ}
\end{equation}
For $\alpha =2$, this is particularly simple:
\begin{equation}
    \Mpop^{(2)} ( U_t^\dagger \sigma_x^{(j)}  U_t )  = t \log( \frac{4}{3+ \cos(4J)}) .
\end{equation}
This has a maximal value for $J= \pi/4$, resulting in $\Mpop^{(2)} ( U_t^\dagger \sigma_x^{(j)}  U_t ) = t$. 

\subsection{Generalization to arbitrary initial (local) operator}
Now consider an arbitrary initial (traceless) local unitary operator on-site $j$
\begin{equation}
    O_j = a_x \sigma_x + a_y \sigma_y+ a_z \sigma_z \label{eq:init_op}
\end{equation}
where from unitarity, $a_x^2 + a_y^2 + a_z^2 = 1$. Subbing this into the Pauli purity, we have that (recalling that $n=2t$ from the light cone)
\begin{equation}
    \begin{split}
        P^{(\alpha)}(U^\dagger_t  O_j U_t) &= \sum_{P \in \mathcal{P}_N}\left(\frac{1}{D}\tr[U^\dagger(a_x \sigma_x + a_y \sigma_y+ a_z \sigma_z)U P]\right)^{2\alpha} \\
    &= \frac{1}{2^{4 \alpha t}}\sum_{P \in \mathcal{P}_N}\left(\tr[U^\dagger a_x \sigma_x U P] + \tr[U^\dagger a_y \sigma_y U P]+\tr[U^\dagger a_z \sigma_z U P]\right)^{2\alpha} \\ 
    &=\frac{1}{2^{4 \alpha t}} \sum_{k_1+k_2+k_3=2\alpha} \sum_{P \in \mathcal{P}_N} \frac{(2\alpha)!}{k_1! k_2! k_3!} (a_x \braket{\sigma_x}_P)^{k_1} (a_y \braket{\sigma_y}_P)^{k_2} (a_z \braket{\sigma_z}_P)^{k_3}. \label{eq:fullsum}
    \end{split}
\end{equation}
For shorthand, we have defined for $a \in \{ x,y,z\}$ (ignoring phases and normalization) 
\begin{equation}
    \braket{\sigma_a}_P := \tr[U^\dagger_t \sigma_a U P].
\end{equation}
Now, we have the expression for $\sum_P \braket{\sigma_x}_P^{2\alpha}$ from Eq.~\eqref{eq:finalXXZ}. In addition, we can compute $\braket{\sigma_z}_P = \tr[\sigma_z P] = \delta_{P, \sigma_z^{(j+t)} } \tr[\id] $, such that $\sum_P \braket{\sigma_z}_P f(P) = f(\sigma_z^{(j+t)} )$ (up to normalization for each). Finally, as $\sigma_y = i \sigma_x \sigma_z $, then $\sum_P \braket{\sigma_y}_P^{2\alpha} = \sum_P \braket{\sigma_x}_{P}^{2\alpha}$ (as the $\sigma_z$ gate commutes with $U_t$ up to a SWAP, and then just permutes the Pauli string $P$ in the full sum). The remaining task is one of combinatorics. 

\begin{figure}[t!]
    \centering
    \includegraphics[width=0.4\linewidth]{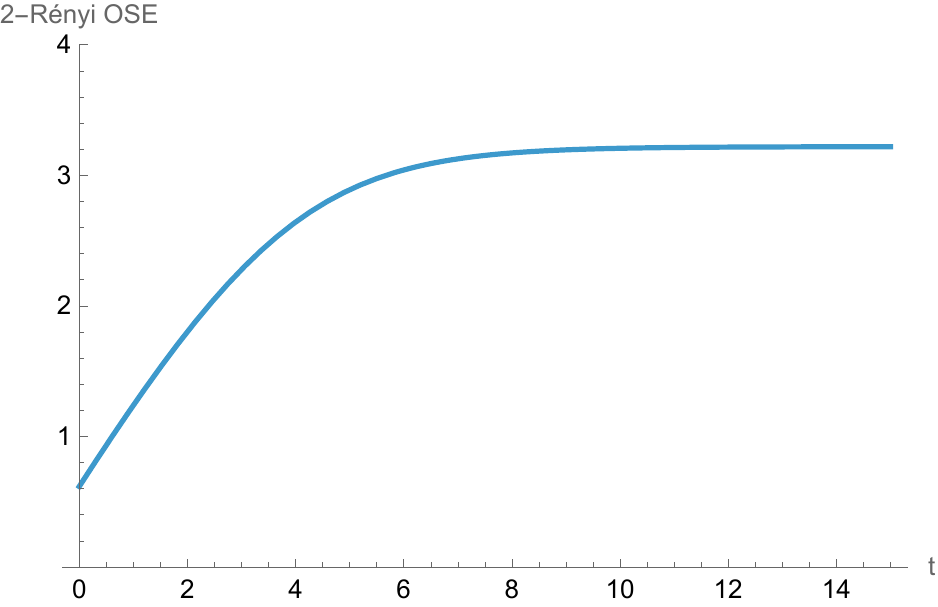}
    \caption{Plot of the growth of $\Mpop^{(2)}(U^\dagger_t O_j U_t)$ in the dual unitary XXZ model for $(a_x,a_y,a_z)= (\sqrt{0.7},\sqrt{0.2},\sqrt{0.1})$ and $J=\pi/8$. The plot asymptotes to the value given in Eq.~\eqref{eq:asymptotic}; $\lim_{t \to \infty}  \Mpop^{(2)}(U^\dagger_t O_j U_t)  \approx 3.22$ for these parameters.
    }
    \label{fig:numerics}
\end{figure}

Consider the cases: 
\begin{enumerate}[(i)]
    \item $k_3 \neq 0, \, k_2 \neq 0, \, k_1 \neq 0$: these terms have at least one factor of $\braket{\sigma_z}_P$, which is zero unless $P=\sigma_z^{(j+t)} $ (identity everywhere except $\sigma_z$ at site $j+t$). The ZX diagram for $\braket{\sigma_x}_{P=\sigma_z^{(j+t)} }$ in this case is
\begin{equation}
    \begin{split}
         \braket{\sigma_x}_{P=\sigma_z^{(j+t)} } &= \tr[U_t^\dagger \sigma_x^{(j)} U \sigma_z^{(j+t)} ] \overset{\eqref{eq:ZZa}}{=} \begin{tikzpicture}
	\begin{pgfonlayer}{nodelayer}
		\node [style=white dot] (15) at (-2.25, 1) {};
		\node [style=white dot] (16) at (-2.25, -1) {};
		\node [style=gray dot] (17) at (-2.25, 0) {};
		\node [style=white dot] (18) at (-1.25, 0) {$\theta$};
		\node [style=white dot] (23) at (-0.25, 1) {};
		\node [style=gray dot] (25) at (-0.25, 0) {};
		\node [style=white dot] (26) at (0.75, 0) {$\theta$};
		\node [style=none] (28) at (4.25, 1) {};
		\node [style=white dot] (32) at (3.25, 1) {};
		\node [style=gray dot] (34) at (3.25, 0) {};
		\node [style=white dot] (35) at (4.25, 0) {$\theta$};
		\node [style=none] (38) at (2, -0.25) {...};
		\node [style=gray dot] (39) at (-3, 1) {$\pi$};
		\node [style=none] (40) at (-4.25, 1) {};
		\node [style=white dot] (43) at (-3.75, 1) {$\pi$};
		\node [style=white dot] (44) at (-0.25, -1) {};
		\node [style=white dot] (45) at (3.25, -1) {};
	\end{pgfonlayer}
	\begin{pgfonlayer}{edgelayer}
		\draw [style=simple] (15) to (17);
		\draw [style=simple] (17) to (18);
		\draw [style=simple] (17) to (16);
		\draw [style=simple] (28.center) to (23);
		\draw [style=simple] (23) to (25);
		\draw [style=simple] (25) to (26);
		\draw [style=simple] (15) to (23);
		\draw [style=simple] (32) to (34);
		\draw [style=simple] (34) to (35);
		\draw [style=simple] (39) to (15);
		\draw [style=simple] (40.center) to (39);
		\draw [style=simple, in=-135, out=-45, loop] (16) to ();
		\draw [style=simple, in=-135, out=-45, loop] (44) to ();
		\draw [style=simple] (25) to (44);
		\draw [style=simple, in=-135, out=-45, loop] (45) to ();
		\draw [style=simple] (34) to (45);
		\draw [style=simple, bend left=150, looseness=0.50] (40.center) to (28.center);
	\end{pgfonlayer}
\end{tikzpicture} \\
         &\overset{(c),(f)}{=}\left( \begin{tikzpicture}
	\begin{pgfonlayer}{nodelayer}
		\node [style=white dot] (15) at (-2.25, 1) {};
		\node [style=white dot] (16) at (-2.25, 0) {};
		\node [style=white dot] (18) at (-1.25, 0) {$\theta$};
		\node [style=white dot] (23) at (-0.25, 1) {};
		\node [style=white dot] (26) at (0.75, 0) {$\theta$};
		\node [style=none] (28) at (4.25, 1) {};
		\node [style=white dot] (32) at (3.25, 1) {};
		\node [style=white dot] (35) at (4.25, 0) {$\theta$};
		\node [style=none] (38) at (2, -0.25) {...};
		\node [style=gray dot] (39) at (-3, 1) {$\pi$};
		\node [style=none] (40) at (-4.25, 1) {};
		\node [style=white dot] (43) at (-3.75, 1) {$\pi$};
		\node [style=white dot] (44) at (-0.25, 0) {};
		\node [style=white dot] (45) at (3.25, 0) {};
	\end{pgfonlayer}
	\begin{pgfonlayer}{edgelayer}
		\draw [style=simple] (28.center) to (23);
		\draw [style=simple] (15) to (23);
		\draw [style=simple] (39) to (15);
		\draw [style=simple] (40.center) to (39);
		\draw [style=simple, bend left=150, looseness=0.50] (40.center) to (28.center);
		\draw [style=simple] (16) to (18);
		\draw [style=simple] (44) to (26);
		\draw [style=simple] (45) to (35);
	\end{pgfonlayer}
\end{tikzpicture} \right)\overset{(id)}{=} \left(\begin{tikzpicture}
	\begin{pgfonlayer}{nodelayer}
		\node [style=white dot] (26) at (-1.25, 0) {$\theta$};
		\node [style=none] (28) at (1.25, 1) {};
		\node [style=white dot] (35) at (1.5, 0) {$\theta$};
		\node [style=none] (38) at (-0.25, 0) {...};
		\node [style=gray dot] (39) at (-0.5, 1) {$\pi$};
		\node [style=none] (40) at (-1.75, 1) {};
		\node [style=white dot] (43) at (-1.25, 1) {$\pi$};
		\node [style=white dot] (44) at (-2.25, 0) {};
		\node [style=white dot] (45) at (0.5, 0) {};
	\end{pgfonlayer}
	\begin{pgfonlayer}{edgelayer}
		\draw [style=simple] (40.center) to (39);
		\draw [style=simple, bend left=135, looseness=0.75] (40.center) to (28.center);
		\draw [style=simple] (44) to (26);
		\draw [style=simple] (45) to (35);
		\draw [style=simple] (39) to (28.center);
	\end{pgfonlayer}
\end{tikzpicture}\right) \propto \tr[\sigma_x \sigma_z] = 0. \label{eq:item1}
    \end{split}
\end{equation}
Therefore, all terms with $k_3 \neq 0$ and $k_1 \neq 0$ are zero. 
\item $k_3 \neq 0,k_2 \neq 0,k_1 = 0$: using that $\sigma_y = i \sigma_x \sigma_z $, and that $ \sigma_z$ commutes with the ZZ gate, we can applying the same steps as Eq.~\eqref{eq:item1} to arrive at 
\begin{equation}
    \braket{\sigma_y}_{P=\sigma_z^{(j+t)} } = \tr[U_t^\dagger \sigma_y^{(j)}  U \sigma_z^{(j+t)} ] \propto \begin{tikzpicture}
	\begin{pgfonlayer}{nodelayer}
		\node [style=none] (28) at (1.5, 0) {};
		\node [style=gray dot] (39) at (1, 0) {$\pi$};
		\node [style=none] (40) at (-1.5, 0) {};
		\node [style=white dot] (43) at (-1, 0) {$\pi$};
		\node [style=white dot] (46) at (0, 0) {$\pi$};
	\end{pgfonlayer}
	\begin{pgfonlayer}{edgelayer}
		\draw [style=simple] (40.center) to (39);
		\draw [style=simple, bend left=105, looseness=0.75] (40.center) to (28.center);
		\draw (28.center) to (39);
	\end{pgfonlayer}
\end{tikzpicture} =  \tr[\sigma_x] = 0.
\end{equation}
\end{enumerate}
From this, we know that the only non-zero term in the sum \eqref{eq:fullsum} involving $\braket{\sigma_z}_P$ is when $k_3= 2\alpha $, which is equal to $a_z^{2\alpha}$ (irrespective of $t$). Eq.~\eqref{eq:fullsum} then reduces to 
\begin{equation}
    P^{(\alpha)}(U\dagger_t O_j U_t)= a_z^{2\alpha} + \frac{1}{2^{4 \alpha t}} \sum_{k=0}^{2\alpha} \sum_{P \in \mathcal{P}_N} \binom{2\alpha}{k} (a_x \braket{\sigma_x}_P)^{k} (a_y \braket{\sigma_y}_P)^{2\alpha-k}.
\end{equation}
As argued below Eq.~\eqref{eq:fullsum}, we can directly evaluate the elements of this sum when $k=2\alpha$ and when $k=0$, which are equal to the final value of the previous section,
\begin{equation}
    P^{(\alpha)}(U^\dagger_t O_j U_t)\overset{\eqref{eq:finalXXZ}}{=} a_z^{2\alpha} + (a_x^{2\alpha} +a_y^{2\alpha})\left( \cos^{2 \alpha }(2 J) + \sin^{2\alpha }(2 J)\right)^t+  \frac{1}{2^{4 \alpha t}} \sum_{k=1}^{2\alpha-1} \sum_{P \in \mathcal{P}_N} \binom{2\alpha}{k} (a_x \braket{\sigma_x}_P)^{k} (a_y \braket{\sigma_y}_P)^{2\alpha-k}.
\end{equation}
To handle these final terms, we consider their ZX diagram, 
\begin{equation}
    \frac{1}{2^{4 \alpha t}} \sum_{P \in \mathcal{P}_N} (\braket{\sigma_x}_P)^{k} (\braket{\sigma_y}_P)^{2\alpha-k} = \input{tn_21.tikz}.\label{eq:full_tn1}
\end{equation}
This is identical to Eq.~\eqref{eq:full_tn} except we have an extra $\sigma_z$ gate ($\pi-$rotation $Z-$spider) on the top wire of the final $2\alpha -k$ copies in replica space. We can apply identical techniques to the previous section, except that the top wire of each replica space has the diagram (c.f. Eq.~\eqref{eq:tn_4}):
\begin{equation}
        \begin{tikzpicture}
	\begin{pgfonlayer}{nodelayer}
		\node [style=none] (38) at (0, 1.25) {...};
		\node [style=none] (63) at (0, -2) {...};
		\node [style=white dot] (210) at (0, 2.75) {};
		\node [style=gray dot] (211) at (-3.25, 1.25) {};
		\node [style=gray dot] (212) at (-1.75, 1.25) {};
		\node [style=gray dot] (213) at (1.75, 1.25) {};
		\node [style=gray dot] (214) at (3.25, 1.25) {};
		\node [style=white dot] (215) at (-3.25, -1.25) {};
		\node [style=white dot] (216) at (-1.75, -1.25) {};
		\node [style=white dot] (217) at (1.75, -1.25) {};
		\node [style=white dot] (218) at (3.25, -1.25) {};
		\node [style=gray dot] (219) at (0, -2.75) {};
		\node [style=white dot] (228) at (-3.75, 0) {};
		\node [style=white dot] (229) at (-2.25, 0) {};
		\node [style=white dot] (230) at (1.25, 0) {};
		\node [style=white dot] (231) at (2.75, 0) {};
		\node [style=none] (232) at (-4.25, -0.25) {};
		\node [style=none] (233) at (-3, -0.25) {};
		\node [style=none] (234) at (0.5, -0.25) {};
		\node [style=none] (235) at (2, -0.25) {};
		\node [style=gray dot] (236) at (-3.5, 0.75) {$\pi$};
		\node [style=gray dot] (237) at (1.5, 0.75) {$\pi$};
		\node [style=gray dot] (238) at (3, 0.75) {$\pi$};
		\node [style=gray dot] (239) at (-2, 0.75) {$\pi$};
		\node [style=white dot] (240) at (1.5, -0.75) {$\pi$};
		\node [style=white dot] (241) at (3, -0.75) {$\pi$};
		\node [style=none] (242) at (-3.5, -2.75) {};
		\node [style=none] (243) at (-1.5, -2.75) {};
		\node [style=none] (244) at (-2.5, -3.25) {$k$};
		\node [style=none] (245) at (0.75, -2.75) {};
		\node [style=none] (246) at (3.75, -2.75) {};
		\node [style=none] (247) at (2.25, -3.25) {$2\alpha-k$};
	\end{pgfonlayer}
	\begin{pgfonlayer}{edgelayer}
		\draw [style=simple] (219) to (217);
		\draw [style=simple] (219) to (218);
		\draw [style=simple] (216) to (219);
		\draw [style=simple] (215) to (219);
		\draw [style=simple] (210) to (211);
		\draw [style=simple] (210) to (212);
		\draw [style=simple] (210) to (213);
		\draw [style=simple] (210) to (214);
		\draw [style=simple, bend right, looseness=1.25] (211) to (215);
		\draw [style=simple, bend left] (211) to (215);
		\draw [style=simple, bend right, looseness=1.25] (212) to (216);
		\draw [style=simple, bend left] (212) to (216);
		\draw [style=simple, bend right, looseness=1.25] (213) to (217);
		\draw [style=simple, bend left] (213) to (217);
		\draw [style=simple, bend right, looseness=1.25] (214) to (218);
		\draw [style=simple, bend left] (214) to (218);
		\draw [style=not edge] (232.center) to (228);
		\draw [style=not edge] (233.center) to (229);
		\draw [style=not edge] (234.center) to (230);
		\draw [style=not edge] (235.center) to (231);
		\draw [style=brace edge] (243.center) to (242.center);
		\draw [style=brace edge] (246.center) to (245.center);
	\end{pgfonlayer}
\end{tikzpicture} \overset{(f),(H),(c)}{=}  \begin{tikzpicture}
	\begin{pgfonlayer}{nodelayer}
		\node [style=gray dot] (258) at (2.75, 0) {};
		\node [style=white dot] (260) at (0, 1.75) {};
		\node [style=white dot] (261) at (0, 0.75) {};
		\node [style=white dot] (262) at (0, -1) {};
		\node [style=white dot] (263) at (0, -2) {};
		\node [style=none, rotate=90] (264) at (0, 0) {...};
		\node [style=none] (269) at (-1, 1.75) {};
		\node [style=none] (270) at (-1, 0.75) {};
		\node [style=none] (271) at (-1, -1) {};
		\node [style=none] (272) at (-1, -2) {};
		\node [style=white dot] (275) at (0.75, -0.75) {$\pi$};
		\node [style=white dot] (276) at (0.75, -1.5) {$\pi$};
		\node [style=none] (277) at (-1.5, 2) {};
		\node [style=none] (278) at (-1.5, 0.5) {};
		\node [style=none] (279) at (-2, 1.25) {$k$};
		\node [style=none] (280) at (-1.5, -0.5) {};
		\node [style=none] (281) at (-1.5, -2.25) {};
		\node [style=none] (282) at (-2.75, -1.25) {$2\alpha-k$};
	\end{pgfonlayer}
	\begin{pgfonlayer}{edgelayer}
		\draw [style=simple] (258) to (261);
		\draw [style=simple] (258) to (260);
		\draw [style=simple] (258) to (262);
		\draw [style=simple] (258) to (263);
		\draw [style=not edge] (269.center) to (260);
		\draw [style=not edge] (270.center) to (261);
		\draw [style=not edge] (271.center) to (262);
		\draw [style=not edge] (272.center) to (263);
		\draw [style=brace edge] (278.center) to (277.center);
		\draw [style=brace edge] (281.center) to (280.center);
	\end{pgfonlayer}
\end{tikzpicture}.
\end{equation}
Subbing this expression and Eq.~\eqref{eq:tn_3} into Eq.~\eqref{eq:full_tn1}
\begin{equation}
\begin{split}
    \frac{1}{2^{4 \alpha t}} \sum_{P \in \mathcal{P}_N} (\braket{\sigma_x}_P)^{k} (\braket{\sigma_y}_P)^{2\alpha-k} &= \begin{tikzpicture}
	\begin{pgfonlayer}{nodelayer}
		\node [style=gray dot] (258) at (4, -0.25) {};
		\node [style=white dot] (260) at (1.25, 1.75) {};
		\node [style=white dot] (261) at (1.25, 0.75) {};
		\node [style=white dot] (262) at (1.25, -1.25) {};
		\node [style=white dot] (263) at (1.25, -2.25) {};
		\node [style=none, rotate=90] (264) at (1.25, -0.25) {...};
		\node [style=gray dot] (280) at (-1.25, 3.5) {};
		\node [style=gray dot] (281) at (-1.25, 1.5) {};
		\node [style=gray dot] (282) at (-1.25, -1) {};
		\node [style=gray dot] (283) at (-1.25, -3) {};
		\node [style=none] (284) at (-1, 0.25) {$t$};
		\node [style=none, rotate=90] (285) at (-1.25, 0.25) {...};
		\node [style=white dot] (286) at (-1, 4.75) {$\theta$};
		\node [style=white dot] (288) at (-1.75, 4.75) {$\bar{\theta}$};
		\node [style=white dot] (289) at (-2.25, 4.25) {$\theta$};
		\node [style=white dot] (290) at (-2.5, 3.5) {$\bar{\theta}$};
		\node [style=white dot] (291) at (-2.5, 2.5) {$\theta$};
		\node [style=white dot] (292) at (-2.5, 2) {$\bar{\theta}$};
		\node [style=white dot] (293) at (-2.5, 1.5) {$\theta$};
		\node [style=white dot] (294) at (-2.5, 1) {$\bar{\theta}$};
		\node [style=white dot] (295) at (-2.5, -0.5) {$\theta$};
		\node [style=white dot] (296) at (-2.5, -1) {$\bar{\theta}$};
		\node [style=white dot] (297) at (-2.5, -1.5) {$\theta$};
		\node [style=white dot] (298) at (-2.5, -2) {$\bar{\theta}$};
		\node [style=white dot] (299) at (-2.5, -3) {$\theta$};
		\node [style=white dot] (300) at (-2.25, -3.75) {$\bar{\theta}$};
		\node [style=white dot] (301) at (-1.75, -4.25) {$\theta$};
		\node [style=white dot] (302) at (-1, -4.25) {$\bar{\theta}$};
		\node [style=none] (303) at (-3.75, 1.75) {$2\alpha$};
		\node [style=none] (304) at (-3.25, 2.75) {};
		\node [style=none] (305) at (-3.25, 0.75) {};
		\node [style=none] (306) at (-3.75, -1.25) {$2\alpha$};
		\node [style=none] (307) at (-3.25, -0.25) {};
		\node [style=none] (308) at (-3.25, -2.25) {};
		\node [style=none] (309) at (-3, -4.75) {$2\alpha$};
		\node [style=none] (310) at (-3.25, -3.5) {};
		\node [style=none] (311) at (-1.75, -5.25) {};
		\node [style=none] (312) at (-3.25, 4.75) {$2\alpha$};
		\node [style=none] (313) at (-2, 5.5) {};
		\node [style=none] (314) at (-3.25, 3.75) {};
		\node [style=white dot] (315) at (2, -1) {$\pi$};
		\node [style=white dot] (316) at (2.25, -1.5) {$\pi$};
		\node [style=none] (317) at (5, 0.75) {$k$};
		\node [style=none] (318) at (4.5, 1.5) {};
		\node [style=none] (319) at (4.5, 0.25) {};
		\node [style=none] (320) at (5.75, -1.25) {$2\alpha-k$};
		\node [style=none] (321) at (4.5, -0.5) {};
		\node [style=none] (322) at (4.5, -2) {};
	\end{pgfonlayer}
	\begin{pgfonlayer}{edgelayer}
		\draw [style=simple] (258) to (261);
		\draw [style=simple] (258) to (260);
		\draw [style=simple] (258) to (262);
		\draw [style=simple] (258) to (263);
		\draw [style=simple] (280) to (260);
		\draw [style=simple] (280) to (261);
		\draw [style=simple] (280) to (262);
		\draw [style=simple] (280) to (263);
		\draw [style=simple] (281) to (260);
		\draw [style=simple] (281) to (261);
		\draw [style=simple] (281) to (262);
		\draw [style=simple] (281) to (263);
		\draw [style=simple] (282) to (260);
		\draw [style=simple] (282) to (261);
		\draw [style=simple] (282) to (262);
		\draw [style=simple] (282) to (263);
		\draw [style=simple] (283) to (260);
		\draw [style=simple] (283) to (261);
		\draw [style=simple] (283) to (262);
		\draw [style=simple] (283) to (263);
		\draw [style=simple] (286) to (280);
		\draw [style=simple] (288) to (280);
		\draw [style=simple] (289) to (280);
		\draw [style=simple] (290) to (280);
		\draw [style=simple] (291) to (281);
		\draw [style=simple] (292) to (281);
		\draw [style=simple] (293) to (281);
		\draw [style=simple] (294) to (281);
		\draw [style=simple] (282) to (295);
		\draw [style=simple] (282) to (296);
		\draw [style=simple] (282) to (297);
		\draw [style=simple] (282) to (298);
		\draw [style=simple] (283) to (299);
		\draw [style=simple] (283) to (300);
		\draw [style=simple] (283) to (301);
		\draw [style=simple] (283) to (302);
		\draw [style=brace edge] (305.center) to (304.center);
		\draw [style=brace edge] (308.center) to (307.center);
		\draw [style=brace edge] (311.center) to (310.center);
		\draw [style=brace edge] (314.center) to (313.center);
		\draw [style=brace edge] (318.center) to (319.center);
		\draw [style=brace edge] (321.center) to (322.center);
	\end{pgfonlayer}
\end{tikzpicture} \overset{(b)}{=} \begin{tikzpicture}
	\begin{pgfonlayer}{nodelayer}
		\node [style=gray dot] (305) at (4.25, 0) {};
		\node [style=white dot] (306) at (1.25, 0) {};
		\node [style=none, rotate=90] (308) at (2.75, -2.25) {...};
		\node [style=gray dot] (309) at (0, 0) {};
		\node [style=gray dot] (352) at (1.25, 0) {};
		\node [style=white dot] (353) at (0, 0) {};
		\node [style=gray dot] (354) at (-1.5, 3.25) {};
		\node [style=gray dot] (355) at (-1.5, 1.25) {};
		\node [style=gray dot] (356) at (-1.5, -1.25) {};
		\node [style=gray dot] (357) at (-1.5, -3.25) {};
		\node [style=none] (358) at (-1.25, 0) {$t$};
		\node [style=none, rotate=90] (359) at (-1.5, 0) {...};
		\node [style=white dot] (360) at (-1.25, 4.5) {$\theta$};
		\node [style=white dot] (361) at (-2, 4.5) {$\bar{\theta}$};
		\node [style=white dot] (362) at (-2.5, 4) {$\theta$};
		\node [style=white dot] (363) at (-2.75, 3.25) {$\bar{\theta}$};
		\node [style=white dot] (364) at (-2.75, 2.25) {$\theta$};
		\node [style=white dot] (365) at (-2.75, 1.75) {$\bar{\theta}$};
		\node [style=white dot] (366) at (-2.75, 1.25) {$\theta$};
		\node [style=white dot] (367) at (-2.75, 0.75) {$\bar{\theta}$};
		\node [style=white dot] (368) at (-2.75, -0.75) {$\theta$};
		\node [style=white dot] (369) at (-2.75, -1.25) {$\bar{\theta}$};
		\node [style=white dot] (370) at (-2.75, -1.75) {$\theta$};
		\node [style=white dot] (371) at (-2.75, -2.25) {$\bar{\theta}$};
		\node [style=white dot] (372) at (-2.75, -3.25) {$\theta$};
		\node [style=white dot] (373) at (-2.5, -4) {$\bar{\theta}$};
		\node [style=white dot] (374) at (-2, -4.5) {$\theta$};
		\node [style=white dot] (375) at (-1.25, -4.5) {$\bar{\theta}$};
		\node [style=none] (376) at (-4, 1.5) {$2\alpha$};
		\node [style=none] (377) at (-3.5, 2.5) {};
		\node [style=none] (378) at (-3.5, 0.5) {};
		\node [style=none] (379) at (-4, -1.5) {$2\alpha$};
		\node [style=none] (380) at (-3.5, -0.5) {};
		\node [style=none] (381) at (-3.5, -2.5) {};
		\node [style=none] (382) at (-3.25, -5) {$2\alpha$};
		\node [style=none] (383) at (-3.5, -3.75) {};
		\node [style=none] (384) at (-2, -5.5) {};
		\node [style=none] (385) at (-3.5, 4.5) {$2\alpha$};
		\node [style=none] (386) at (-2.25, 5.25) {};
		\node [style=none] (387) at (-3.5, 3.5) {};
		\node [style=white dot] (388) at (2.75, -1) {$\pi$};
		\node [style=white dot] (389) at (2.75, -3) {$\pi$};
		\node [style=none] (390) at (2.75, -1.5) {$2\alpha-k$};
		\node [style=none, rotate=90] (391) at (2.75, 1.25) {...};
		\node [style=none] (392) at (2.75, 2) {$k$};
	\end{pgfonlayer}
	\begin{pgfonlayer}{edgelayer}
		\draw [style=simple, bend left=90, looseness=2.75] (306) to (305);
		\draw [style=simple, in=-90, out=-90, looseness=3.25] (306) to (305);
		\draw [style=simple, in=135, out=45, looseness=1.25] (306) to (305);
		\draw [style=simple, bend right=45, looseness=1.50] (306) to (305);
		\draw [style=simple] (306) to (309);
		\draw [style=simple] (360) to (354);
		\draw [style=simple] (361) to (354);
		\draw [style=simple] (362) to (354);
		\draw [style=simple] (363) to (354);
		\draw [style=simple] (364) to (355);
		\draw [style=simple] (365) to (355);
		\draw [style=simple] (366) to (355);
		\draw [style=simple] (367) to (355);
		\draw [style=simple] (356) to (368);
		\draw [style=simple] (356) to (369);
		\draw [style=simple] (356) to (370);
		\draw [style=simple] (356) to (371);
		\draw [style=simple] (357) to (372);
		\draw [style=simple] (357) to (373);
		\draw [style=simple] (357) to (374);
		\draw [style=simple] (357) to (375);
		\draw [style=simple] (354) to (353);
		\draw [style=simple] (355) to (353);
		\draw [style=simple] (356) to (353);
		\draw [style=simple] (357) to (353);
		\draw [style=brace edge] (378.center) to (377.center);
		\draw [style=brace edge] (381.center) to (380.center);
		\draw [style=brace edge] (384.center) to (383.center);
		\draw [style=brace edge] (387.center) to (386.center);
	\end{pgfonlayer}
\end{tikzpicture} \\
    &\overset{(f),(\pi)}{=} \begin{tikzpicture}
	\begin{pgfonlayer}{nodelayer}
		\node [style=gray dot] (305) at (4.25, 0) {};
		\node [style=white dot] (306) at (1.25, 0) {};
		\node [style=gray dot] (309) at (0, 0) {};
		\node [style=gray dot] (352) at (1.25, 0) {};
		\node [style=white dot] (353) at (0, 0) {};
		\node [style=gray dot] (354) at (-1.5, 3.25) {};
		\node [style=gray dot] (355) at (-1.5, 1.25) {};
		\node [style=gray dot] (356) at (-1.5, -1.25) {};
		\node [style=gray dot] (357) at (-1.5, -3.25) {};
		\node [style=none] (358) at (-1.25, 0) {$t$};
		\node [style=none, rotate=90] (359) at (-1.5, 0) {...};
		\node [style=white dot] (360) at (-1.25, 4.5) {$\theta$};
		\node [style=white dot] (361) at (-2, 4.5) {$\bar{\theta}$};
		\node [style=white dot] (362) at (-2.5, 4) {$\theta$};
		\node [style=white dot] (363) at (-2.75, 3.25) {$\bar{\theta}$};
		\node [style=white dot] (364) at (-2.75, 2.25) {$\theta$};
		\node [style=white dot] (365) at (-2.75, 1.75) {$\bar{\theta}$};
		\node [style=white dot] (366) at (-2.75, 1.25) {$\theta$};
		\node [style=white dot] (367) at (-2.75, 0.75) {$\bar{\theta}$};
		\node [style=white dot] (368) at (-2.75, -0.75) {$\theta$};
		\node [style=white dot] (369) at (-2.75, -1.25) {$\bar{\theta}$};
		\node [style=white dot] (370) at (-2.75, -1.75) {$\theta$};
		\node [style=white dot] (371) at (-2.75, -2.25) {$\bar{\theta}$};
		\node [style=white dot] (372) at (-2.75, -3.25) {$\theta$};
		\node [style=white dot] (373) at (-2.5, -4) {$\bar{\theta}$};
		\node [style=white dot] (374) at (-2, -4.5) {$\theta$};
		\node [style=white dot] (375) at (-1.25, -4.5) {$\bar{\theta}$};
		\node [style=none] (376) at (-4, 1.5) {$2\alpha$};
		\node [style=none] (377) at (-3.5, 2.5) {};
		\node [style=none] (378) at (-3.5, 0.5) {};
		\node [style=none] (379) at (-4, -1.5) {$2\alpha$};
		\node [style=none] (380) at (-3.5, -0.5) {};
		\node [style=none] (381) at (-3.5, -2.5) {};
		\node [style=none] (382) at (-3.25, -5) {$2\alpha$};
		\node [style=none] (383) at (-3.5, -3.75) {};
		\node [style=none] (384) at (-2, -5.5) {};
		\node [style=none] (385) at (-3.5, 4.5) {$2\alpha$};
		\node [style=none] (386) at (-2.25, 5.25) {};
		\node [style=none] (387) at (-3.5, 3.5) {};
		\node [style=white dot] (388) at (2.75, 1) {$\pi$};
	\end{pgfonlayer}
	\begin{pgfonlayer}{edgelayer}
		\draw [style=simple, in=120, out=60, looseness=1.25] (306) to (305);
		\draw [style=simple, bend right=60, looseness=1.25] (306) to (305);
		\draw [style=simple] (306) to (309);
		\draw [style=simple] (360) to (354);
		\draw [style=simple] (361) to (354);
		\draw [style=simple] (362) to (354);
		\draw [style=simple] (363) to (354);
		\draw [style=simple] (364) to (355);
		\draw [style=simple] (365) to (355);
		\draw [style=simple] (366) to (355);
		\draw [style=simple] (367) to (355);
		\draw [style=simple] (356) to (368);
		\draw [style=simple] (356) to (369);
		\draw [style=simple] (356) to (370);
		\draw [style=simple] (356) to (371);
		\draw [style=simple] (357) to (372);
		\draw [style=simple] (357) to (373);
		\draw [style=simple] (357) to (374);
		\draw [style=simple] (357) to (375);
		\draw [style=simple] (354) to (353);
		\draw [style=simple] (355) to (353);
		\draw [style=simple] (356) to (353);
		\draw [style=simple] (357) to (353);
		\draw [style=brace edge] (378.center) to (377.center);
		\draw [style=brace edge] (381.center) to (380.center);
		\draw [style=brace edge] (384.center) to (383.center);
		\draw [style=brace edge] (387.center) to (386.center);
	\end{pgfonlayer}
\end{tikzpicture} \overset{(id),(f)}{=}  \begin{tikzpicture}
	\begin{pgfonlayer}{nodelayer}
		\node [style=white dot] (306) at (1.25, 0) {};
		\node [style=gray dot] (309) at (0, 0) {};
		\node [style=gray dot] (352) at (1.25, 0) {};
		\node [style=white dot] (353) at (0, 0) {};
		\node [style=gray dot] (354) at (-1.5, 3.25) {};
		\node [style=gray dot] (355) at (-1.5, 1.25) {};
		\node [style=gray dot] (356) at (-1.5, -1.25) {};
		\node [style=gray dot] (357) at (-1.5, -3.25) {};
		\node [style=none] (358) at (-1.25, 0) {$t$};
		\node [style=none, rotate=90] (359) at (-1.5, 0) {...};
		\node [style=white dot] (360) at (-1.25, 4.5) {$\theta$};
		\node [style=white dot] (361) at (-2, 4.5) {$\bar{\theta}$};
		\node [style=white dot] (362) at (-2.5, 4) {$\theta$};
		\node [style=white dot] (363) at (-2.75, 3.25) {$\bar{\theta}$};
		\node [style=white dot] (364) at (-2.75, 2.25) {$\theta$};
		\node [style=white dot] (365) at (-2.75, 1.75) {$\bar{\theta}$};
		\node [style=white dot] (366) at (-2.75, 1.25) {$\theta$};
		\node [style=white dot] (367) at (-2.75, 0.75) {$\bar{\theta}$};
		\node [style=white dot] (368) at (-2.75, -0.75) {$\theta$};
		\node [style=white dot] (369) at (-2.75, -1.25) {$\bar{\theta}$};
		\node [style=white dot] (370) at (-2.75, -1.75) {$\theta$};
		\node [style=white dot] (371) at (-2.75, -2.25) {$\bar{\theta}$};
		\node [style=white dot] (372) at (-2.75, -3.25) {$\theta$};
		\node [style=white dot] (373) at (-2.5, -4) {$\bar{\theta}$};
		\node [style=white dot] (374) at (-2, -4.5) {$\theta$};
		\node [style=white dot] (375) at (-1.25, -4.5) {$\bar{\theta}$};
		\node [style=none] (376) at (-4, 1.5) {$2\alpha$};
		\node [style=none] (377) at (-3.5, 2.5) {};
		\node [style=none] (378) at (-3.5, 0.5) {};
		\node [style=none] (379) at (-4, -1.5) {$2\alpha$};
		\node [style=none] (380) at (-3.5, -0.5) {};
		\node [style=none] (381) at (-3.5, -2.5) {};
		\node [style=none] (382) at (-3.25, -5) {$2\alpha$};
		\node [style=none] (383) at (-3.5, -3.75) {};
		\node [style=none] (384) at (-2, -5.5) {};
		\node [style=none] (385) at (-3.5, 4.5) {$2\alpha$};
		\node [style=none] (386) at (-2.25, 5.25) {};
		\node [style=none] (387) at (-3.5, 3.5) {};
		\node [style=white dot] (388) at (4, 0) {$\pi$};
		\node [style=white dot] (389) at (3, 0) {};
	\end{pgfonlayer}
	\begin{pgfonlayer}{edgelayer}
		\draw [style=simple] (306) to (309);
		\draw [style=simple] (360) to (354);
		\draw [style=simple] (361) to (354);
		\draw [style=simple] (362) to (354);
		\draw [style=simple] (363) to (354);
		\draw [style=simple] (364) to (355);
		\draw [style=simple] (365) to (355);
		\draw [style=simple] (366) to (355);
		\draw [style=simple] (367) to (355);
		\draw [style=simple] (356) to (368);
		\draw [style=simple] (356) to (369);
		\draw [style=simple] (356) to (370);
		\draw [style=simple] (356) to (371);
		\draw [style=simple] (357) to (372);
		\draw [style=simple] (357) to (373);
		\draw [style=simple] (357) to (374);
		\draw [style=simple] (357) to (375);
		\draw [style=simple] (354) to (353);
		\draw [style=simple] (355) to (353);
		\draw [style=simple] (356) to (353);
		\draw [style=simple] (357) to (353);
		\draw [style=brace edge] (378.center) to (377.center);
		\draw [style=brace edge] (381.center) to (380.center);
		\draw [style=brace edge] (384.center) to (383.center);
		\draw [style=brace edge] (387.center) to (386.center);
		\draw [bend left=45, looseness=1.25] (352) to (389);
		\draw [bend right=45, looseness=1.25] (352) to (389);
		\draw (389) to (388);
	\end{pgfonlayer}
\end{tikzpicture} \\
    &\overset{(H)}{=}  \begin{tikzpicture}
	\begin{pgfonlayer}{nodelayer}
		\node [style=white dot] (306) at (1.25, 0) {};
		\node [style=gray dot] (309) at (0, 0) {};
		\node [style=gray dot] (352) at (1.25, 0) {};
		\node [style=white dot] (353) at (0, 0) {};
		\node [style=gray dot] (354) at (-1.5, 3.25) {};
		\node [style=gray dot] (355) at (-1.5, 1.25) {};
		\node [style=gray dot] (356) at (-1.5, -1.25) {};
		\node [style=gray dot] (357) at (-1.5, -3.25) {};
		\node [style=none] (358) at (-1.25, 0) {$t$};
		\node [style=none, rotate=90] (359) at (-1.5, 0) {...};
		\node [style=white dot] (360) at (-1.25, 4.5) {$\theta$};
		\node [style=white dot] (361) at (-2, 4.5) {$\bar{\theta}$};
		\node [style=white dot] (362) at (-2.5, 4) {$\theta$};
		\node [style=white dot] (363) at (-2.75, 3.25) {$\bar{\theta}$};
		\node [style=white dot] (364) at (-2.75, 2.25) {$\theta$};
		\node [style=white dot] (365) at (-2.75, 1.75) {$\bar{\theta}$};
		\node [style=white dot] (366) at (-2.75, 1.25) {$\theta$};
		\node [style=white dot] (367) at (-2.75, 0.75) {$\bar{\theta}$};
		\node [style=white dot] (368) at (-2.75, -0.75) {$\theta$};
		\node [style=white dot] (369) at (-2.75, -1.25) {$\bar{\theta}$};
		\node [style=white dot] (370) at (-2.75, -1.75) {$\theta$};
		\node [style=white dot] (371) at (-2.75, -2.25) {$\bar{\theta}$};
		\node [style=white dot] (372) at (-2.75, -3.25) {$\theta$};
		\node [style=white dot] (373) at (-2.5, -4) {$\bar{\theta}$};
		\node [style=white dot] (374) at (-2, -4.5) {$\theta$};
		\node [style=white dot] (375) at (-1.25, -4.5) {$\bar{\theta}$};
		\node [style=none] (376) at (-4, 1.5) {$2\alpha$};
		\node [style=none] (377) at (-3.5, 2.5) {};
		\node [style=none] (378) at (-3.5, 0.5) {};
		\node [style=none] (379) at (-4, -1.5) {$2\alpha$};
		\node [style=none] (380) at (-3.5, -0.5) {};
		\node [style=none] (381) at (-3.5, -2.5) {};
		\node [style=none] (382) at (-3.25, -5) {$2\alpha$};
		\node [style=none] (383) at (-3.5, -3.75) {};
		\node [style=none] (384) at (-2, -5.5) {};
		\node [style=none] (385) at (-3.5, 4.5) {$2\alpha$};
		\node [style=none] (386) at (-2.25, 5.25) {};
		\node [style=none] (387) at (-3.5, 3.5) {};
		\node [style=white dot] (388) at (3.25, 0) {$\pi$};
		\node [style=white dot] (389) at (2.25, 0) {};
	\end{pgfonlayer}
	\begin{pgfonlayer}{edgelayer}
		\draw [style=simple] (306) to (309);
		\draw [style=simple] (360) to (354);
		\draw [style=simple] (361) to (354);
		\draw [style=simple] (362) to (354);
		\draw [style=simple] (363) to (354);
		\draw [style=simple] (364) to (355);
		\draw [style=simple] (365) to (355);
		\draw [style=simple] (366) to (355);
		\draw [style=simple] (367) to (355);
		\draw [style=simple] (356) to (368);
		\draw [style=simple] (356) to (369);
		\draw [style=simple] (356) to (370);
		\draw [style=simple] (356) to (371);
		\draw [style=simple] (357) to (372);
		\draw [style=simple] (357) to (373);
		\draw [style=simple] (357) to (374);
		\draw [style=simple] (357) to (375);
		\draw [style=simple] (354) to (353);
		\draw [style=simple] (355) to (353);
		\draw [style=simple] (356) to (353);
		\draw [style=simple] (357) to (353);
		\draw [style=brace edge] (378.center) to (377.center);
		\draw [style=brace edge] (381.center) to (380.center);
		\draw [style=brace edge] (384.center) to (383.center);
		\draw [style=brace edge] (387.center) to (386.center);
		\draw (389) to (388);
	\end{pgfonlayer}
\end{tikzpicture} = 0. \label{eq:hey}
\end{split}
\end{equation}
Here, we have followed a similar method to Eq.~\eqref{eq:bigone}, and in the end used that 
\begin{equation}
    \begin{tikzpicture}
	\begin{pgfonlayer}{nodelayer}
		\node [style=white dot] (0) at (0.5, 0) {$\pi$};
		\node [style=white dot] (1) at (-0.5, 0) {};
	\end{pgfonlayer}
	\begin{pgfonlayer}{edgelayer}
		\draw (1) to (0);
	\end{pgfonlayer}
\end{tikzpicture}
 = (\bra{0} + \bra{1})(\ket{0} - \ket{1}) = 0. 
\end{equation}
Notice that the proof of Eq.~\eqref{eq:hey} did not depend on the value of $k \neq 0,2\alpha$. Therefore, we arrive at the final expression of 
\begin{equation}
     P^{(\alpha)}(U^\dagger_t O_j U_t)= a_z^{2\alpha} + (a_x^{2\alpha} +a_y^{2\alpha})\left( \cos^{2 \alpha }(2 J) + \sin^{2\alpha }(2 J)\right)^t.
\end{equation}
{Therefore, the final OSE is equal to
\begin{equation}
     \Mpop^{(\alpha)}(U^\dagger_t O_j U_t)= \frac{1}{1-\alpha} \left( \log \left( a_z^{2\alpha} + (a_x^{2\alpha} +a_y^{2\alpha})\left( \cos^{2 \alpha }(2 J) + \sin^{2\alpha }(2 J)\right)^t \right)  \right),\label{eq:xxz_final}
\end{equation}
where we keep in mind that $a_z^2=1-a_x^2-a_y^2$ (see below Eq.~\eqref{eq:init_op}).
Then for $0 < a_z < 1$, $ \Mpop^{(\alpha)}(U^\dagger_t O_j U_t)$ is an approximately linearly growing function of $t$ before saturating to the constant,
\begin{equation}
    \lim_{t \to \infty}  \Mpop^{(\alpha)}(U^\dagger_t O_j U_t) = \frac{\alpha \log(a_z^2)}{1-\alpha} \label{eq:asymptotic}
\end{equation}}
See Fig.~\ref{fig:numerics} for a plot of Eq.~\eqref{eq:xxz_final} for example parameters. 

{We can also take the replica limit of Eq.~\eqref{eq:xxz_final} to find the Shannon entropy OSE 
\begin{align}
    \lim_{\alpha \to 1}\Mpop^{(\alpha)}(O_U) =& -2 \Big(a_x^2 \log(a_x) + a_y^2 \log(a_y) + a_z^2 \log(a_z)  \\
    &+ t  (a_x^2 + a_y^2) (\cos(2 J)^2 \log(\cos(2 J)) + \log(\sin(2 J)) \sin(2 J)^2)\Big) \\
    =&c(J) (a_x^2 + a_y^2) t + d(a_x,a_y)
\end{align}
For $J\neq0,\frac{\pi}{4}$ this is a well-defined linearly increasing function,  with gradient $0<c(J) \leq 1$ and constant offset $d(a_x,a_y)$. Note that the maximal growth is achieved for $J=\pi/8$ in which case $c(\pi/8)=1) $.}

\end{document}